\documentclass[a4paper,fleqn,usenatbib]{mnras}

\usepackage{mathptmx}
\usepackage[T1, T4]{fontenc}
\usepackage{ae,aecompl}

\usepackage{graphicx}	% Including figure files
\usepackage{amsmath}	% Advanced maths commands
\usepackage{amssymb}	% Extra maths symbols
\usepackage{textcomp}
\usepackage{color}
\usepackage{multicol}
\usepackage{longtable}
\usepackage{rotating}
\usepackage{pdflscape}
\usepackage{natbib}
\usepackage{enumitem}
\usepackage{appendix}
\usepackage{url}
\usepackage{float}
\usepackage{tabularx}
\usepackage{booktabs}
\usepackage[caption = false]{subfig}

\title[Merger histories of brightest group galaxies]{Merger histories of brightest group galaxies from MUSE stellar kinematics}

\author[Loubser et al.]{S. I. Loubser$^{1,2}$\thanks{E-mail:Ilani.Loubser@nwu.ac.za (SIL)}, P. Lagos$^{3,1}$, A. Babul$^{4}$, E. O'Sullivan$^{5}$, S. L. Jung$^{6}$, V. Olivares$^{7,8}$,
\newauthor{and K. Kolokythas$^{1}$}\\
$^{1}$Centre for Space Research, North-West University, Potchefstroom 2520, South Africa\\
$^{2}$National Institute for Theoretical and Computational Sciences (NITheCS), Potchefstroom 2520, South Africa\\
$^{3}$Instituto de Astrof\'isica e Ci\^encias do Espa\c{c}o, Universidade do Porto, CAUP, Rua das Estrelas, 4150-762 Porto, Portugal\\
$^{4}$Department of Physics and Astronomy, University of Victoria, Victoria, BC, V8W 2Y2, Canada\\
$^{5}$Center for Astrophysics | Harvard $\&$ Smithsonian, 60 Garden Street, Cambridge, MA 02138, USA\\
$^{6}$Research School of Astronomy $\&$ Astrophysics, Australian National University, Canberra, ACT 2611, Australia\\
$^{7}$LERMA, Observatoire de Paris, PSL Research Univ., CNRS, Sorbonne Univ., 75014 Paris, France\\
$^{8}$Department of Physics and Astronomy, University of Kentucky, 505 Rose Street, Lexington, KY 40506, USA}

\date{Accepted 2022 June 24. Received 2022 June 24; in original form 2022 March 11}

\pubyear{2022}

\begin{document}

\label{firstpage}
\pagerange{\pageref{firstpage}--\pageref{lastpage}}
\maketitle

\begin{abstract}
Using Multi-Unit Spectroscopic Explorer (MUSE) spectroscopy, we analyse the stellar kinematics of 18 brightest group early-type (BGEs) galaxies, selected from the Complete Local-Volume Groups Sample (CLoGS). We analyse the kinematic maps for distinct features, and measure specific stellar angular momentum within one effective radius ($\lambda_{e}$). We classify the BGEs as fast (10/18) or slow (8/18) rotators, suggesting at least two different evolution paths. We quantify the anti-correlation between higher-order kinematic moment $h_{3}$ and V/$\sigma$ (using the $\xi_{3}$ parameter), and the kinematic misalignment angle between the photometric and kinematic position angles (using the $\Psi$ parameter), and note clear differences between these parameter distributions of the fast and slow rotating BGEs. We find that all 10 of our fast rotators are aligned between the morphological and kinematical axis, consistent with an oblate galaxy shape, whereas the slow rotators are spread over all three classes: oblate (1/8), triaxial (4/8), and prolate (3/8). We place the results into context using known radio properties, X-ray properties, and observations of molecular gas. We find consistent merger histories inferred from observations for the fast-rotating BGEs, indicating that they experienced gas-rich mergers or interactions, and these are very likely the origin of the cold gas. Observational evidence for the slow rotators are consistent with gas-poor mergers. For the slow rotators with cold gas, all evidence point to cold gas cooling from the intragroup medium. 
\end{abstract}


%%%%%%%%%%%%%%%%% BODY OF PAPER %%%%%%%%%%%%%%%%%%

\section{Introduction}
\label{Section:introduction}

%Spatially-resolved kinematics.\\
Studying the build-up of stellar mass of galaxies is a pivotal part of our understanding of their evolutionary paths. This can be done through cosmological simulations (e.g.,\ \citealt{Springel2005, Delucia2007, Naab2014, Schaye2015, Oppenheimer2021}), or through studies of stellar dynamics in present-day galaxies to infer their individual assembly history (e.g., \citealt{Bender1994, Emsellem2011, Cappellari2016, VandeSande2017, FalconBarroso2017}). Detailed dynamical studies, in particular when combined with simulations, provide insight into the three-dimensional intrinsic shape of galaxies, which can not be determined through photometric studies alone. Disturbances in the stellar kinematics are also long-lived compared to merger signatures visible from imaging \citep{Cox2006, Glazebrook2013, Nevin2021}. Integral Field Spectroscopy (IFS) allows us to study the internal structure of galaxies in detail, as this presents the two-dimensional view of the stellar kinematics necessary to understand the assembly history and substructure of individual galaxies that may distinguish different formation and evolution scenarios (e.g. \citealt{Emsellem2007, Emsellem2011, Sanchez2012, Loubser2013, Fogarty2015, Mentz2016, Brough2017, Tsatsi2017, Krajnovic2018, Graham2018, Li2018obs, Pinna2019}, and many more).  

%Groups.\\
In particular, we focus on the central, brightest members of galaxy groups, how they evolve, and the corresponding signatures in the galaxy or group properties \citep{LeBrun2014, Liang2016, O'Sullivan2017, Oppenheimer2021, Jung2022}. With up to half of galaxies residing in groups in the local Universe \citep{Eke2004, Robotham2011}, we also need to understand the influence of the group environment on galaxy evolution. Because of their location at the bottom of their host halo's gravitational potential well, we expect brightest group galaxies to experience multiple mergers and tidal encounters with other group member galaxies over their evolution. The merger driven size growth (for both central group and cluster galaxies) is supported by the observed rate of mergers from galaxy pair counts and identified interacting galaxies (e.g.\ \citealt{Fakhouri2010, Groenewald2017, Banks2021}), as well as the signatures of recent accretion events in the halos of galaxies observed through deep imaging (e.g.\ \citealt{Mihos2005, Mancillas2019, Nevin2021, Yoon2022}). Simulations suggest that these interactions are responsible for the stellar mass assembly of the brightest group galaxies, and may also induce kinematic transformations (\citealt{Pillepich2018, Tacchella2019, Jackson2020, Jung2022}, also see review by \citealt{Oppenheimer2021}). In this paper, we study brightest group galaxies classified as early-type galaxies (Brightest Group Ellipticals, BGEs) from the Complete Local-Volume Groups Sample (CLoGS) sample (as described in detail below)\footnote{See \citet{O'Sullivan2017} for the detailed selection criteria of the CLoGS sample. Of particular interest for this paper is that the groups were selected to have an early-type galaxy as the central dominant galaxy, but were not required to be X-ray bright.}.

%Merger histories from kinematics
Early-type galaxies can be separated into two kinematic classes, slow and fast rotating galaxies, each class exhibiting very different dynamical properties suggesting that there are at least two evolution paths that give rise to this bimodality. A number of cosmological zoom-in simulations were developed to understand the mechanisms that result in the formation of slow and fast rotators, e.g.\ \citet{Bois2011, Khochfar2011, Naab2014, Penoyre2017, Lagos2018, Choi2018, Schulze2018, Pillepich2019, Frigo2019, WaloMartin2020, Pulsoni2020, Lagos2020}. Slow rotators, particularly the most massive galaxies, appear to be the evolutionary endpoint for galaxies that have ceased star formation and experienced at least one major dry merger (or multiple minor dry mergers, e.g.\ \citealt{Taranu2013}), while fast rotators represent an earlier stage of evolution. However, some studies find more complex formation mechanisms for massive early-type galaxies than the two unique formation histories suggested above (e.g.  \citealt{Jesseit2009}, \citealt{Bois2010}, \citealt{Naab2014}, \citealt{Jung2022}).

The different evolutionary histories resulting in slow and fast rotators are seemingly characterized by different numbers of mergers experienced, the merger mass ratio, timing, gas fractions, and the configuration of merger orbits \citep{Naab2014, Penoyre2017}. However, the details still do not agree, with the added complication that it also depends on the cooling and thermal stability of the gas surrounding the galaxies, as well as star formation and AGN feedback models adopted by the simulations \citep{Naab2017}. In addition to the details of the formation of slow and fast rotators, some other questions about the kinematic properties also remain, e.g.\ is the distribution of rotation bimodal (slow vs fast rotators) with different formation histories, or a continuous transition from one class into another (see discussion in \citealt{VandeSande2021a})? 

Another outstanding question is whether the same processes are responsible for the morphological (i.e.\ quenching of star formation) and kinematic transformation in galaxies. One common conclusion among simulations is that even if a slow rotator remnant is formed after a merger, continuous accretion and star formation can rebuild the galaxy disk and turn the galaxy into a fast rotator \citep{Naab2014, Penoyre2017, WaloMartin2020, Lagos2020}. This suggests that quenching either prior or during the kinematic transformation is required to more effectively decrease $\lambda$ and produce a slow rotator \citep{Lagos2020}. \citet{Lagos2020} find, in the EAGLE simulations, that in most slow rotators quenching precedes kinematic transformation by $\sim$2 Gyrs. Although they find several trends between different types of mergers and slow rotator kinematic properties, they could not identify a single galaxy property that can unambiguously indicate a given assembly history.

Furthermore intrinsic properties, that can not be measured directly, inferred from observed properties e.g.\ galaxy shape inversion from kinematic misalignment measurements, are often not unique \citep{Weijmans2014, Li2018sim, Bassett2019}. However, when combined with higher-order kinematics, as well as properties of the galaxies (and in our case also the properties of the group halo's) as observed in other wavelengths (e.g.\ X-ray or radio observations), we have a more complete view of galaxies' evolutionary paths. This is particularly important for the central galaxies in galaxy groups as there is no clear dominant transformation mechanism that galaxies are subjected to, but rather multiple secular and external mechanisms \citep{Lovisari2021, Kleiner2021}. Hence, we focus on a sample of BGEs selected from a survey of galaxy groups well-studied through multi-wavelength observations.  

%Advantage of CLoGS.\\
The Complete Local-Volume Groups Sample (CLoGS, \citealt{O'Sullivan2017}) is an optically-selected sample of 53 nearby groups for which X-ray (\textit{Chandra \& XMM-Newton}, \citealt{O'Sullivan2017}), radio (GMRT $\&$ VLA, \citealt{Kolokythas2018, Kolokythas2019}) and mm (IRAM-30m $\&$ APEX, \citealt{O'Sullivan2015, O'Sullivan2018}) data have been collected. Stellar kinematics from archival long-slit spectra from the Hobby-Ebberly Telescope for a sub-sample of 23 CLoGS BGEs were presented in \citet{Loubser2018}, and archival \textit{GALEX} and \textit{WISE} photometry for all 53 BGEs in \citet{Kolokythas2022}. This rich multi-wavelength dataset places us in a unique position to study various properties of BGEs that serve as signatures of their evolution and merger histories, and interpret the influence of the environment. 

%MUSE data. This paper.\\
This paper is part of a series analysing 18 Multi-Unit Spectroscopic Explorer \citep[MUSE;][]{Bacon2010} cubes of CLoGS BGEs. \citet{Olivares2022} present a detailed analysis of the ionised gas kinematics, \citet{Lagos2022} present an analysis of gas ionisation mechanisms and chemical abundances, and here we present detailed stellar kinematics. Dynamical modelling, and stellar population analyses will be presented in future papers. Section \ref{Section:data} contains a summary of the data and data reduction, and the details of the spatial binning and the stellar template fitting to derive the kinematics. Section \ref{features} contains the kinematic maps, and analysis of the distinct kinematical features. In Section \ref{lambdaR}, we measure the effective kinematic parameters, and classify the BGEs as fast or slow rotators. Section \ref{higherorder} presents the higher-order kinematics (including the $\xi_{3}$ parameter). Section \ref{Psi} presents the kinematic position angle and the misalignment with the photometric position angle ($\Psi$), as well as the misalignment between the stellar and gas kinematics. Section \ref{FormationScenarios} contains a discussion on the merger histories inferred from our stellar kinematic results, and combines our results with the multi-wavelength data, before we conclude in Section \ref{Con}.

\section{Data and analysis}
\label{Section:data}

\subsection{MUSE observations and data reduction}

The observations and data reduction are fully described in \citet{Lagos2022} and \citet{Olivares2022}, and we only summarise the most relevant information here. Our sample consists of 18 BGEs from the CLoGS sample \citep{O'Sullivan2017}, presented in Table \ref{table1}. We have three objects, NGC 410, NGC 777 and NGC 1060, in common with the MASSIVE sample \citep{Ma2014} and we compare some of our derived properties in Appendix \ref{Section:comparison}. The observations were made using the MUSE IFS, on the Very Large Telescope (VLT). The observations were made over a 1$'\times$1$'$ Field of View (FoV) sampled by 24 spectrographs with a spectral coverage between $\sim$4800 and 9300 \AA, a spectral resolution of $\sim$2.6 \AA\ and spatial sampling of 0.2$''$ per spaxel, leading to a total of $\sim$100,000 spectra per exposure. For each galaxy at least three exposures, with a few arcsecond dither pattern, were taken. The data were reduced using the MUSE pipeline (v2.6.2), and the data cubes were corrected for Galactic extinction using the reddening function given by \cite{Cardelli1989}.

\subsection{Photometric parameters}

We define the effective radius ($R_{e}$) similar to comparable studies e.g.\ \citet{Cappellari2011} (ATLAS3D) and \citet{Ma2014} (MASSIVE), based on the half-light radii from the 2MASS \citep{Skrutskie2006} extended source catalogue XSC (\citealt{Jarrett2000}, specifically the parameters j\textunderscore r\textunderscore eff, h\textunderscore r\textunderscore eff, and k\textunderscore r\textunderscore eff). This radius is derived from the 2MASS surface brightness profile in each of the three bands from the semi-major axis of the ellipse that encloses half of the total light. For $R_{e}$, we use the median value in the three bands\footnote{\citet{Ma2014} showed that $R_{e}$ calculated this way correlates well with $R_{e}$ derived from optical (SDSS DR8), although there is a systematic offset such that at $\sim$1 kpc, the optical $R_{e}$ is a factor of $\sim$1.2 larger than the 2MASS $R_{e}$ (their equation 4). We use the 2MASS $R_{e}$ to be able to compare to ATLAS3D \citep{Cappellari2011} and M3G \citep{Krajnovic2018}. Where we compare our derived values to MASSIVE in table \ref{table:comparison}, we use the 2MASS (and not SDSS) $R_{e}$ from \citet{Ma2014}.}: \begin{equation} R_{e} = \rm median(\text{j\textunderscore r\textunderscore eff,h\textunderscore r\textunderscore eff,k\textunderscore r\textunderscore eff})\sqrt{\text{sup\textunderscore ba}} \end{equation} where sup\textunderscore ba is the minor-to-major axis ratio measured from the 2MASS 3-band co-added image at the 3$\sigma$ isophote. With the exception of NGC 5846 (0.8$R_{e}$), all our maps cover at least 1.5$R_{e}$. 

\begin{figure*}
   \includegraphics[scale=0.64, angle=-90, trim=10 25 10 0, clip]{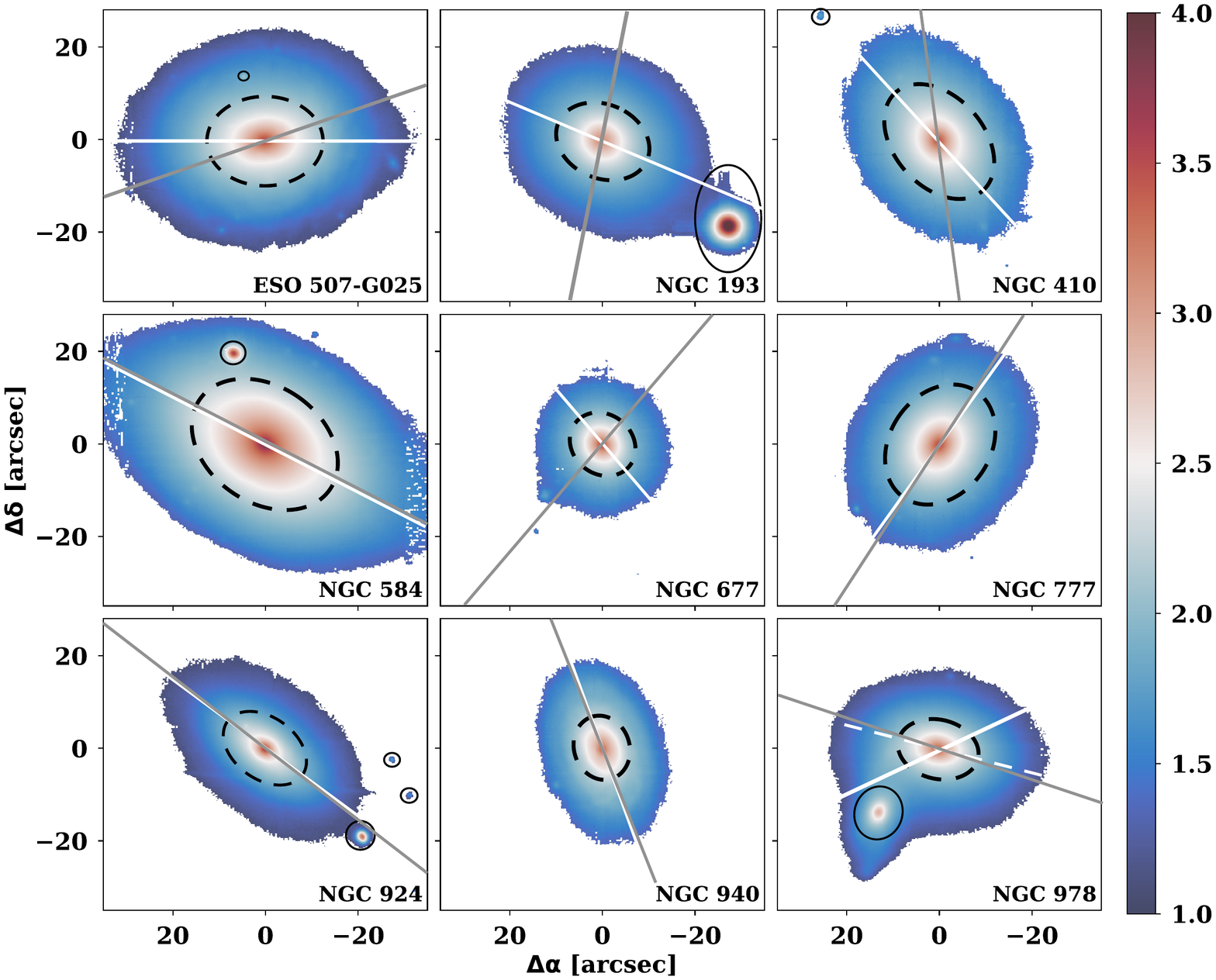}
   \caption{MUSE white-light images (showing log(flux) in arbitrary units). The PA$_{\rm phot}$ is anticlockwise (East of North). North is up, East is to the left. We indicate PA$_{\rm phot}$ with white lines (in NGC 978 we use the solid line for the 2MASS PA$_{\rm phot}$, and the dashed line for our PA$_{\rm phot}$). We use grey lines for the fitted PA$_{\rm kin}$ (see Section \ref{Psi}). We indicate $R_{e}$ with a black ellipse (dashed line). We indicate close companions or objects in the line-of-sight with a black solid line or ellipse.}
\label{fig:Flux}
\end{figure*}

\begin{figure*}
   \includegraphics[scale=0.64, angle=90, trim=10 25 10 0, clip]{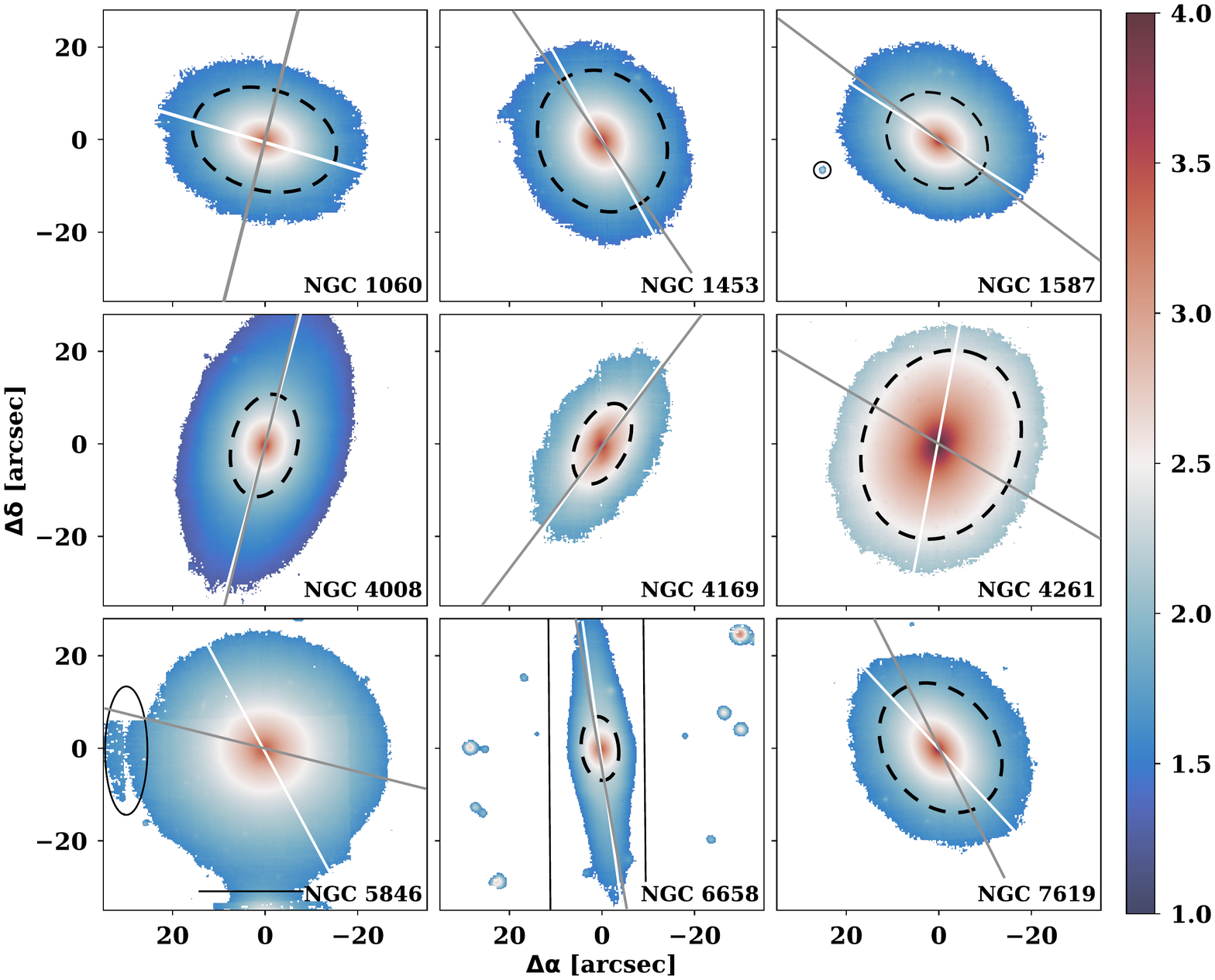}
   \contcaption{MUSE white-light images (showing log(flux) in arbitrary units). The PA$_{\rm phot}$ is anticlockwise (East of North). North is up, East is to the left. We indicate PA$_{\rm phot}$ with white lines. We use grey lines for the fitted PA$_{\rm kin}$ (see Section \ref{Psi}). We indicate $R_{e}$ with a black ellipse (dashed line). We indicate close companions or objects in the line-of-sight with a black solid line or ellipse (for NGC 5846 where these areas fall within the region where the effective parameters were measured, these areas were masked during the measurements).}
\label{fig:Flux2}
\end{figure*}

We also use the total $K$-band luminosity (M$_{K}$) obtained from the 2MASS XSC, and we perform Galactic extinction and other corrections as described in detail in \citet{Loubser2018} (their Section 3.2). We take the ellipticity ($\epsilon$) as 1 -- sup\textunderscore ba, and we also obtain the photometric position angle PA$_{\rm phot}$ from the 2MASS XSC as measured from the ``total" isophote from the 3-band co-added image. We have a $\pm$0.1 fiducial uncertainty on the ellipticity and $\pm$3 degree fiducial uncertainty on the PA$_{\rm phot}$. All the photometric parameters are also presented in Table \ref{table1}. PA$_{\rm phot}$ from the 2MASS XSC agrees with our MUSE white-light images (obtained by adding the MUSE cube along the wavelength axis) as shown in Figure \ref{fig:Flux}, except for NGC 978 where 2MASS gives 115$\degr$ (white solid line), but we find PA$_{\rm phot}$ $\sim$75$\degr$ (white dashed line). This is presumably because the source was blended with the close companion in the 2MASS measurements, therefore we use the latter (75$\degr$).

%------------------------------------------------------------------------------------
\begin{table*}
\caption{General parameters of our sample. Here Col. (1) gives the galaxy name, Cols. (2) and (3) the coordinates, and Col. (4) the redshift from NED. Col. (5) and (6) give $R_{e}$ (see Section \ref{Section:data}), to convert from (5) to (6) we assume $H_{0}$ = 67.8 km/sec/Mpc, $\Omega_{\rm matter}$ = 0.308, $\Omega_{\rm vacuum}$ = 0.692 (these parameters have a negligible effect on the physical scales of the galaxies due to their close proximity). Col (7) give the spatial coverage of our data, in terms of $R_{e}$, along the photometric semi-major axis. Col (8) gives the total, Galactic extinction corrected M$_{K}$ as calculated in \citet{Loubser2018} (see their Section 3.2 for details). Col (9) gives the ellipticity ($\epsilon$), and Col (10) is PA$_{\rm phot}$ as measured from the 2MASS ``total" isophote (East of North).}      
\label{table1}      
\centering                         
\begin{tabular}{l c c c r r c c c r} 
\hline
Name  & RA      & DEC     &  $z$  & 2MASS $R_{e}$  & 2MASS $R_{e}$ & Coverage & M$_{K}$ & $\epsilon\pm$0.10 & PA$_{\rm phot}\pm$3.0 \\
            &      (J2000) & (J2000) &   &  (arcsec) & (kpc) & ($\times$ $R_{e}$) & (mag) & & (deg) \\
(1)      &  (2)       & (3)         &  (4)           & (5)   & (6)  & (7) & (8) & (9) & (10) \\
\hline 
ESO 507-G025 & 12h51m31.8s & -26d27m07s &  0.0108 & 13.34 & 3.33 & 2.5 & --24.71$\pm$0.03 & 0.24 & 90.0 \\
NGC 193 & 00h39m18.6s & +03d19m52s &  0.0147 & 11.84 & 3.39 & 2.0 & --24.65$\pm$0.03 & 0.28 & 70.0 \\
NGC 410 & 01h10m58.9s & +33d09m07s &  0.0177 & 16.77 & 5.87 & 1.7 & --25.76$\pm$0.02 & 0.26 & 40.0 \\
NGC 584 &01h31m20.7s & -06d52m05s &  0.0060 & 19.55 & 2.09 & 1.7 & --24.22$\pm$0.03 & 0.34 & 62.5 \\
NGC 677 &01h49m14.0s & +13d03m19s &  0.0170 & 9.34 & 3.15 & 1.8 & --24.87$\pm$0.03 & 0.10 & 35.0 \\
NGC 777 &02h00m14.9s & +31d25m46s &  0.0167 & 14.57 & 4.87 & 1.7 & --25.61$\pm$0.02 & 0.10 & 145.0 \\
NGC 924 &02h26m46.8s & +20d29m51s &  0.0149 & 8.00 & 2.37 & 3.5 & --24.37$\pm$0.03 & 0.44 & 55.0 \\
NGC 940 &02h29m27.5s & +31d38m27s &  0.0171 & 7.61 & 2.61 & 2.8 & --24.83$\pm$0.03 & 0.30 & 15.0 \\
NGC 978 &02h34m47.6s & +32d50m37s &  0.0158 & 9.70 & 3.08 & 2.4 & --24.53$\pm$0.03 & 0.26 & 75 \\
NGC 1060 &02h43m15.0s & +32d25m30s &  0.0173 & 16.76 & 5.85 & 1.5 & --25.97$\pm$0.02 & 0.14 & 70.0 \\
NGC 1453 &03h46m27.2s & -03d58m08s &  0.0130 & 16.01 & 4.23 & 1.5 & --25.48$\pm$0.02 & 0.14 & 25.0 \\
NGC 1587 &04h30m39.9s & +00d39m42s &  0.0123 & 12.89 & 3.30 & 2.1 & --25.00$\pm$0.03 & 0.16 & 60.0 \\
NGC 4008 &11h58m17.0s & +28d11m33s &  0.0121 & 12.32 & 3.40 & 2.9 & --24.56$\pm$0.03 & 0.42 & 165.0 \\
NGC 4169 &12h12m18.8s & +29d10m46s &  0.0126 & 7.99 & 2.29 & 2.9 & --24.41$\pm$0.03 & 0.44 & 150.0 \\
NGC 4261 &12h19m23.2s & +05d49m31s &  0.0074 & 22.36 & 4.05 & 1.5 & --25.47$\pm$0.03 & 0.16 & 172.5\\
NGC 5846 &15h06m29.3s & +01d36m20s &  0.0057 & 32.02 & 4.32 & 0.8 & --25.11$\pm$0.02 & 0.08 & 27.5\\
NGC 6658 &18h33m55.6s & +22d53m18s &  0.0142 & 7.01 & 2.04 & 4.7 & --24.25$\pm$0.03 & 0.76 & 5.0 \\
NGC 7619 &23h20m14.5s & +08d12m22s &  0.0125 & 14.84 & 3.55 & 1.8 & --25.28$\pm$0.02 & 0.18 & 40.0 \\
\hline                                   
\end{tabular}
\end{table*}

%------------------------------------------------------------------------------------

\subsection{Voronoi binning}

We use the Galaxy IFU Spectroscopy Tool (GIST) \citep{Bittner2019}, which uses the well-known pPXF \citep{Cappellari2004, Cappellari2017} and GandALF \citep{Sarzi2006} routines, to bin the spectra and extract stellar kinematics. We logarithmically rebin the spectra (along the wavelength axis) to have constant bins in velocity. 

The higher-order moments, in particular, depend on a high signal-to-noise ratio (S/N), and the data cubes were tessellated using the Voronoi-binning method \citep{Cappellari2003} to achieve adequate S/N (minimum S/N$\sim$80) per spatial bin as measured from the continuum between 6000 and 6200 \AA. 

\subsection{Kinematics extraction}

\begin{figure}
\centering
\includegraphics[scale=0.19]{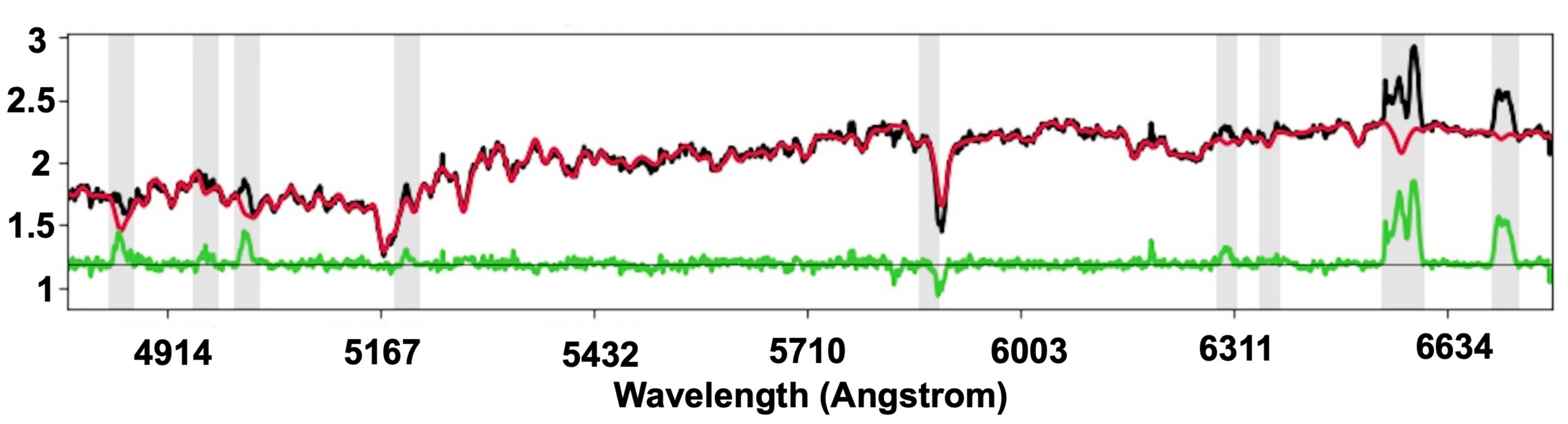}\\
   \caption{Example of an extracted central spectrum (4800 to 6800 \AA{}) for ESO 507-G025. The observed spectrum is indicated in black (in arbitrary flux units), and the best-fitting stellar template in red (with the residual of the two shown in green). Grey bands indicate potential emission-lines or possible sky-line residuals masked during the fit.}
\label{fig:Spectra}
\end{figure}

We use pPXF (through GIST) to derive the line-of-sight velocity distribution (LOSVD) described by the velocity $V$, velocity dispersion $\sigma$, and the higher-order Gauss-Hermite moments $h_{3}$ and $h_{4}$ \citep{Gerhard1993, vanderMarel1993}. We use the E-MILES model library \citep{Vazdekis2015}, covering the wavelength range used here at a spectral resolution of $\sim$ 2.51 \AA{} \citep{FalconBarroso2011}. An example of a spectrum extracted from a central bin in ESO 507-G025 is shown in Figure \ref{fig:Spectra}. We also masked potential emission lines or possible sky-line residuals during the fit, as illustrated by the grey bands in Figure \ref{fig:Spectra}. Our stellar kinematic maps and measurements agree with those from \citet{Olivares2022} who used the Indo-US library \citep{Valdes2004}. We limit the fit to a wavelength range of 4800 to 6800 \AA{}, to avoid any potentially strong telluric and sky residuals. We also test the effect of wavelength range on the measured kinematics, and found that at S/N = 80, there are negligible differences between using 4800 to 6800 \AA{} and slightly shorter wavelength ranges.  We correct for the wavelength-dependent line-spread-function of the MUSE observations by adopting the prescription from \citet{Bacon2017}. 

In our stellar template fitting, we include an additive polynomial of the 4th order to account for any potential deviations in the continuum shape between the stellar template and observed galaxy spectra, or flux calibration differences \citep{Bittner2019}. We initially use the default value for the penalisation in pPXF \citep{Cappellari2004}, and then use 100 Monte-Carlo simulations to compute errors on the kinematic parameters without the penalisation term. The standard deviation of the measured kinematic parameters is used as errors (see full description of the GIST kinematic prosedure in \citealt{Bittner2019}). 

\section{Stellar kinematics results} 
\label{results}

We show maps of $V$ for each of the galaxies in Figure \ref{fig:Kin}. We show the rest of the maps ($\sigma$, $h_{3}$, $h_{4}$, and $\lambda$) for ESO 507-G025 in Appendix \ref{OtherKINmaps} as an example, and the maps for the other 17 BGEs are included as supplementary information. The isophotes are based on flux from the MUSE cube and displayed in steps of 0.5 magnitude. We discuss distinct features from the $V$ and $\sigma$ maps in Section \ref{features}. We discuss the global extracted properties and classifications in Section \ref{lambdaR}, and the higher order kinematics ($h_{3}$ and $h_{4}$) in Section \ref{higherorder}. We measure the projected angular momentum of each galaxy using the dimensionless parameter $\lambda$ \citep{Emsellem2007}. The $\lambda$ parameter is used to quantify the dynamical importance of rotation relative to random motion in a galaxy, as discussed in detail in Section \ref{lambdaR}. 

\begin{figure*}
\captionsetup[subfloat]{farskip=-2pt,captionskip=-3pt}
\centering
\subfloat{\includegraphics[width=6cm,height=5cm]{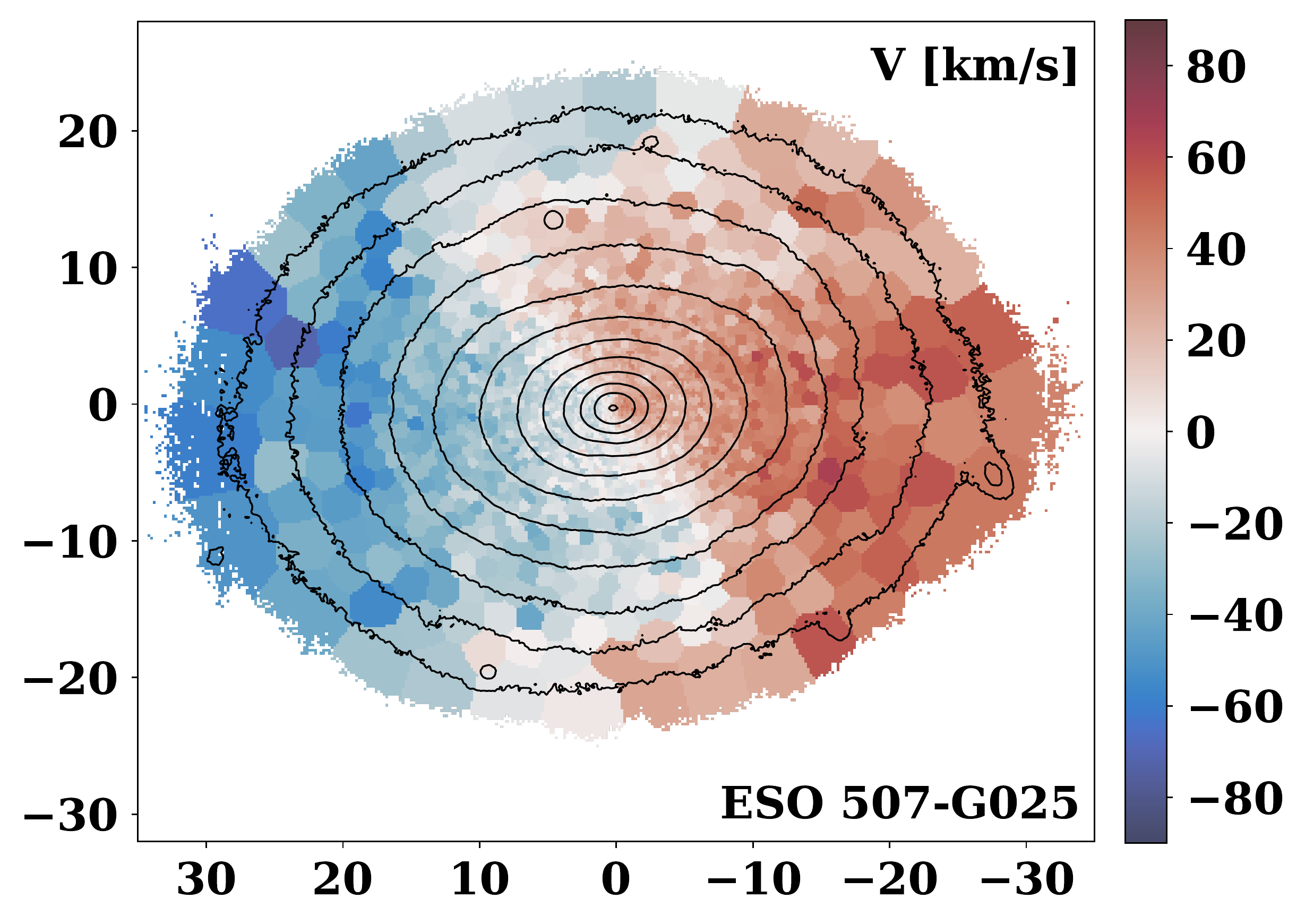}}
\subfloat{\includegraphics[width=6cm,height=5cm]{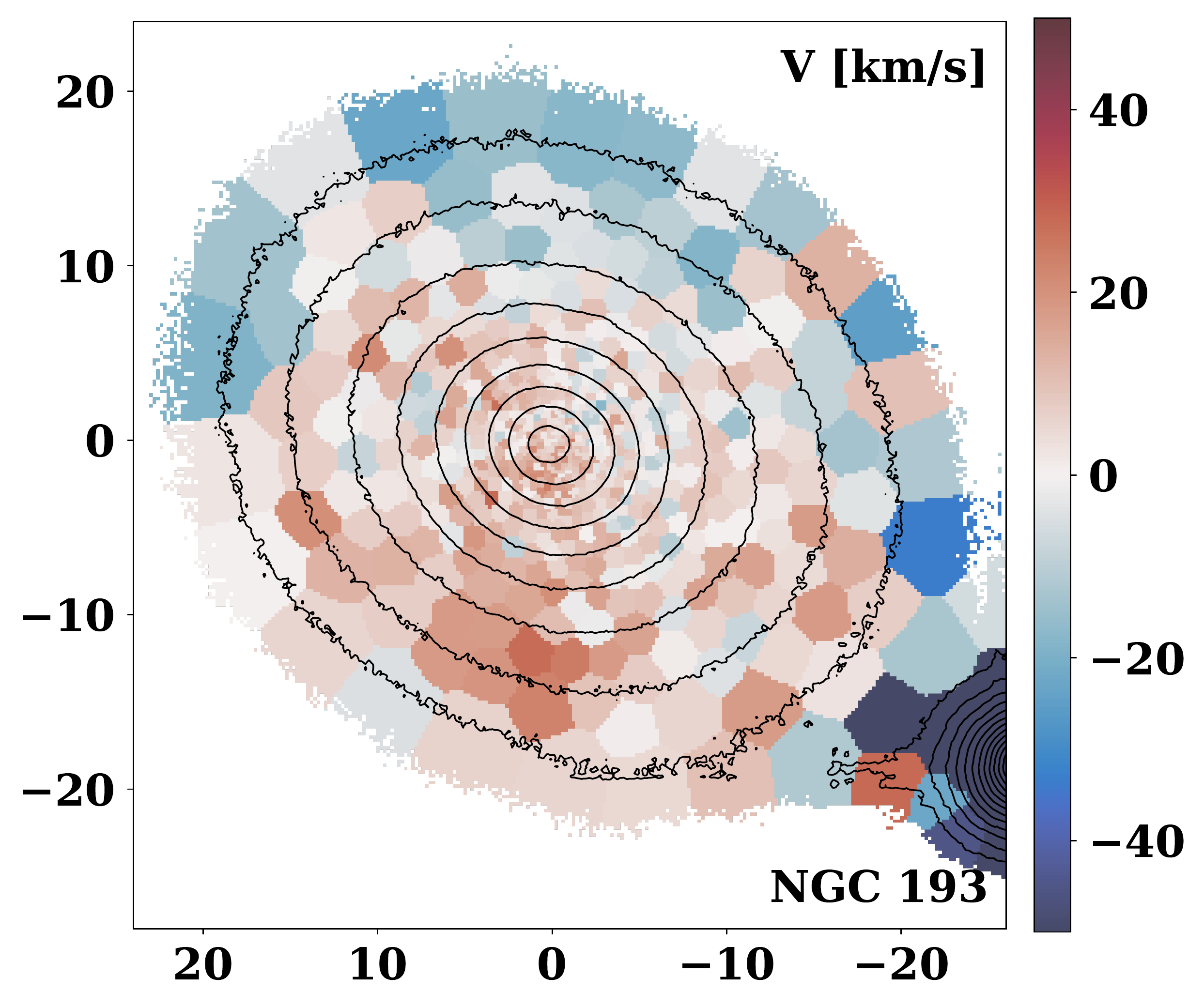}}
\subfloat{\includegraphics[width=6cm,height=5cm]{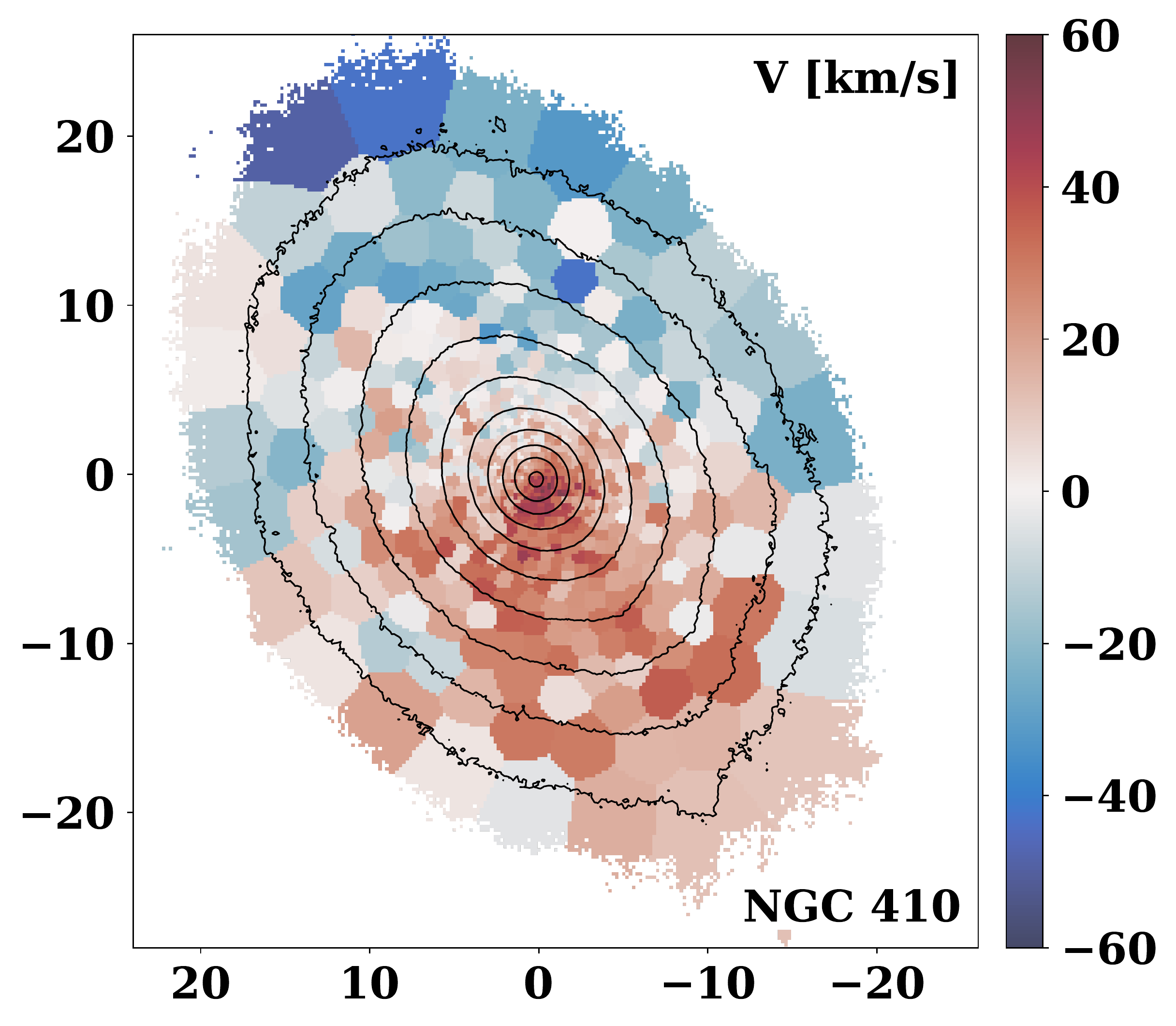}}\\
\subfloat{\includegraphics[width=6cm,height=5cm]{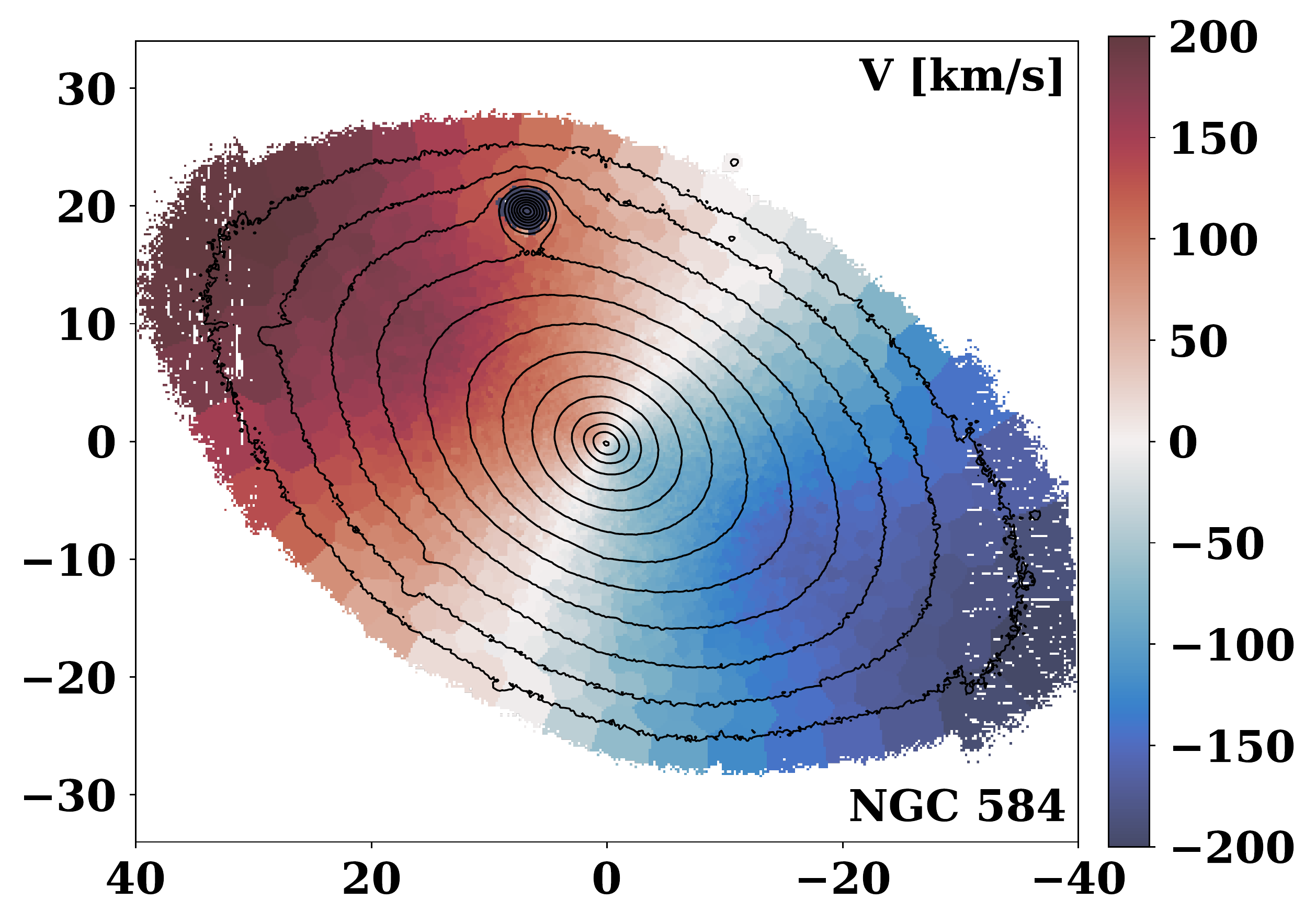}}
\subfloat{\includegraphics[width=6cm,height=5cm]{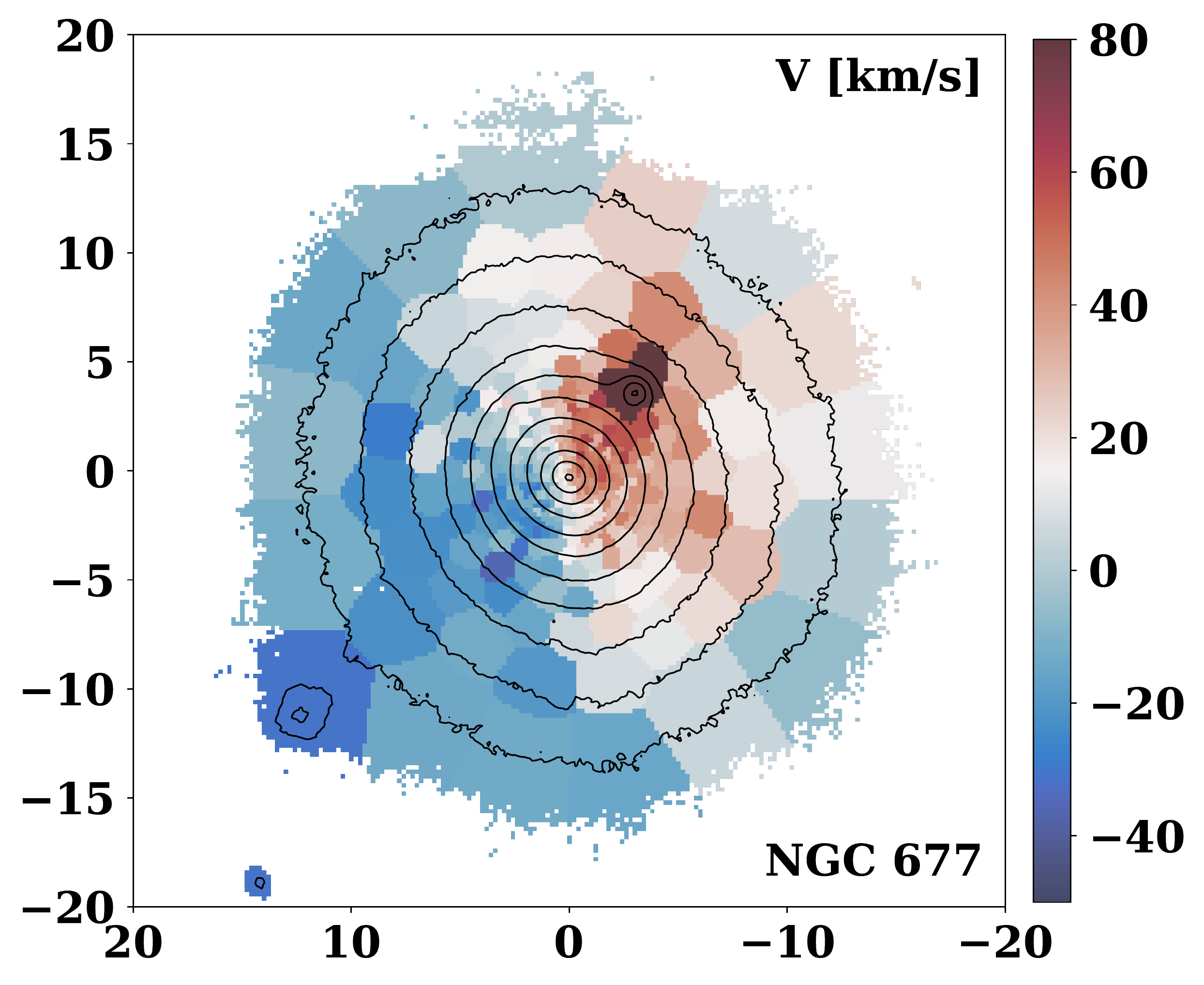}}
\subfloat{\includegraphics[width=6cm,height=5cm]{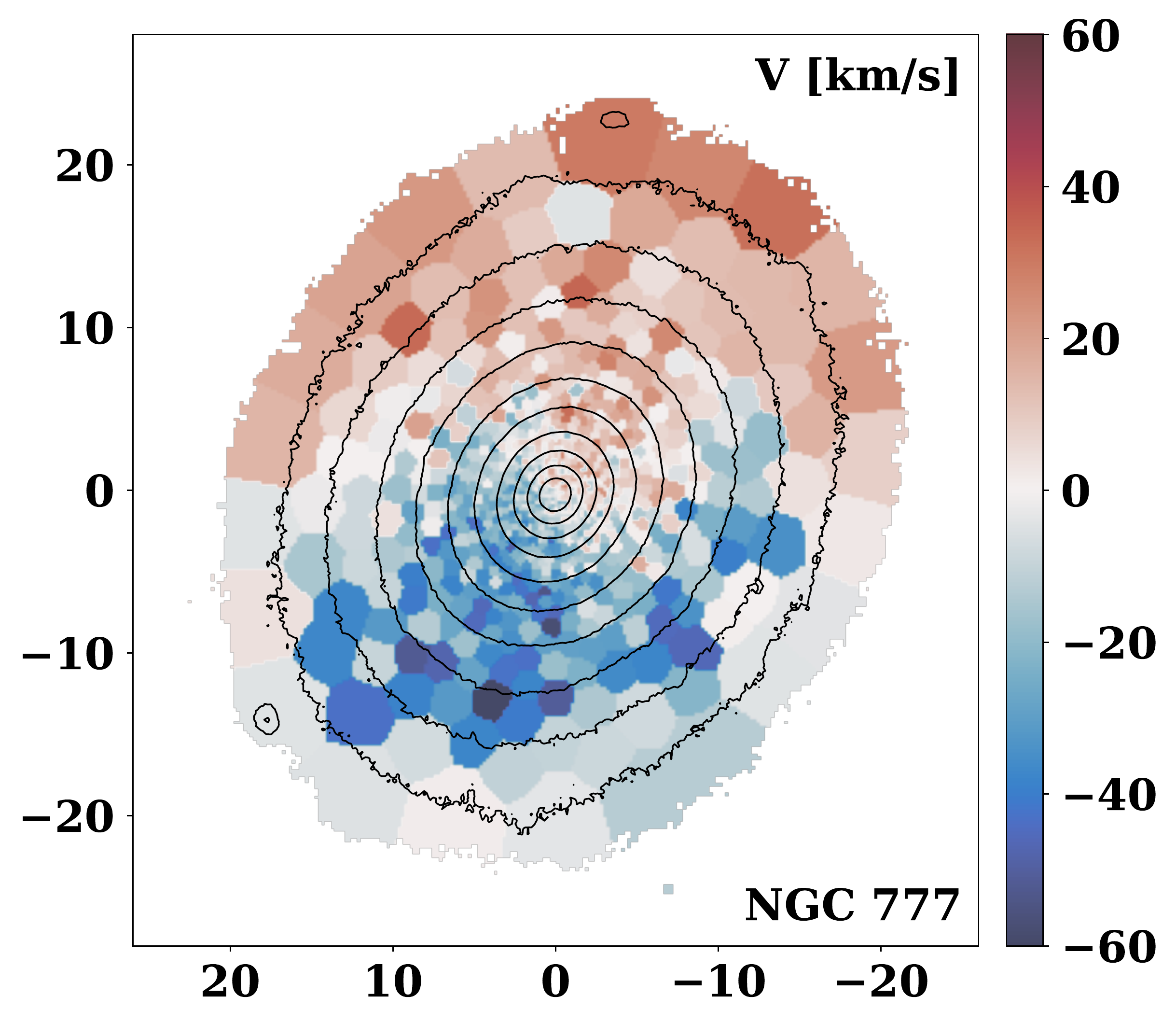}}\\
\subfloat{\includegraphics[width=6cm,height=5cm]{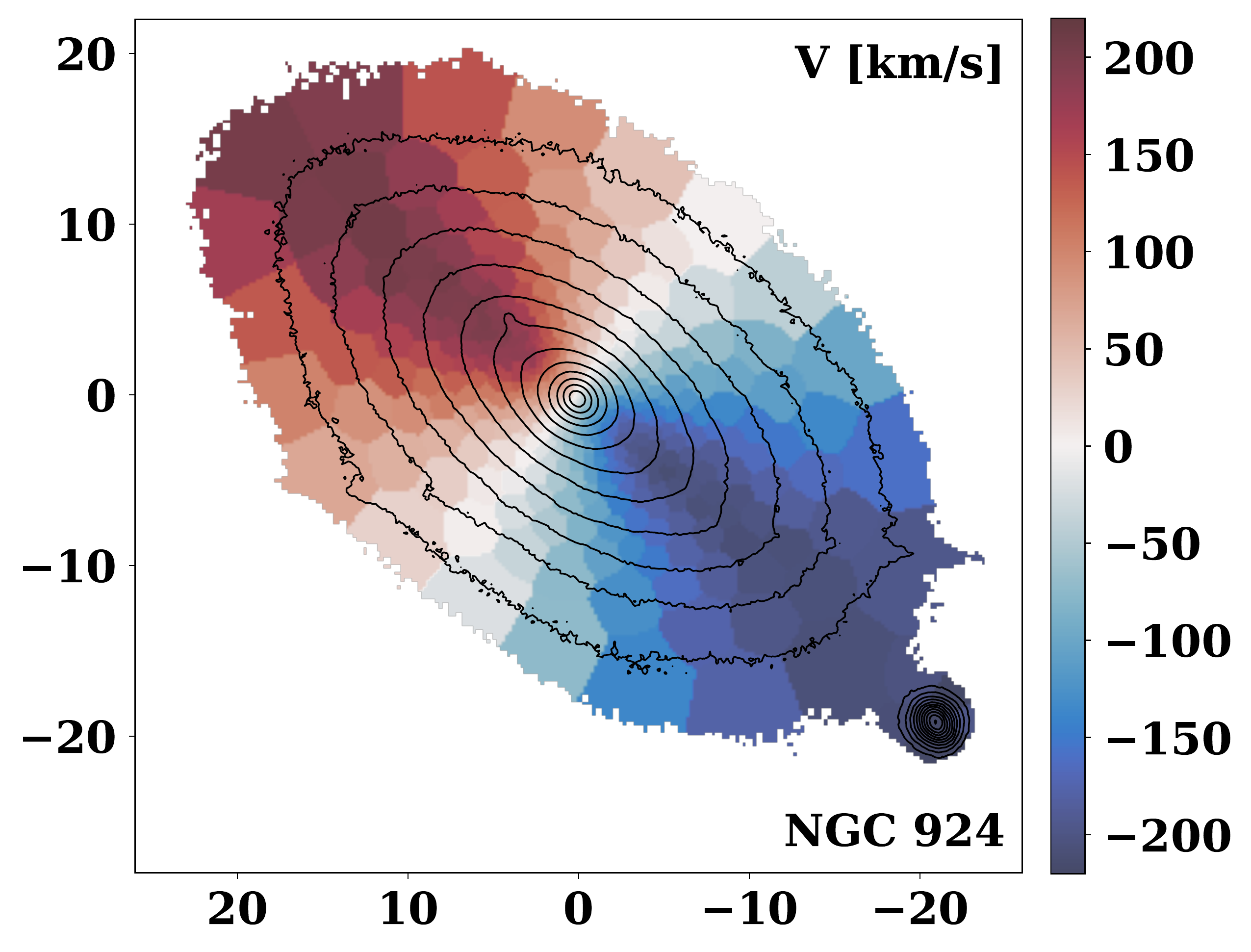}}
\subfloat{\includegraphics[width=6cm,height=5cm]{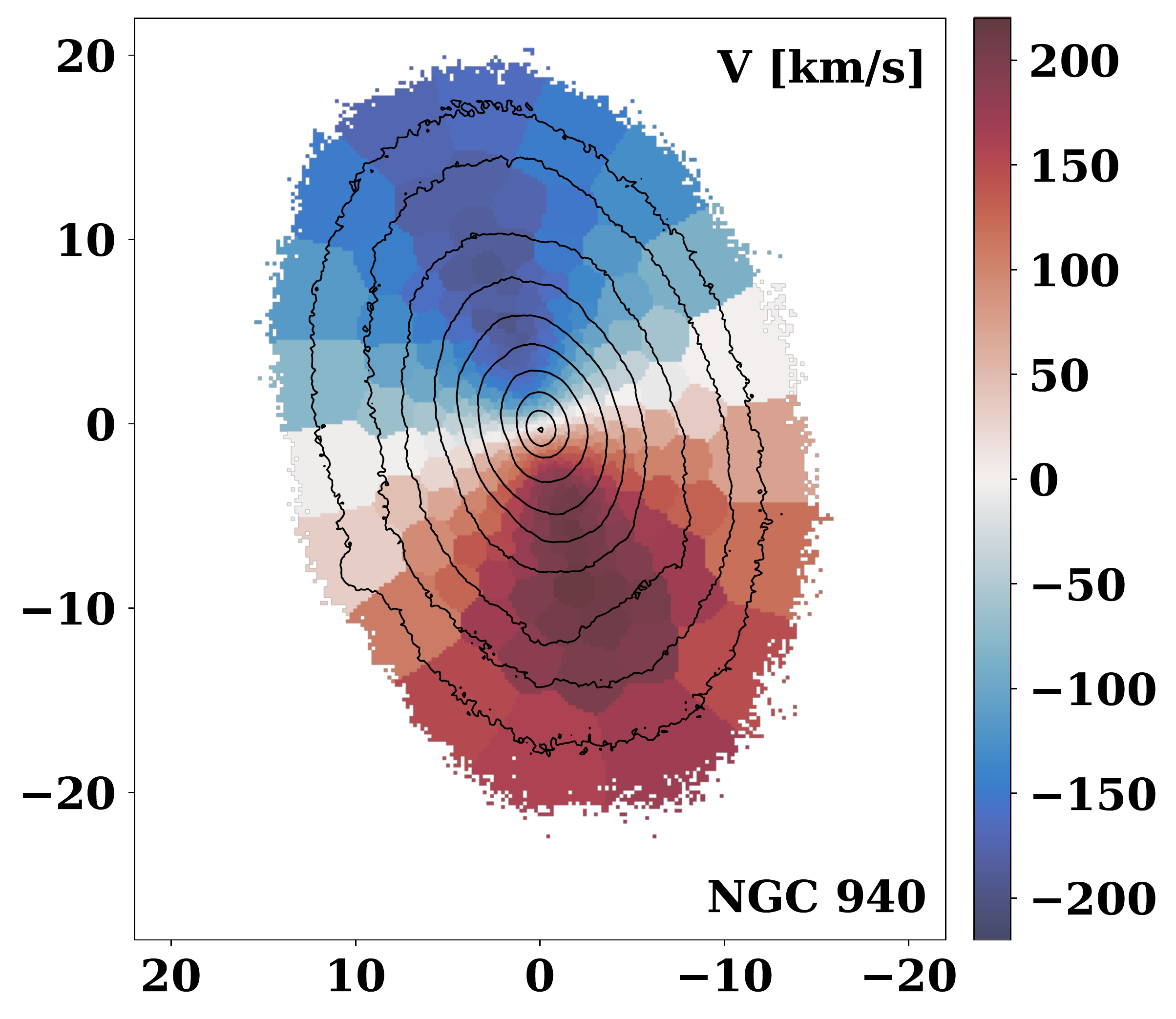}}
\subfloat{\includegraphics[width=6cm,height=5cm]{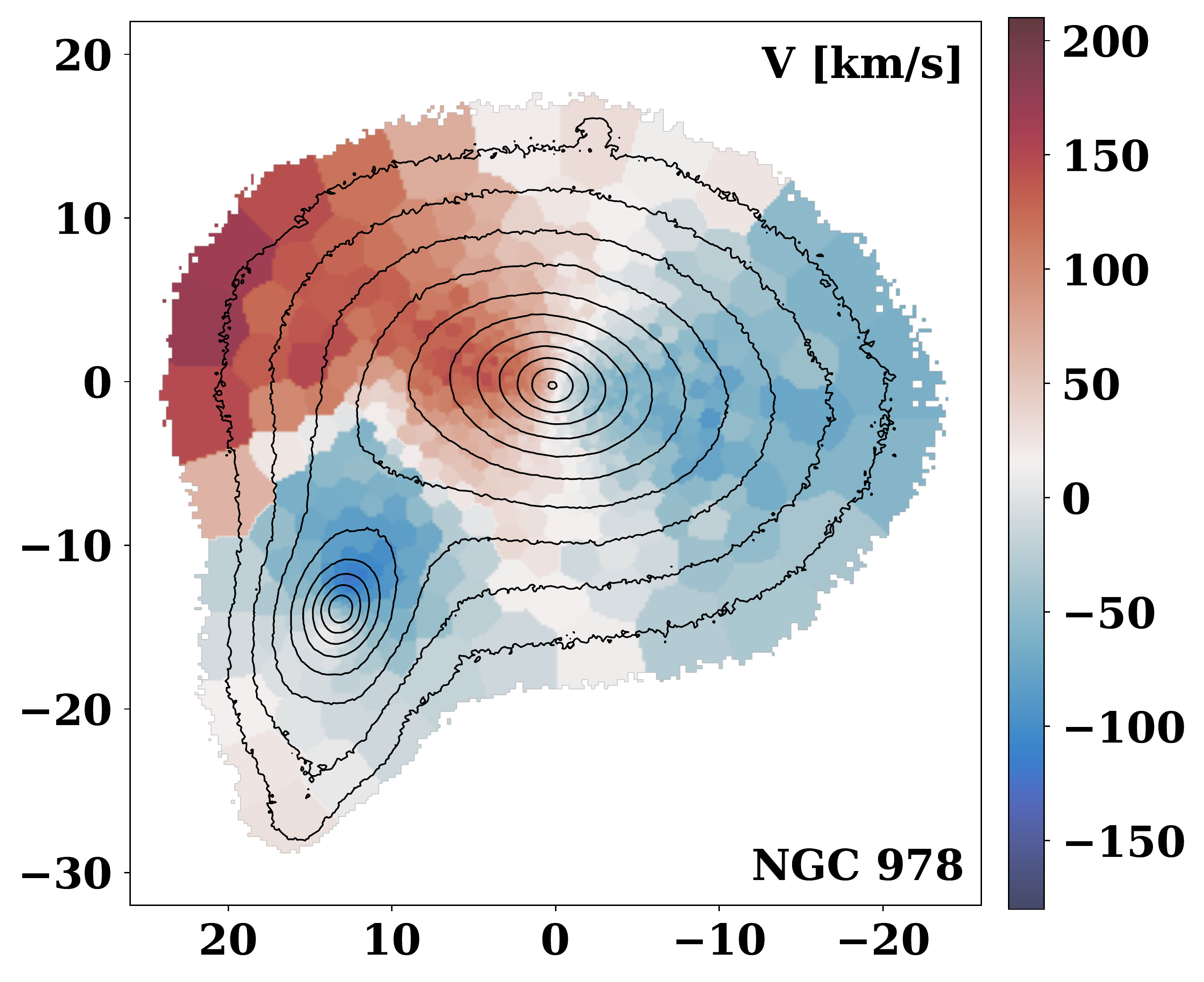}}\\
   \caption{Maps of the stellar velocity field ($V$). The orientation of the maps is the same as in Figure \ref{fig:Flux}. All spatial scales are in arseconds. The isophotes are based on flux from the MUSE cube and displayed in steps of 0.5 magnitude. }
\label{fig:Kin}
\end{figure*}

\begin{figure*}
\captionsetup[subfloat]{farskip=-2pt,captionskip=-3pt}
\centering
\subfloat{\includegraphics[width=6cm,height=5cm]{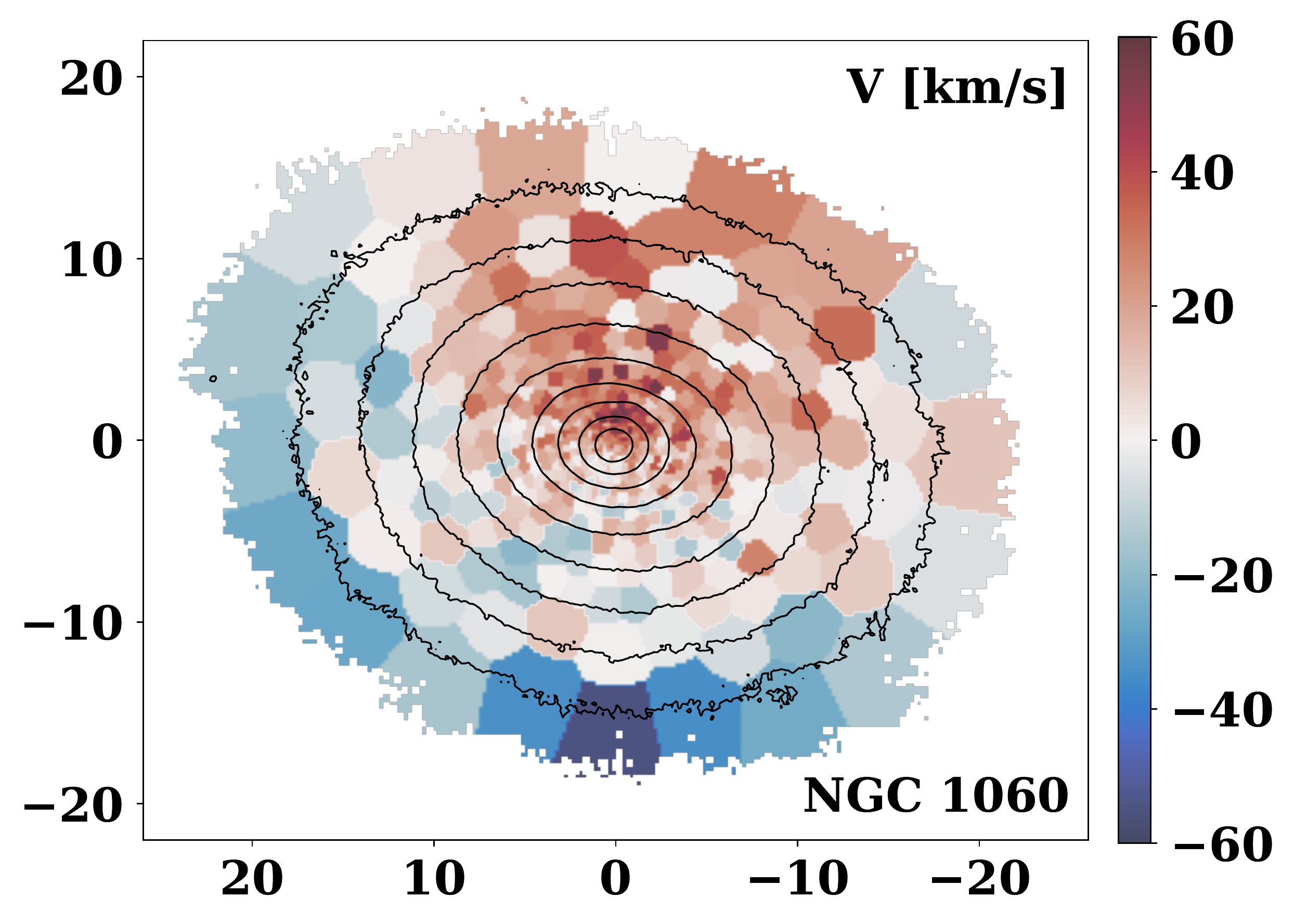}}
\subfloat{\includegraphics[width=6cm,height=5cm]{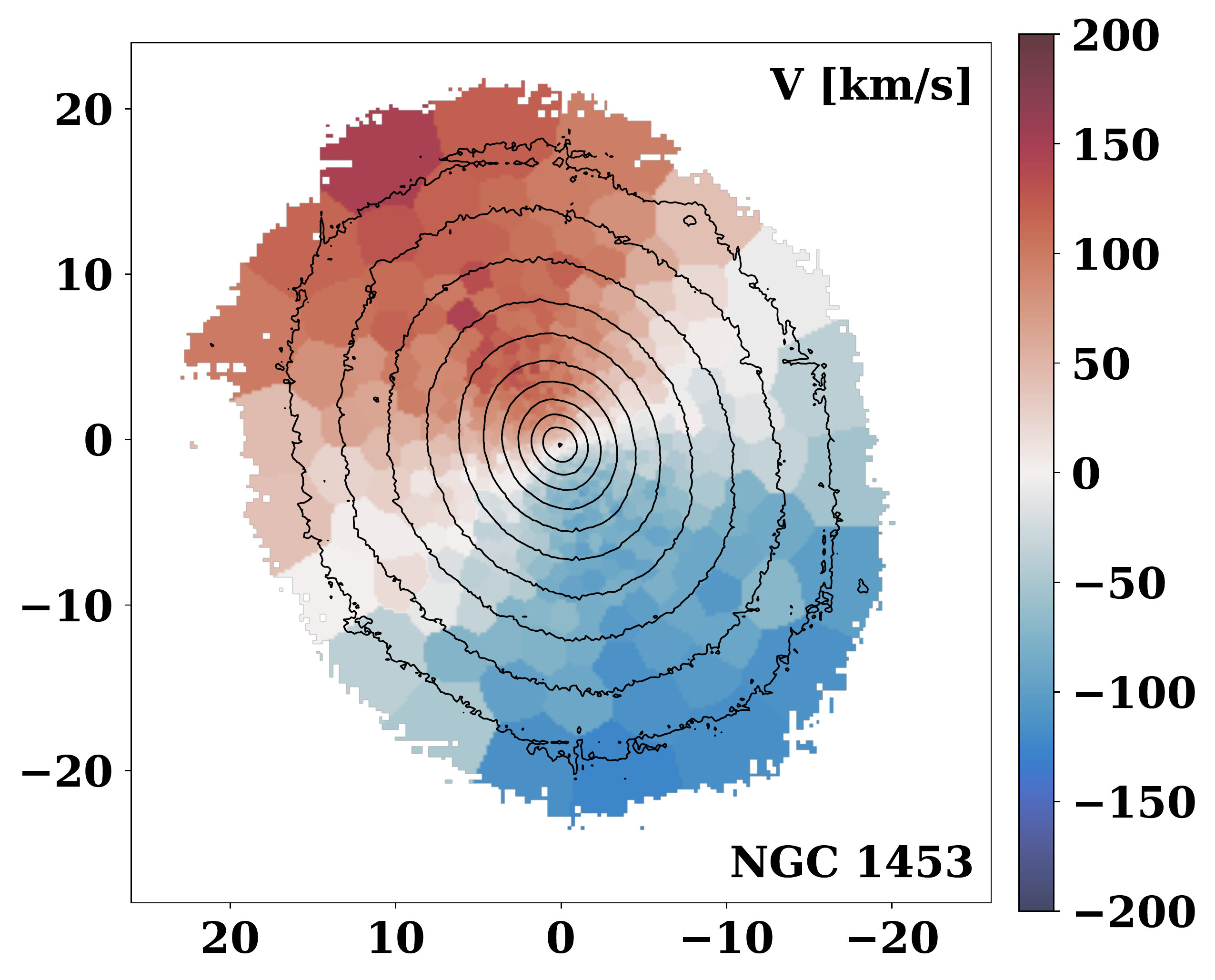}}
\subfloat{\includegraphics[width=6cm,height=5cm]{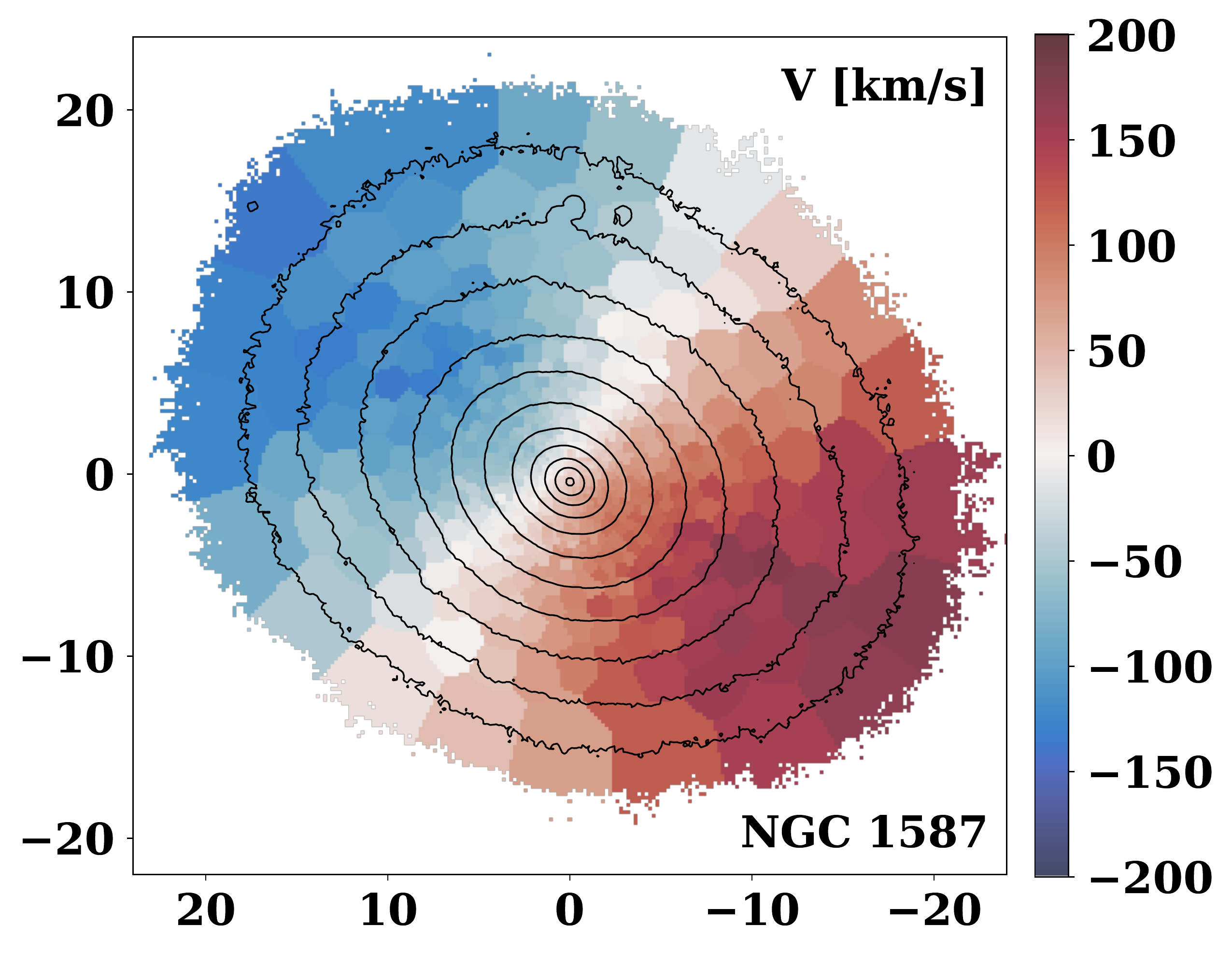}}\\
\subfloat{\includegraphics[width=6cm,height=5cm]{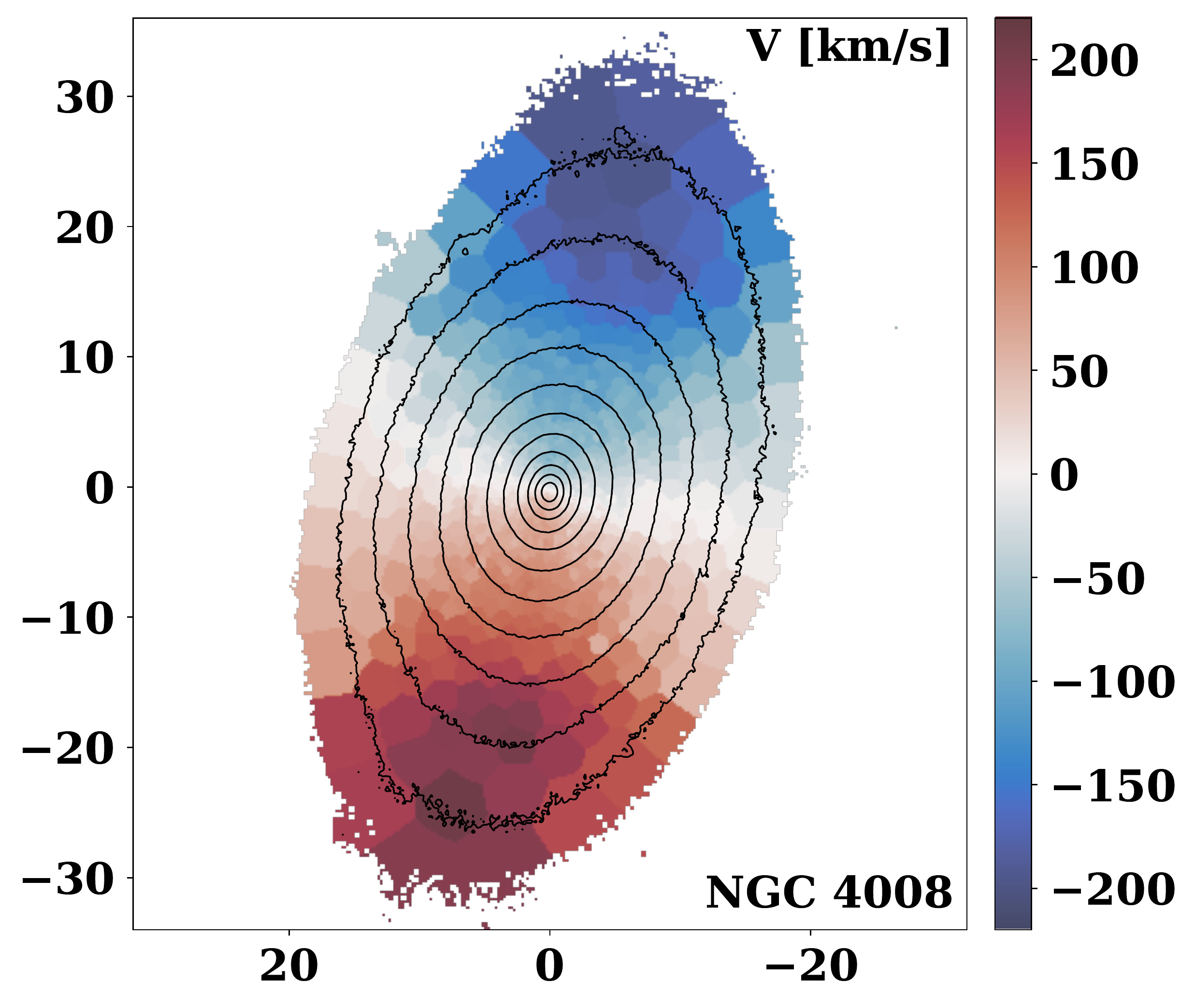}}
\subfloat{\includegraphics[width=6cm,height=5cm]{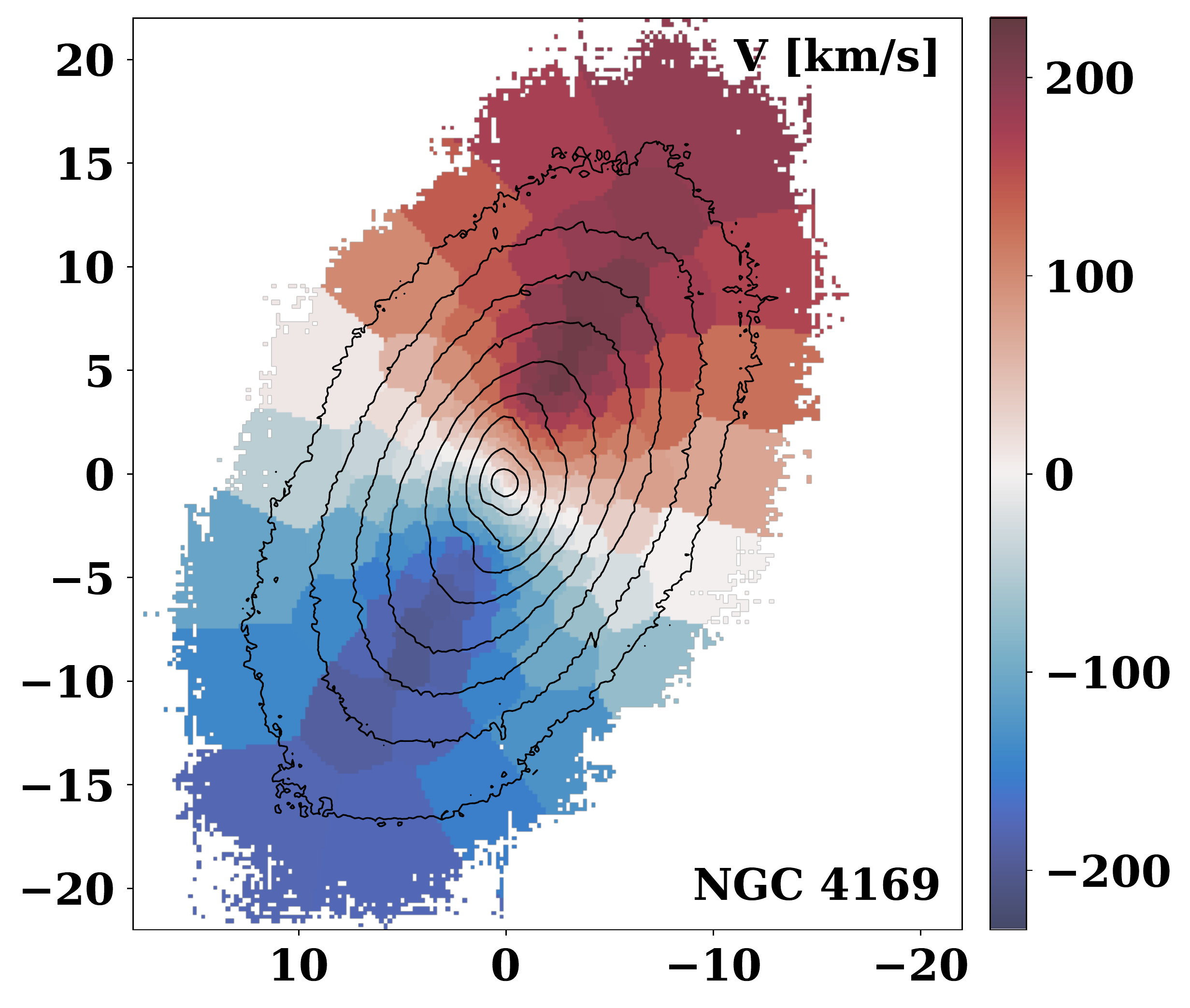}}
\subfloat{\includegraphics[width=6cm,height=5cm]{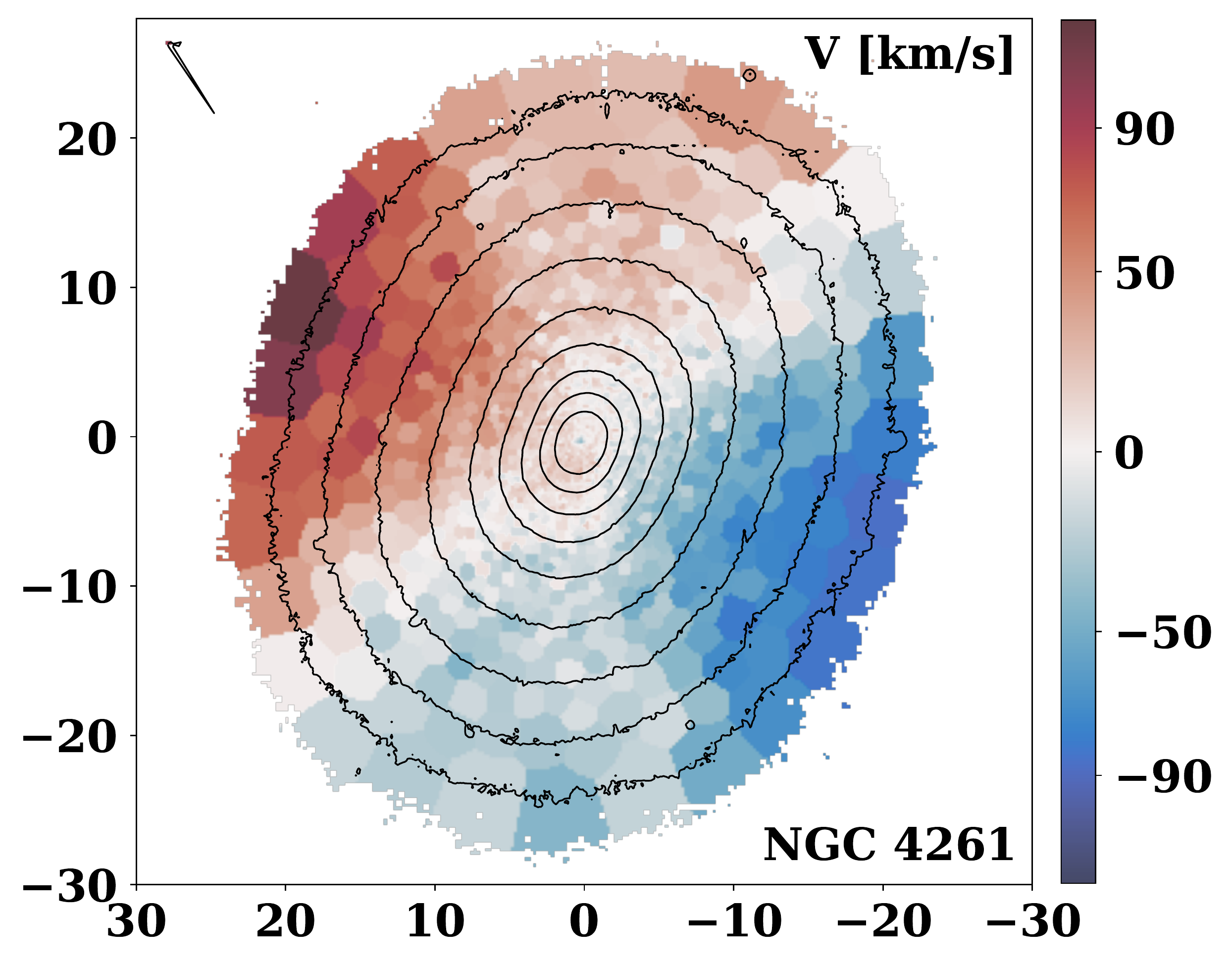}}\\
\subfloat{\includegraphics[width=6cm,height=5cm]{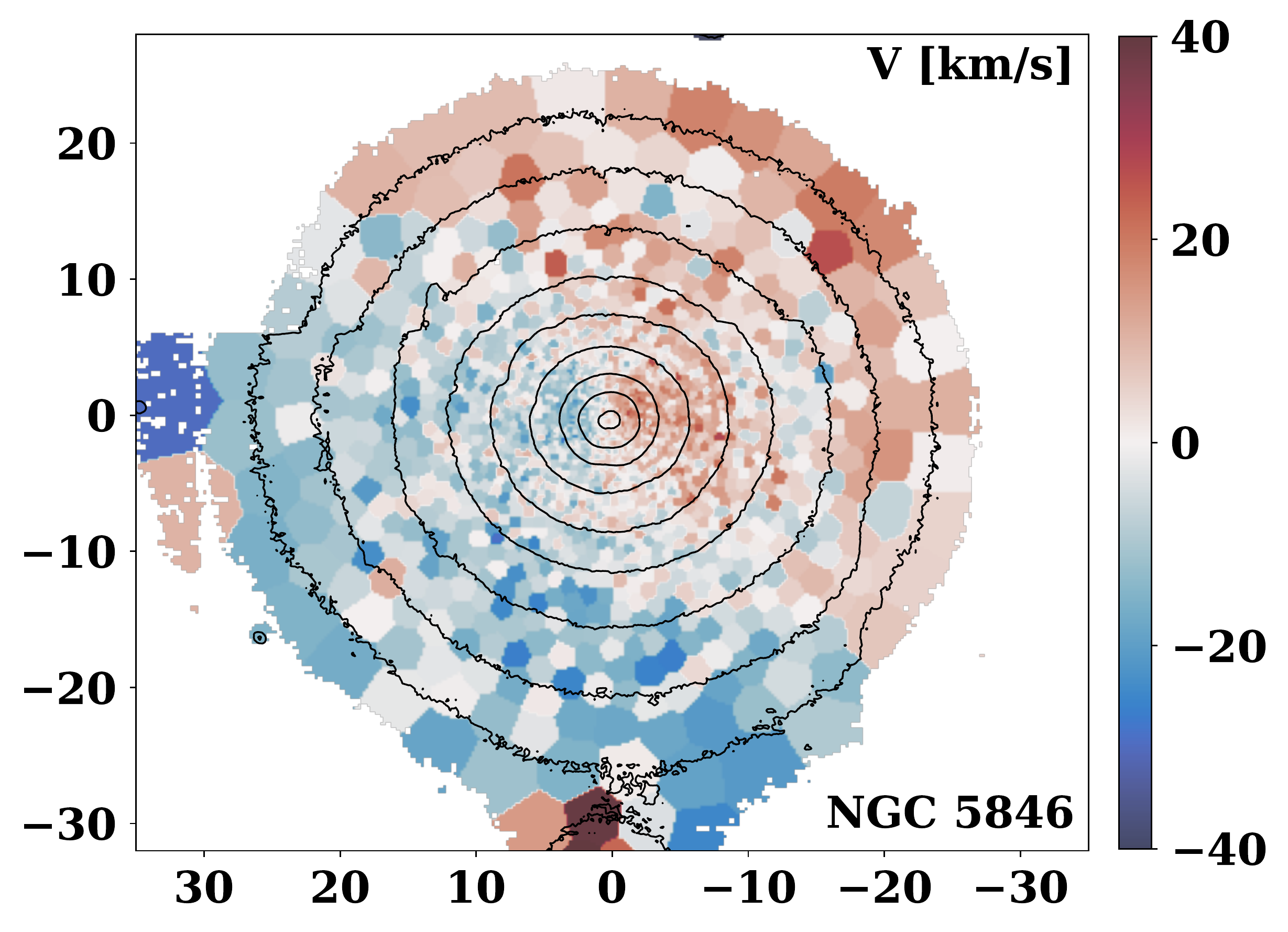}}
\subfloat{\includegraphics[width=6cm,height=5cm]{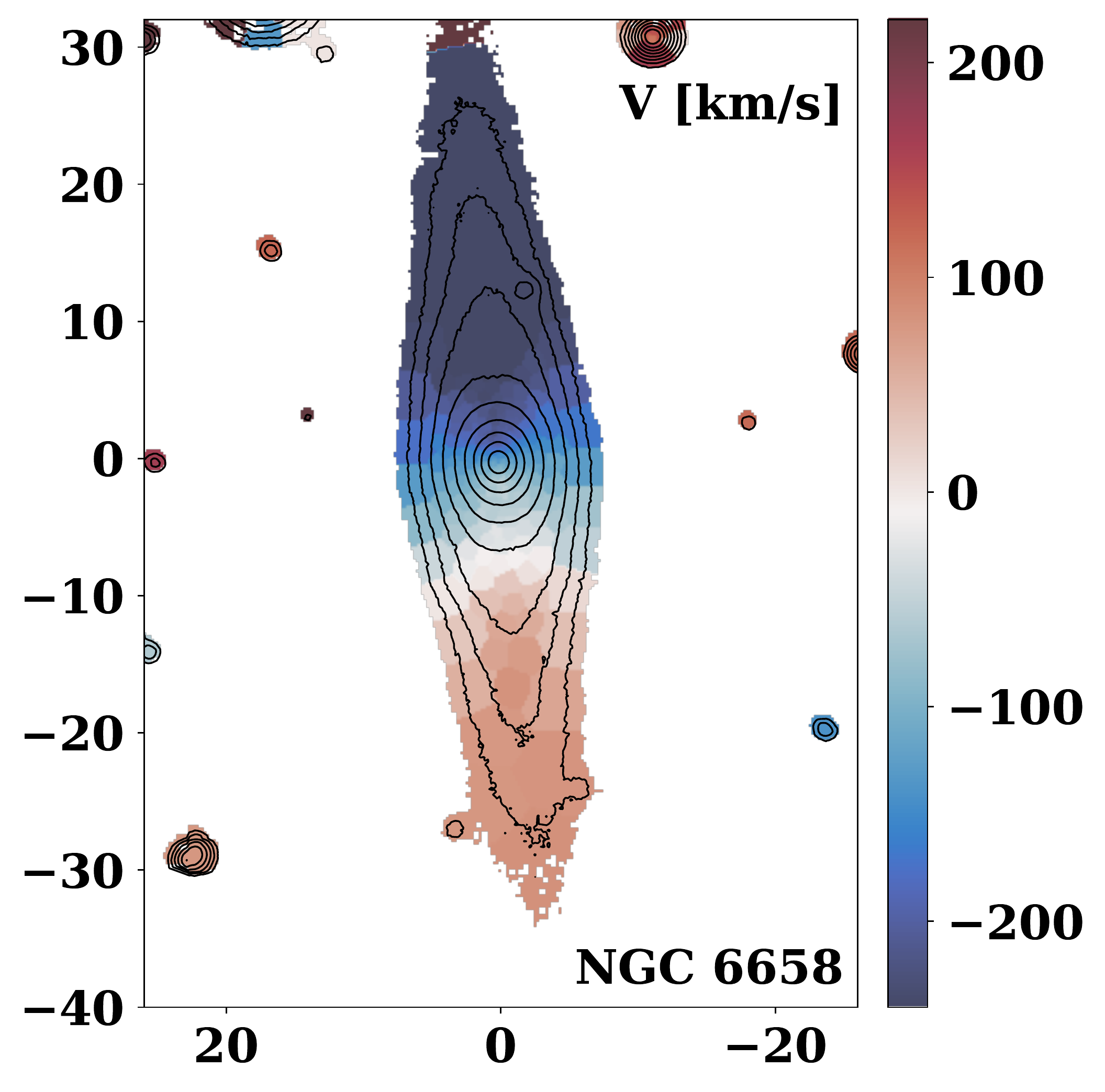}}
\subfloat{\includegraphics[width=6cm,height=5cm]{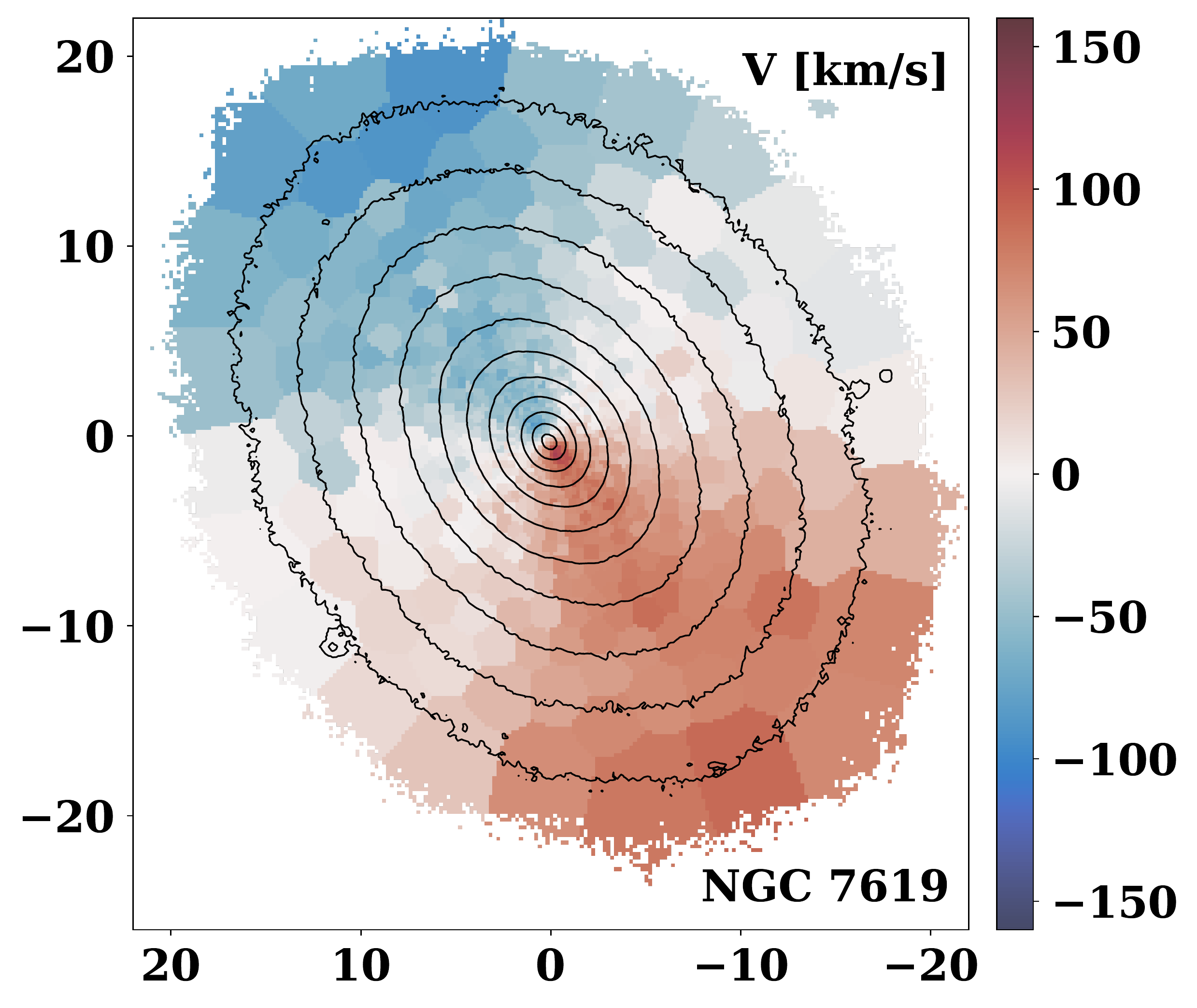}}
   \contcaption{Maps of the stellar velocity field ($V$). The orientation of the maps is the same as in Figure \ref{fig:Flux}. All spatial scales are in arseconds. The isophotes are based on flux from the MUSE cube and displayed in steps of 0.5 magnitude. }
\label{fig:Kin}
\end{figure*}

\subsection{Distinct features from the $V$ and $\sigma$ kinematic maps} 
\label{features}
 
We visually classify each galaxy according to its stellar velocity field (Figure \ref{fig:Kin}). Various kinematic features are evident upon visual inspection of the $V$ and $\sigma$ kinematic maps:

\begin{enumerate}

\item \textit{Rotation}

Despite being massive early-type galaxies, all our BGEs show some level of rotation, as the MUSE data quality is such that even very little net rotation can be detected (e.g.\ NGC 193, NGC 5846). At least half of the BGEs show regular rotation with peak velocity in excess of 100 km s$^{-1}$. A very small number of massive galaxies show such rotation (see e.g.\ 2 out of 41 brightest cluster galaxies, hereafter BCGs, in \citealt{Loubser2008}), but it is far more prevalent in galaxies at lower masses \citep{Brough2017}. 

We make a visual distinction between ``regular rotator" (regular stellar velocity field) or ``non-regular rotator" (complex stellar velocity field). An illustration of the visual classification of regular and non-regular rotators can be seen in Figure 4 of \citet{Cappellari2016}\footnote{We perform a qualitative classification from the map following the illustration in Figure 4 of \citealt{Cappellari2016}. We do a quantitative classification in Section \ref{lambdaR}.}. We find the majority (10/18) of our BGEs to be regular rotators (RR): NGC 584, NGC 924, NGC 940, NGC 978, NGC 1453, NGC 1587, NGC 4008, NGC 4169, NGC 6658, and NGC 7619. This is somewhat in contrast to e.g. \citet{Krajnovic2018} who find regular rotation only in their satellite galaxies and not their BCGs in their M3G study (MUSE Most Massive Galaxies; PI: Emsellem),
although their central galaxies are more massive than the ones studied here (see Figure \ref{fig:Mass_size}). Non-regular rotation (NRR) is the most common characteristic of the \citet{Krajnovic2018} M3G BCGs, whereas we find it for only 8/18 BGEs: ESO 507-G025, NGC 193, NGC 410, NGC 677, NGC 777, NGC 1060, NGC 5846, as well as NGC 4261 (which shows non-regular rotation around the major axis, as reported and discussed by \citealt{Davies1986}).

The regular or non-regular appearance of the velocity field of early-type galaxies directly relates to the presence (or absence) of an embedded disc, and also to the measured angular momentum \citep{Krajnovic2011, Emsellem2011, Krajnovic2020}. Regular and non-regular rotation has also been used to classify fast rotators (regular kinematics) and slow rotators (non-regular kinematics), e.g.\ in \citet{Emsellem2011}. We show that all our regular rotators are fast rotators and all the non-regular rotators are slow rotators in Section \ref{lambdaR}.

\item \textit{Kinematically distinct cores (KDC):}

The existence of a KDC is indicated by an abrupt change in the kinematic position angle over a small radial range (e.g. \citealt{Loubser2008}). KDCs are fairly rare in massive elliptical galaxies, \citet{Krajnovic2018} find two amongst their 14 massive galaxies (BCGs) in the M3G sample. \citet{Ene2018, Ene2020} also find two amongst 90 massive galaxies in their MASSIVE sample. We find no KDCs in this sample of BGEs.

\item \textit{Radial velocity dispersion ($\sigma$) profiles:}

The most massive elliptical galaxies, BCGs, often exhibit velocity dispersion profiles that rise from the centre of the galaxy to the outer parts \citep{Loubser2008, Loubser2018, Loubser2020}. This is less common amongst BGEs, see e.g.\ Figure 3 in \citet{Loubser2018} which show only one (NGC 2768) out of 23 BGEs from the CLoGS sample studied using long-slit spectra, compared to the majority of the BCGs studied in the same analysis. NGC 2768 is not included in our MUSE subsample of CLoGS BGEs. In fact, in this study, none of the BGEs show rising velocity dispersion profiles (also see simulations in \citealt{Jung2022}). 

\end{enumerate}

\begin{table*}
\caption{Derived properties of our sample. RR or NRR refer to regular or non-regular rotation as visually classified in Section \ref{features}. The kinematic position angle, PA$_{\rm kin}$, is measured East of North to the maximum velocity.  All other stellar kinematic parameters ($\Psi$, (V/$\sigma)_{e}$, $\lambda_{e}$, $\langle h_{4} \rangle_{e}$, and $\xi_{3}$), as well as the slow (SR) or fast (FR) rotator classification, are derived as described in the text. *For NGC 5846, we use values within 0.8$R_{e}$.}      
\label{table2}      
\centering                         
\begin{tabular}{l r r r r r c r r} 
\hline
Name     & Rotation & PA$_{\rm kin}$ & $\Psi$ & (V/$\sigma)_{e}$ & $\lambda_{e}$ & SR/FR &  $\langle h_{4} \rangle_{e}$ & $\xi_{3}$  \\
         & RR/NRR  & (deg) & (deg) &  &  &  &  & \\
         (1)      &  (2)   &  (3)       & (4)         &  (5)           & (6)   & (7)  & (8) & (9) \\
\hline 
ESO 507-G025 & NRR & 289.0 $\pm$ 11.8 & 19.0 $\pm$ 12.2 & 0.080 &  0.085 & Slow & 0.037 &  --0.84 \\
NGC 193      & NRR & 167.3 $\pm$ 58.2 & 82.7 $\pm$ 58.3  & 0.035 &  0.030 & Slow & 0.035 & --0.35  \\
NGC 410     &  NRR  & 189.1 $\pm$ 20.0 & 30.9 $\pm$ 20.2 & 0.054 &  0.046 & Slow & 0.065 & --1.03  \\
NGC 584     &  RR  & 61.8 $\pm$ 0.9 & 0.7 $\pm$ 3.1 & 0.338 &  0.377 & Fast & 0.024 & --4.19 \\
NGC 677      & NRR  & 140.0 $\pm$ 5.5 & 75.0 $\pm$ 6.3 & 0.092  & 0.090 & Slow & 0.059 & --0.83  \\
NGC 777      & NRR  & 329.1 $\pm$ 20.9 & 4.1 $\pm$ 21.1 & 0.038 &  0.038 & Slow & 0.078 & --0.16 \\
NGC 924     & RR  & 52.7 $\pm$ 1.8 & 2.3 $\pm$ 3.5 & 0.374 &  0.497 & Fast & 0.056 & --11.95 \\
NGC 940     & RR  & 16.4 $\pm$ 0.9 & 1.4 $\pm$ 3.1 & 0.663 &  0.643 & Fast & 0.074 & --7.91 \\
NGC 978     & RR  & 70.9 $\pm$ 2.7 & 4.1 $\pm$ 4.0 & 0.213 &  0.255 & Fast & 0.052 & --4.74 \\
NGC 1060     & NRR & 343.6 $\pm$ 23.6 & 86.4 $\pm$ 23.8 & 0.049 &  0.038 & Slow & 0.065 & --0.77 \\
NGC 1453     & RR & 32.7 $\pm$ 3.6 & 7.7 $\pm$ 4.7 & 0.165 &  0.182 & Fast & 0.034 & --2.40 \\
NGC 1587     & RR & 232.7 $\pm$ 3.6 & 7.3 $\pm$ 4.7 & 0.187 &  0.225 & Fast & 0.041 & --2.97 \\
NGC 4008     & RR & 167.3 $\pm$ 2.7 & 2.3 $\pm$ 4.0 & 0.213 &  0.249 & Fast  & 0.036 & --2.16 \\
NGC 4169     & RR & 329.1 $\pm$ 2.7 & 0.9 $\pm$ 4.0 & 0.503 &  0.539 & Fast & 0.061 & --9.04 \\
NGC 4261     & NRR & 60.0 $\pm$ 6.4 & 67.5 $\pm$ 7.1 & 0.044 &  0.046 & Slow & 0.028 & --0.19 \\
NGC 5846     & NRR & 256.4 $\pm$ 34.5 & 48.9 $\pm$ 34.6 & *0.039 &  *0.032 & Slow & *0.046 & --0.63 \\
NGC 6658     & RR & 188.2 $\pm$ 0.9 & 3.2 $\pm$ 3.1 & 0.263 &  0.332 & Fast & 0.042 & --2.41 \\
NGC 7619     & RR & 205.5 $\pm$ 3.6 &  14.5 $\pm$ 4.7 & 0.161 &  0.142 & Fast & 0.029 & --2.55 \\
\hline                                   
\end{tabular}
\end{table*}

\subsection{Effective kinematic parameters and fast vs slow rotator classification}
\label{lambdaR}

For early-type galaxies, visual morphology alone is not sufficient to infer whether a galaxy formed via gas-rich or gas-poor mergers. A more robust method is the fast/slow rotator classification, separating the two different classes of early-type galaxies \citep{Emsellem2007, Cappellari2016, Graham2018}.

Before we make any measurements of the effective kinematic parameters, we correct for systemic velocity by subtracting the median of the velocity ($V$) measurements over all spatial bins from every individual bin.

\subsubsection{The ratio of ordered versus random motions ($V/\sigma$)}

The ratio of ordered versus random motions ($V/\sigma$) within one effective radius ($R_{e}$) is determined as in \citet{Cappellari2007}:
\begin{equation}
\left(\frac{V}{\sigma}\right)^2_{e} = \frac{\Sigma F_{i} V_{i}^2}{\Sigma F_{i} \sigma_{i}^2}
\end{equation}
where $F_{i}$ is the flux in each bin. We sum only bins within an ellipse of semi-major axis corresponding to one effective radius. These measurements are presented in Table \ref{table2}. Our measurements for the galaxies that we have in common with the MASSIVE sample agree very well with theirs (see Appendix \ref{Section:comparison} for the comparison). We use $V/\sigma$ further in Section \ref{higherorder}, where we interpret anti-correlations between $V/\sigma$ and $h_{3}$. 

\subsubsection{Classification according to the spin parameter ($\lambda_{e}$)}

Following \citet{Emsellem2007}, the $\lambda_{e}$ parameter is measured as
\begin{equation}
\label{eqn:lambda} \lambda_{e} = \frac{\Sigma F_{i} R_{i} | V_{i} |}{\Sigma F_{i} R_{i} \sqrt{V_{i}^2 + \sigma_{i}^2}} \end{equation}
with the galactocentric radius $R_{i}$, the flux $F_{i}$, the radial velocity $V_{i}$, and the velocity dispersion $\sigma_{i}$ for each individual spatial bin within $R_{e}$ . We determine $\lambda_{e}$ using an ellipse\footnote{We use the intrinsic radius (semi-major axis) instead of the projected (circular) radius here, as the intrinsic radius follows the light profile of the galaxy more accurately. \citet{VandeSande2017} quantify the possible effects by measuring $\lambda_{e}$ using both methods. For round objects ($\epsilon<0.4$), the effect is especially small ($<0.01$). Our galaxies are nearby and we can also assume $\lambda_{e}$ to be unaffected by seeing (see \citealt{Graham2018}).} of semi-major axis corresponding to one effective radius. 

\begin{figure}
   \includegraphics[scale=0.33]{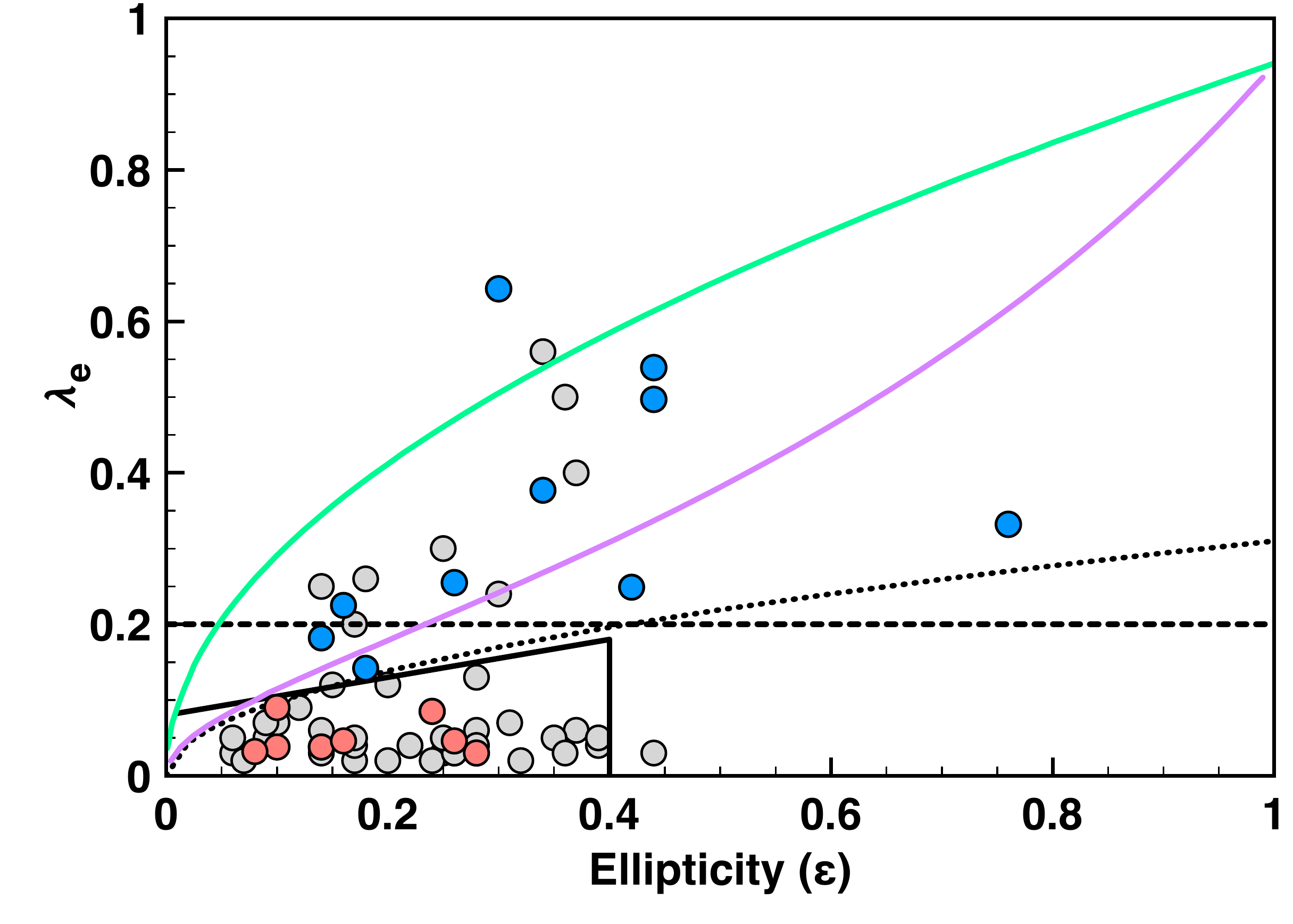}
   \caption{We plot $\lambda_{e}$ vs $\epsilon$ for the BGEs (blue and red symbols, for fast and slow rotators). Here, we also indicate the 41 most massive galaxies from the MASSIVE sample \citep{Veale2017a} with grey circles. We indicate the \citet{Emsellem2011} fast vs slow rotator classification with a black dotted line, the \citet{Lauer2012} classification with a black dashed line, and the \citet{Cappellari2016} classification with a black solid line. The green line is the prediction for an edge-on isotropic rotator from \citet{Binney2005} (Equation 14 in \citealt{Cappellari2016}), and the magenta line is the edge-on relation from \citet{Cappellari2007} (Equation 11 in  \citealt{Cappellari2016}).}
\label{fig:Lambda_eps}
\end{figure}

We classify our galaxies into fast or slow rotators, as this division connects with two dominant channels of galaxy formation (as reviewed in \citealt{Cappellari2016}). Following \citet{Emsellem2007} and \citet{Emsellem2011}, we use the spin parameter approximation $\lambda_{e}$ to separate fast-rotating galaxies from slow-rotating galaxies by classifying galaxies above $\lambda_{e} = 0.31 \sqrt \epsilon$ as fast rotators and below as slow rotators (see Table \ref{table2}).\footnote{We use $\lambda_{e}$ to quantitatively classify the BGEs into fast and slow rotators, and since our fast/slow rotation classification exactly agrees with our qualitative regular/non-regular rotator classification from the velocity maps, any uncertainty on $R_{e}$ (propagated to $\lambda_{e}$) does not affect our main conclusions.} We plot $\lambda_{e}$ vs $\epsilon$ in Figure \ref{fig:Lambda_eps} (the BGEs are indicated by blue and red symbols, for fast and slow rotators). Here, we also indicate the 41 most massive galaxies from the MASSIVE sample \citep{Veale2017a} with grey circles. We indicate the \citet{Emsellem2011} classification with a black dotted line, the \citet{Lauer2012} classification ($\lambda_{e}=0.2$) with a black dashed line, and the \citet{Cappellari2016} classification with a black solid line ($\lambda_{e} < 0.08 + \epsilon/4$ where $\epsilon<0.4$). We use the \citet{Emsellem2011} classification here (but we get the same result if we use the \citet{Cappellari2016} classification), and the results correspond exactly to the regular/non-regular rotation classification in that all regular rotating BGEs are fast rotators, and all non-regular rotating BGEs are slow rotators. The $\lambda_{e}$ parameter is also related to $(V/\sigma)_{e}$ (see equation B1 in \citealt{Emsellem2011}).

Figure \ref{fig:Lambda_eps} also show that all the BGEs classified as non-regular rotators are all at $\epsilon < 0.3$. This is in agreement with ATLAS$^{\rm 3D}$, SAMI and CALIFA kinematic results \citep{Krajnovic2011, Fogarty2015, Cappellari2016}, and indicates that non-regular rotators are quite close to spherical and only weakly triaxial \citep{Cappellari2016}.

\subsubsection{Fast vs slow rotators as a function of luminosity and mass}
\label{Stelmass}

For early-type galaxies, the ratio f(SR)/f(FR) between the fraction of slow and fast-rotating galaxies seem to be a function (albeit not a simple one) of environment, as well as a function of luminosity and stellar mass. Slow rotators live predominantly in high density environments, such as massive groups or clusters \citep{Cappellari2011b}. The ratio of fast and slow rotators is nearly constant outside clusters, whereas it becomes a strong function of the galaxy number density inside clusters \citep{D'Eugenio2013, Houghton2013, Scott2014, Fogarty2014}. This suggests that the formation of slow rotators must be linked to the clusters themselves \citep{Cappellari2016}. 

The fraction of slow rotators also strongly correlate with luminosity and mass \citep{Emsellem2011, VandeSande2017, Veale2017a, Greene2017, Graham2018}. These correlations are best studied with samples which cover a wide range in mass, morphology and environment \citep{Brough2017, Wang2020, VandeSande2021b}, nonetheless we illustrate the distribution of the fast and slow-rotating BGEs as a function of luminosity and mass below.

\begin{figure}
   \includegraphics[scale=0.33]{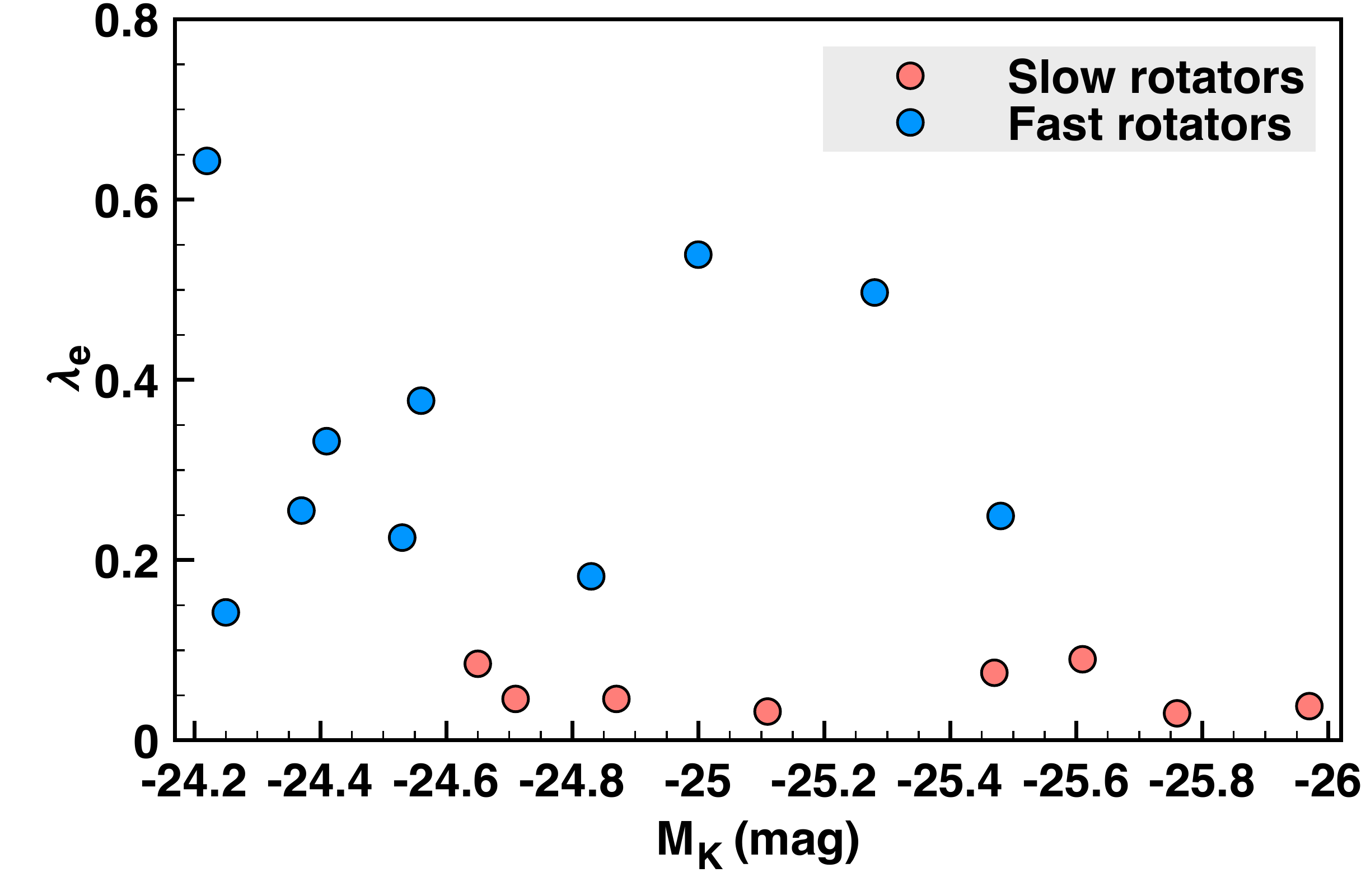}
   \caption{The distribution of $\lambda_{e}$ with M$_{K}$, separating the fast (blue) and slow rotators (red).}
\label{fig:Lambda_MK}
\end{figure}

We plot $\lambda_{e}$ vs M$_{K}$ in Figure \ref{fig:Lambda_MK} separating the fast (blue) and slow rotators (red). We can see that the six least luminous BGEs are fast rotators, and the three most luminous BGEs are slow rotators (NGC 410, NGC 777, and NGC 1060), and an overlap between the two classes over a broad range of luminosities in between. This trend, that towards higher stellar mass the fraction of fast rotating galaxies decreases, agrees with previous studies. \citet{Veale2017a} studied the fraction of slow rotators in bins of luminosity for MASSIVE and ATLAS3D galaxies, and show that the fraction increases dramatically from 10 per cent at M$_{K} \sim -22$ mag to 90 per cent at M$_{K} \sim -26$ mag. A similar trend is shown for the ATLAS3D sample in \citet{Emsellem2011}.
 
The parameter $\lambda_{e}$ is a projected quantity, and we adopt a similar approach to \citet{Brough2017} to estimate the effects of inclination for our early-type galaxies. If we use $\lambda_{e}/\sqrt\epsilon$ (as $\epsilon$ is also estimated as a global parameter) as an approximate correction on $\lambda_{e}$ and investigate $\lambda_{e}/\sqrt\epsilon$ vs M$_{K}$ instead, it does not affect our conclusion that the least luminous BGEs are more likely to have a high spin parameter, and the most luminous BGEs more likely to have a lower spin parameter.

\begin{figure}
   \includegraphics[scale=0.33]{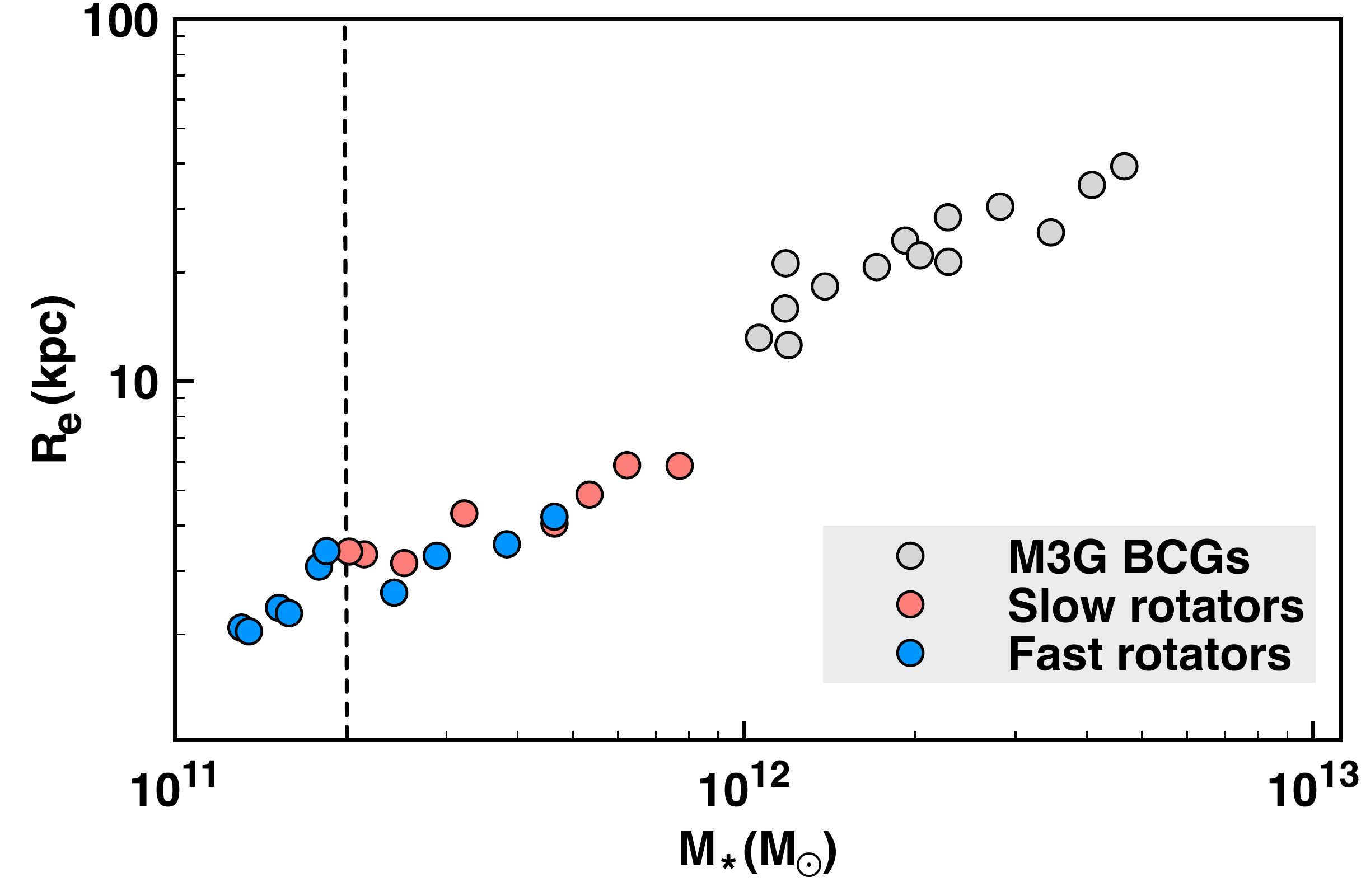}
   \caption{The mass (M$_{*}$) -- size ($R_{e}$) diagram. We indicate the fast (blue) and slow (red) rotating BGEs. We also show the BCGs from the M3G sample \citep{Krajnovic2018}. We also indicate the characteristic mass at 2 $\times 10^{11}$ M$_{\sun}$ (above which passive slow rotators dominate).}
\label{fig:Mass_size}
\end{figure}

The distribution of galaxies in the size-mass plane is a well-studied scaling relation \citep{VanderWel2014, VandeSande2017, Krajnovic2018, WaloMartin2020}. We illustrate our fast and slow rotating BGEs on the mass (M$_{*}$) -- size ($R_{e}$) diagram in Figure \ref{fig:Mass_size}. We estimate the stellar mass of the BGEs using the widely-used conversion between $K$-band luminosity and stellar mass for early-type galaxies from the ATLAS3D sample \citep{CappellariSingle2013}:
\begin{equation}
\log_{10}(\rm M_{*}) = 10.58 - 0.44(\rm M_{K} + 23)
\end{equation}

We indicate the fast (blue) and slow (red) rotating BGEs. We also show the BCGs from the M3G sample \citep{Krajnovic2018}. We also indicate the characteristic mass at 2 $\times 10^{11}$ M$_{\sun}$ (above which passive slow rotators with cores dominate, see \citealt{Krajnovic2020}), and find that all of our slow rotators are above this limit.

\subsection{Higher-order kinematics}
\label{higherorder}

Observational (e.g.\ \citealt{VandeSande2017}) as well as cosmological hydrodynamical zoom-in studies (e.g.\ \citealt{Naab2014}) have shown that the evolutionary history of galaxies cannot be accurately constrained from the spin parameter alone, and should be complimented with the higher-order kinematic measurements preferably from integral field spectroscopy. The higher-order stellar kinematic moments $h_{3}$ and $h_{4}$, defined as the deviations from a Gaussian line-of-sight velocity distribution (LOSVD), are parameterized with Gauss-Hermite polynomials (\citealt{vanderMarel1993}; \citealt{Gerhard1993}).

\subsubsection{Gauss-Hermite fourth moment ($h_{4}$)}

We derive the luminosity-weighted $\langle h_{4} \rangle_{e}$ values within the effective radius, and list them in Table \ref{table2}. We also compare our measurements to MASSIVE in Table \ref{table:comparison} for our three galaxies in common, and find that our values are slightly higher. But similar to \citet{Veale2017a} and \citet{Loubser2020}, we find positive $\langle h_{4} \rangle_{e}$ values for all our galaxies. These positive values can either have a physical origin, e.g.\ a superposition of components with different LOSVDs (see \citealt{Gerhard1993, Bender1994, Baes2005}), or can be the result of an observational bias or template mismatch, e.g.\ in galaxies with a young stellar population component (see discussion in section 3.2.7 of \citealt{VandeSande2017}, and section 3.2 in \citealt{Loubser2020}). Given the current unknown nature of the positive $h_{4}$ values, we do not interpret them further.

\subsubsection{Gauss-Hermite third moment ($h_{3}$)}

Correlations between the $h_{3}$ and the line-of-sight velocity ($V$) in spatially-resolved measurements allow us to probe the eccentricity of the stellar orbits \citep{Bender1994, Bureau2005}. 

\begin{figure}
   \includegraphics[scale=0.33]{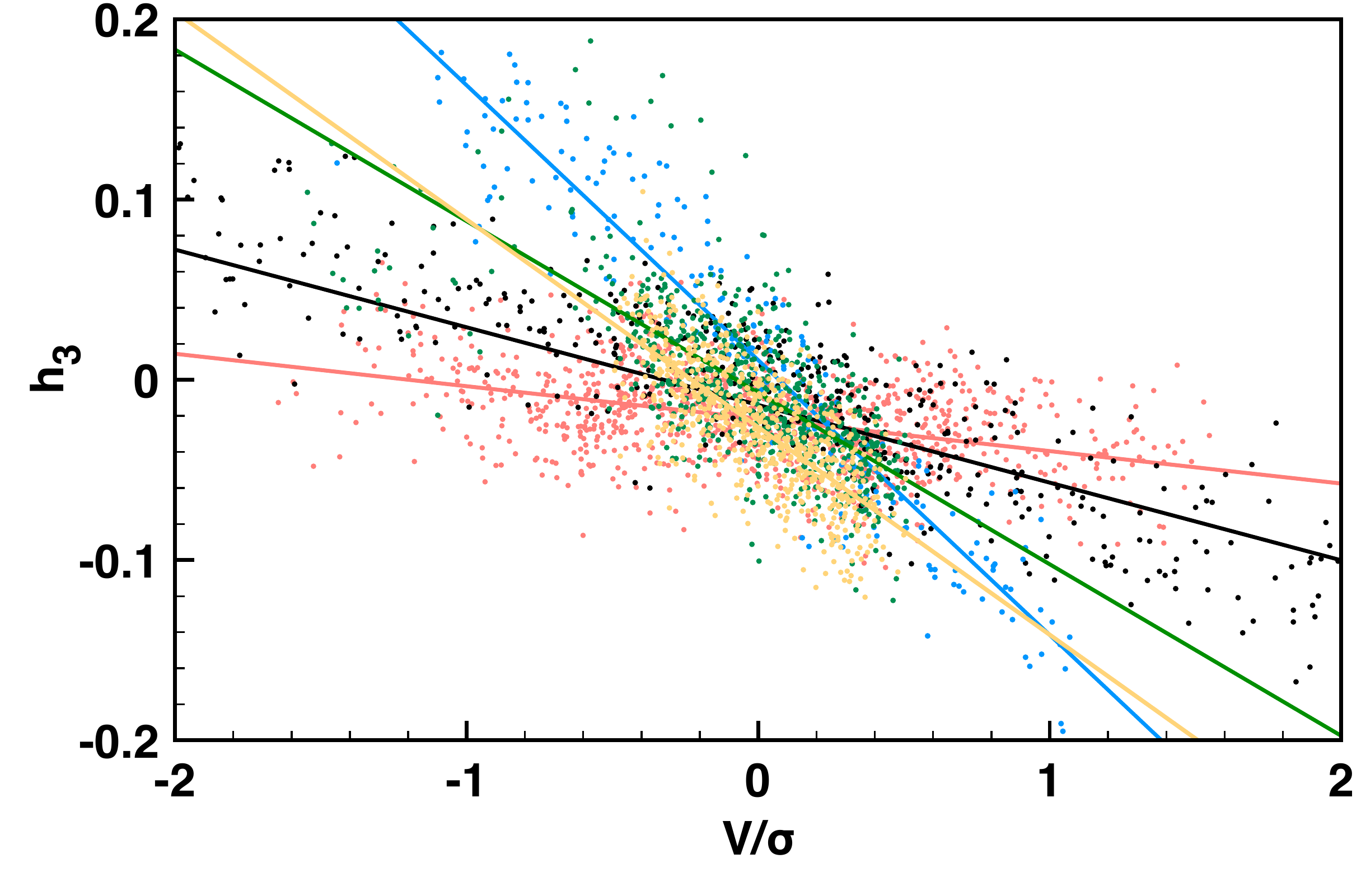}\\
   \includegraphics[scale=0.33]{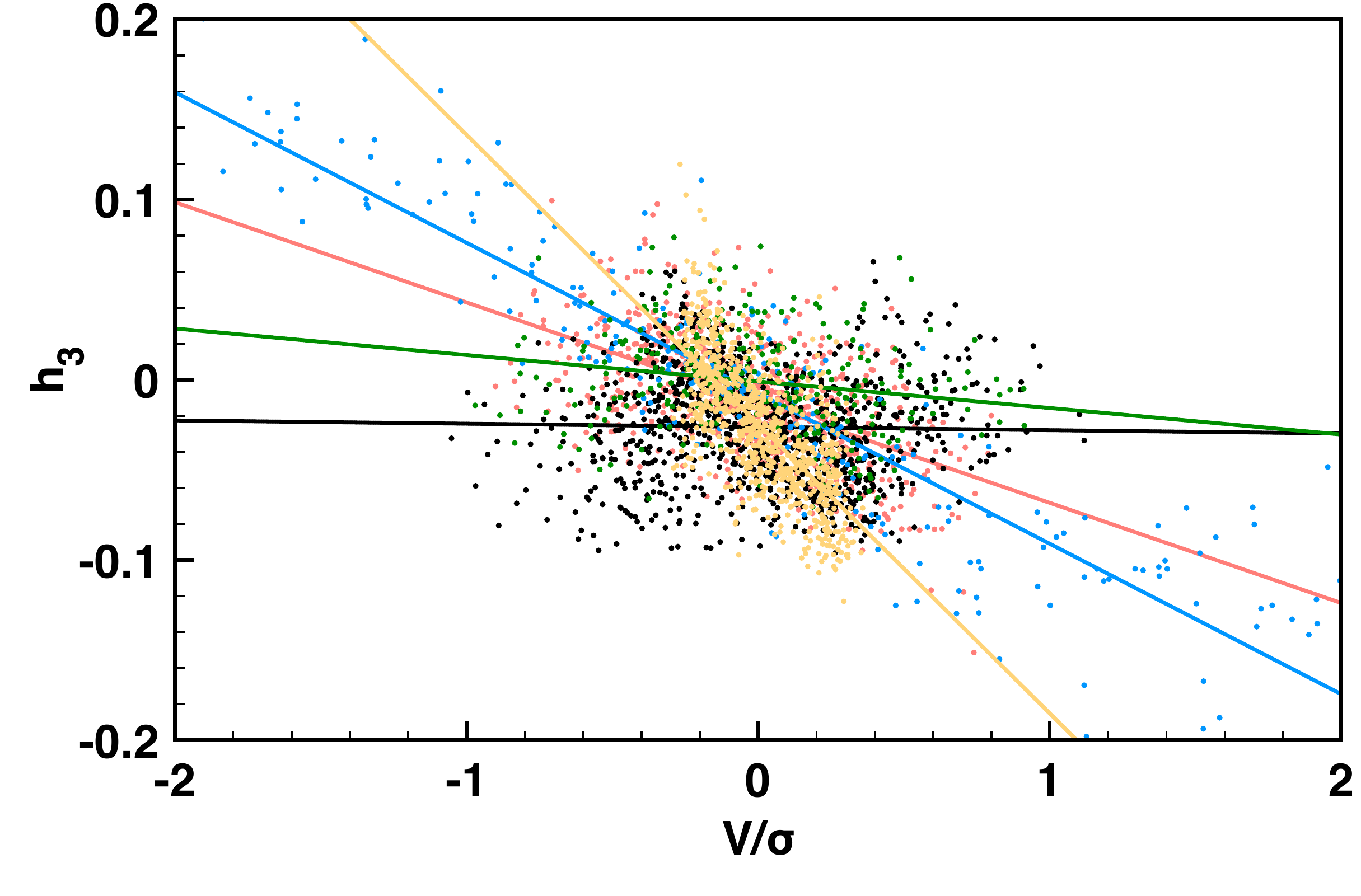}\\
   \caption{Spatially-resolved $V/\sigma$ and $h_{3}$ measurements from each of the fast rotators in our sample (coded by colour). Top: Red -- NGC 584, Black -- NGC 924, Blue -- NGC 940, Green -- NGC 978, Yellow -- NGC 1453. Bottom: Red -- NGC 1587, Black -- NGC 4008, Blue -- NGC 4169, Green -- NGC 6658, Yellow -- NGC 7619.}
\label{fig:h3_first}
\end{figure}

Figure \ref{fig:h3_first} show how the spatially-resolved $V/\sigma$ and $h_{3}$ measurements are anti-correlated within each of the fast rotators in our sample (coded by colour), as expected for galaxies with embedded disk-like components \citep{Naab2001, Jesseit2007, Hoffman2009, Hoffman2010, Naab2014, Schulze2018}. For each galaxy, the straight line shows our best fit to the relation between $V/\sigma$ and $h_{3}$ (for a flux-weighted parametrisation of the slope, see $\xi_{3}$ below). Such an anti-correlation (with a slope of approximately --0.1) is expected from projection effects \citep{Bender1994}. Similar to other studies (e.g. \citealt{Krajnovic2008, Krajnovic2011, Veale2017a, VandeSande2017}), we also find weaker anti-correlations between the $V/\sigma$ and $h_{3}$ measurements for our slow rotators, but steeper than for the fast rotators. 

\subsubsection{Kinematic asymmetry parameter ($\xi_{3}$)}

\begin{figure}
\centering
   \includegraphics[scale=0.33]{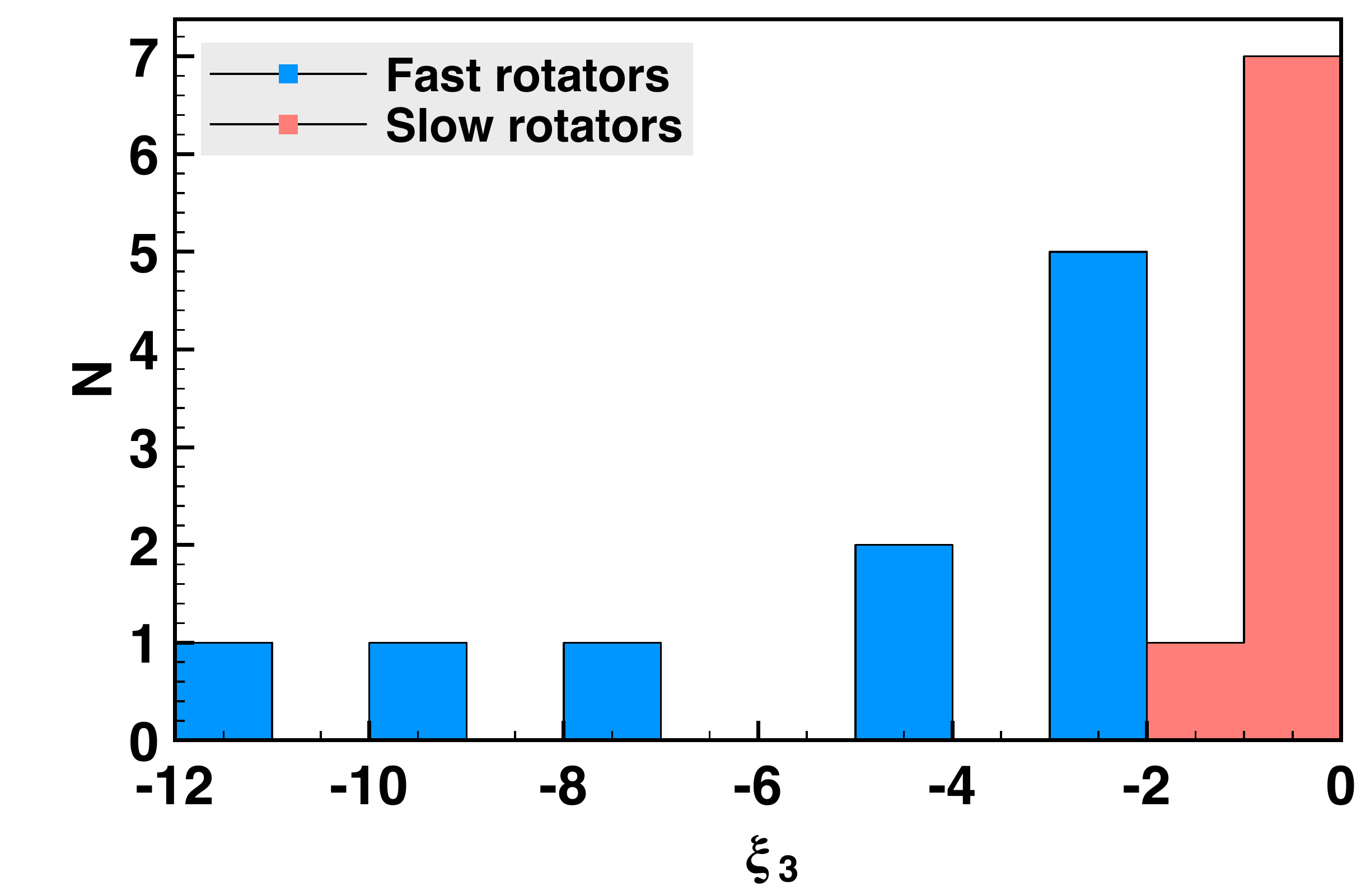}
   \caption{Distribution of the kinematic asymmetry parameter ($\xi_{3}$) for our fast (blue) and slow (red) rotators.}
\label{fig:histo}
\end{figure}

We quantify the differences between the $h_{3}$ and $V/\sigma$ anti-correlations of the fast and slow rotators shown in Table \ref{table2}. Following \citet{Frigo2019}, we use:

\begin{equation}
\xi_{3} = \frac{\Sigma_{i}F_{i}h_{3,i}(V_{i}/\sigma_{i})}{\Sigma_{i}F_{i}h_{3,i}^{2}}
\end{equation}
where $i$ in this case is each spatial bin (summed out to $R_{e}$). This global parameter measures the slope of the $h_{3} - V/\sigma$ anti-correlation, and when $h_{3}$ and $V/\sigma$ are fully anti-correlated, the correlation is given by $h_{3} = (1/\xi_{3})V/\sigma$ \citep{Krajnovic2020}.

\citet{Frigo2019} present examples of various $h_{3}$ and $V/\sigma$ distributions with and without correlations, and their corresponding $\xi_{3}$ parameters. Fast-rotating galaxies have $\xi_{3} < -3$, while slow rotators have $\xi_{3}$ close to 0. From Figure \ref{fig:histo}, we can see a clear difference between the $\xi_{3}$ distributions of our fast and slow-rotating BGEs, but large variations between individual fast-rotating galaxies. The lack of overlap between the two distributions indicates a clear difference of the orientations in the $h_{3} - V/\sigma$ plane. 

%--------------------------------------------------------------------------------------------------------------------

\section{Stellar kinematics position angle and misalignment}
\label{Psi}

In this section, we discuss the misalignment angle ($\Psi$) between the stellar kinematics PA (PA$_{\rm kin}$) and the photometric PA (PA$_{\rm phot}$), and use it as an indicator of galaxy shape for the BGEs. We also investigate the misalignment ($\Psi_{\rm gas}$) between the stellar kinematics (PA$_{\rm kin}$) and the gas kinematics (PA$_{\rm gas}$).

\subsection{Kinematic position angle (PA$_{\rm kin}$) and misalignment angle ($\Psi$) with the photometric position angle (PA$_{\rm phot}$)}

We use the fit\textunderscore kinematic\textunderscore pa\footnote{http://www-astro.physics.ox.ac.uk/$\sim$mxc/software/} routine as described in Appendix C of \citet{Krajnovic2006} to measure the global kinematic position angle, PA$_{\rm kin}$, from the stellar velocity map of each BGE. The routine uses the observed velocity map to generate a model map for each possible PA$_{\rm kin}$ value, then determines the angle that minimizes the $\chi^{2}$ between the observed map and the model map, finding the value of PA$_{\rm kin}$. PA$_{\rm kin}$ is measured East of North (anti-clockwise) to the maximum velocity. We masked close companions or objects in the line-of-sight as indicated in Figure \ref{fig:Flux}, and we use 100 steps to calculate the model. In galaxies where we do not detect large net-streaming motions (the non-regular/slow rotators), the uncertainty on PA$_{\rm kin}$ is large. 

From the photometric position angle PA$_{\rm phot}$ and the kinematic position angle PA$_{\rm kin}$, we find the kinematic misalignment angle ($\Psi$) following \citet{Franx1991}:
\begin{equation}
\sin \Psi = | \sin (\rm PA_{\rm phot} - \rm PA_{\rm kin}) |
\end{equation}
While the range of PA$_{\rm phot}$ is 0$\degr$ to 180$\degr$ and the range of PA$_{\rm kin}$ is 0$\degr$ to 360$\degr$, the misalignment angle by construction, is restricted to be between 0$\degr$ and 90$\degr$. Thus, the misalignment of the two position angles is measured regardless of the sense of the galaxy rotation. Our measurements are listed in Table \ref{table2}. For the uncertainty on $\Psi$, we add the uncertainties on PA$_{\rm phot}$ and PA$_{\rm kin}$ in quadrature. 

\begin{figure}
   \includegraphics[scale=0.33]{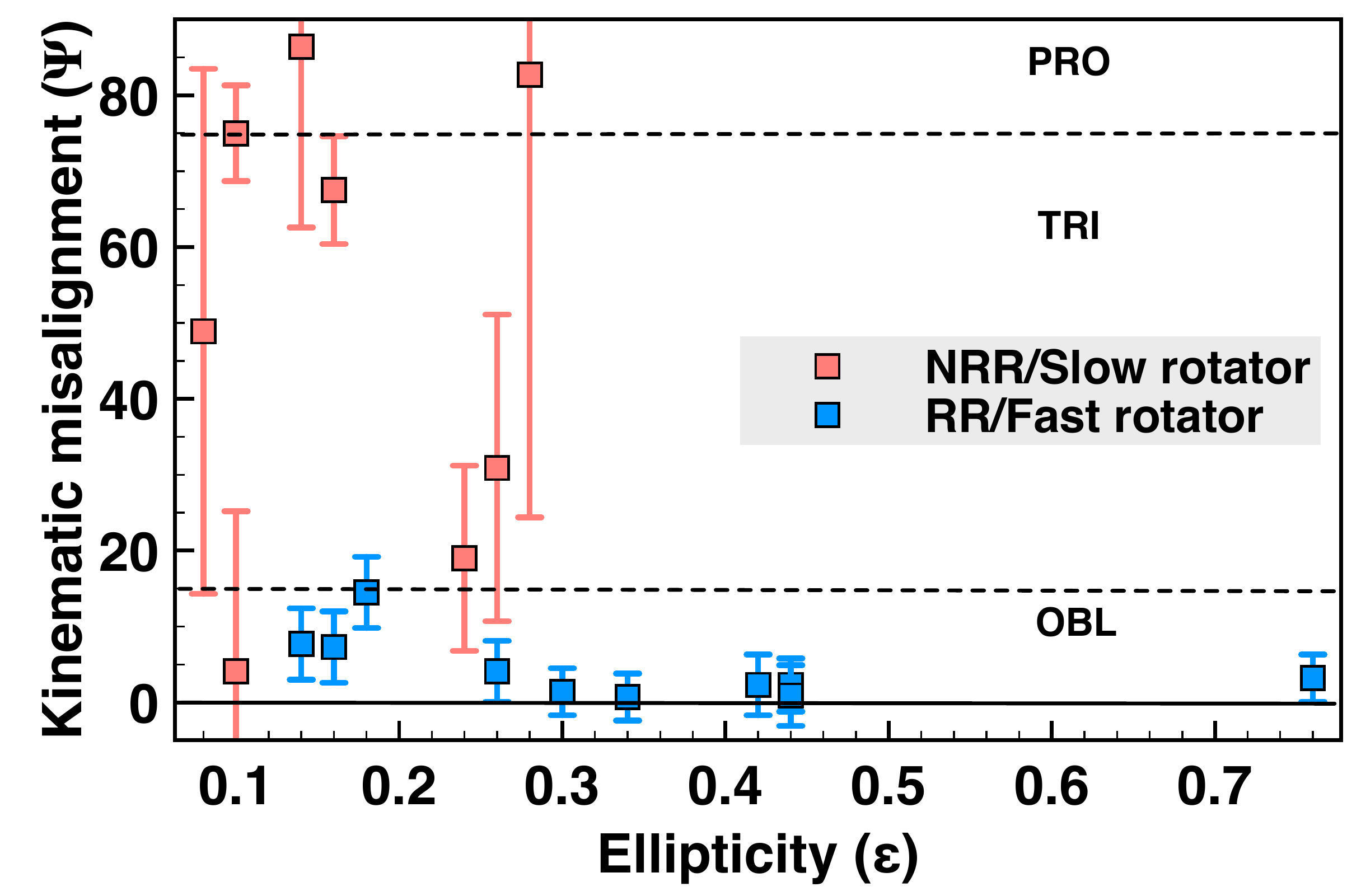}\\
    \includegraphics[scale=0.33]{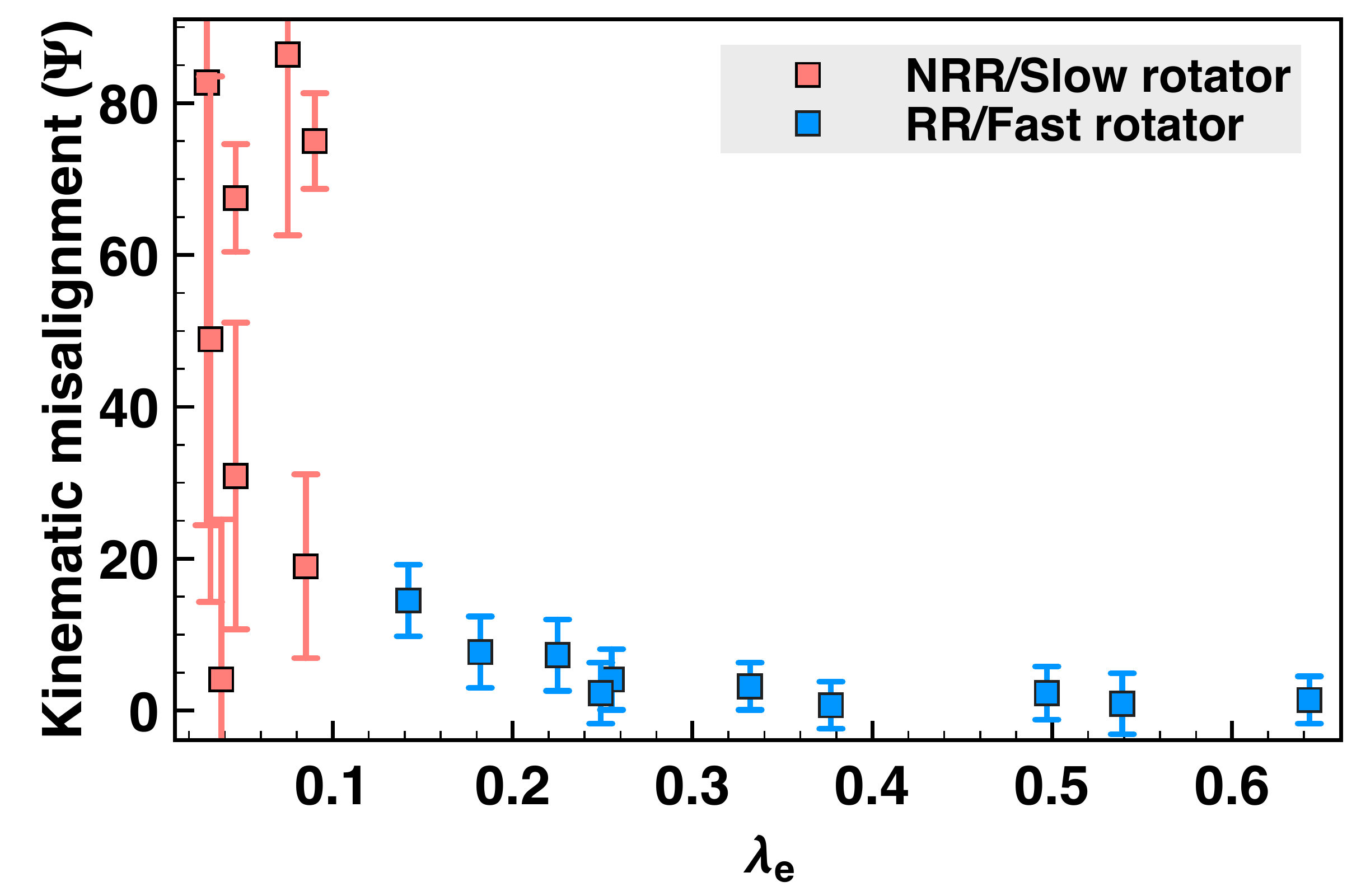}\\
   \caption{Top: Kinematic misalignment angle ($\Psi$) vs ellipticity ($\epsilon$) for the regular rotating (RR, also corresponding to fast rotator) and non-regular rotating (NRR, also corresponding to slow rotator) BGEs. Bottom: Kinematic misalignment angle ($\Psi$) vs $\lambda_{e}$ for the fast (blue) and slow rotating (red) BGEs.}
\label{fig:Psi_eps}
\end{figure}

\subsection{$\Psi$ as an indicator of galaxy shape}
\label{galaxyshape}

Analytical galaxy potentials suggest a correspondence between galaxy intrinsic shape and the intrinsic misalignment angle between kinematic and morphological axes (\citealt{Franx1991}, and subsequent studies).

We present the distribution of $\Psi$ as a function of the galaxy projected ellipticity ($\epsilon$) in Figure \ref{fig:Psi_eps} (top). Following \citet{Krajnovic2018}, we draw two horizontal lines at 15$\degr$ and 75$\degr$ to separate oblate, triaxial and prolate geometries. The figure shows that all BGEs classified as regular rotators (RR, or fast rotators, shown in blue) fall in the oblate category. They are spread over a range of $\epsilon$, and have very well defined PA$_{\rm kin}$ (and therefore $\Psi$). The BGEs classified as non-regular rotators (NRR, slow rotators, shown in red) are all at $\epsilon < 0.3$ (also see Section \ref{lambdaR}), have large uncertainties on PA$_{\rm kin}$, and seem to be spread over all three classes: oblate (1/8), triaxial (4/8), and prolate (3/8). The bottom panel in Figure \ref{fig:Psi_eps} clearly illustrates the difference in the misalignment angle vs $\lambda_{e}$ distributions for the fast and slow rotating BGEs.

Although the recovered intrinsic shape distributions from such methods are not unique \citep{Franx1991}, results showing that all the fast rotators have kinematic axes aligned with photometric axes suggest that they are nearly axisymmetric \citep{Cappellari2007, Emsellem2007, Krajnovic2011, Fogarty2015, Li2018obs}. In this case, the apparent angular momentum coincides with the intrinsic angular momentum, $\Psi$ = 0 \citep{Franx1991}. We find here that all 10 of our fast rotators are aligned ($\Psi < 15 \degr$). \citet{Ene2018} found that 91 per cent of their fast rotators in the MASSIVE sample are aligned. Studies of lower mass early-type galaxies find similar trends e.g. 83 per cent of the fast rotators are aligned in the SAMI survey \citep{Fogarty2015}, and 96 per cent of the fast rotators are aligned in ATLAS3D \citep{Krajnovic2011}. 

For slow rotators, the two axes can be misaligned \citep{Weijmans2014, Foster2017, Li2018obs}, showing the triaxial nature of galaxies \citep{Franx1991}. Triaxial galaxies can exhibit any value of $\Psi$, while prolate galaxies with significant rotation would have $\Psi$ closer to 90 degrees \citep{Schechter1979, Davies1988}. Galaxies with significant misalignment (typically galaxies more massive than 10$^{11}$ M$_{\sun}$, \citealt{Krajnovic2018}) have ``prolate-like" rotation ($\Psi > 75 \degr$, also called ``minor axis" rotation), that is rotation around the major axis.

\citet{Krajnovic2018} find that a third (5/14) of their M3G BCG galaxies show prolate-like rotation ($\Psi > 75 \degr$), whereas studies of the MASSIVE galaxy sample find $< 20$ per cent prolate-like \citep{Ene2020}. \citet{Tsatsi2017} find that 27 per cent of CALIFA galaxies and 23 per cent of ATLAS3D galaxies ($M_{*} > 10^{11.3}$ M$_{\sun}$) show prolate-like rotation. Here 3 out of the 18 BGEs (NGC 193, NGC 677, and NGC 1060) show $\Psi > 75 \degr$ (``minor axis" or ``prolate-like" rotation). Our sample is small in number, and the comparison is complicated by the fact that misaligned galaxies have large uncertainties on $\Psi$, but the fraction of prolate-like galaxies broadly agrees with that from other samples of massive early-type galaxies.

Even though recent works have interpreted the existence of these massive galaxies with 90-degrees misalignment as circumstantial evidence for prolate galaxies (e.g.\ in M3G \citealt{Krajnovic2018}), not all intrinsically prolate galaxies are misaligned. For example, the Illustris simulations found intrinsically prolate galaxies to range from showing no rotation to being kinematically aligned \citep{Li2018sim, Bassett2019}. However, galaxies with prolate-like rotation cannot be oblate spheroids.

\subsection{Misalignment between gas and stars ($\Psi_{\rm gas}$)}
\label{gasvsstars}

The misalignment of cold gas and ionised gas with the stellar rotation in galaxies can suggest clues to the origin of accreted gas. We therefore use the gas kinematic position angle (PA$_{\rm gas}$) from \citet{Olivares2022}, and the stellar PA$_{\rm kin}$ measured here, to derive the misalignment between gas and stars ($\Psi_{\rm gas}$). Both the PA$_{\rm gas}$ and PA$_{\rm kin}$ were measured from North (anti-clockwise) to maximum velocity, and the difference between the two angles is defined as $\Psi_{\rm gas}$ measured between 0$\degr$ (fully aligned) to 180$\degr$ (fully counter-aligned). 

Similar to previous studies \citep{Lagos2015, Davis2016, Bryant2019, Xu2022}, we assume that galaxies with a misalignment angle of less than 30$\degr$ are aligned, galaxies with $30\degr < \Psi_{\rm gas} < 150\degr$ are misaligned, and galaxies with $150\degr < \Psi_{\rm gas} < 180\degr$ could be considered to be counter-rotating. However, this cut-off is rather arbitrary, and in addition, the measurement errors (specifically PA$_{\rm kin}$ for slow rotators) can be large, and the PAs are projected values, together resulting in an uncertainty on the $\Psi_{\rm gas}$ misalignment. For a large sample of SAMI galaxies, \citet{Bryant2019} note that inspection of galaxies with PA offsets between 30 -- 40$\degr$ reveal that some are genuinely misaligned while for others, the errors on the fits mean that they can also be in agreement with a PA offset below 30$\degr$. They suggest to also consider a PA offset cut-off of 40$\degr$ to account for the possible contamination. We present this $\Psi_{\rm gas}$ classification in Table \ref{Auxprop}, and we indicate the two BGEs with $\Psi_{\rm gas}$ between 30 -- 40$\degr$. The exact value of this cut-off (30$\degr$ or 40$\degr$) do not impact the main conclusions of this paper. Our gas/stellar misalignment classifications agree with those of \citet{Olivares2022} within the errors, given that PA$_{\rm kin}$ were independently measured, and from stellar velocity maps constructed using different minimum S/N level spatial bins. 

\section{Discussion: Merger histories from stellar kinematics and multi-wavelength data}
\label{FormationScenarios}

% Summary of fast and slow rotator findings
We discuss the merger histories that can be inferred from the results presented in Sections \ref{results} and \ref{Psi} in Section \ref{mergerhistories} by comparing to results from simulations. In Section \ref{Aux}, we discuss whether this is consistent with the formation histories derived from all the multi-wavelength observations available.

\subsection{Merger histories from stellar kinematics}
\label{mergerhistories}

\subsubsection{Anti-correlations between $h_{3}$ and $V/\sigma$}

\citet{Naab2014} divide the 44 central galaxies from their hydrodynamical simulations into six different classes. Class A and B are fast rotators ($\lambda_{e} \sim 0.3 - 0.6$) that experienced gas-rich mergers, and have a strong anti-correlation between $V/\sigma$ and $h_{3}$. Class C, E, and F are slow rotators ($\lambda_{e} \sim 0.05 - 0.2$) that had late gas-rich mergers or gas-poor minor and/or major mergers, and show a steep relation between $h_{3}$ and $V/\sigma$. Class D consists of galaxies with a late gas-poor merger (intermediate $\lambda_{e} \sim 0.1 - 0.3$), roughly half of which would be classified as fast rotators. These galaxies do not show a strong anti-correlation between $h_{3}$ and $V/\sigma$, or any other signs of embedded disk components (see also \citealt{Naab2001, Jesseit2007}, and discussion in \citealt{VandeSande2017}). \citet{VandeSande2017} do not find evidence for a significant population of fast-rotating galaxies without a strong anti-correlation between $h_{3}$ and $V/\sigma$ in the SAMI survey. Here, we find that all of the $h_{3}$ versus V/$\sigma$ anti-correlation slopes of our fast rotating BGEs are around $\sim$ --0.1 as expected, with the steepest being NGC 940 (--0.15) and NGC 7619 (--0.16). NGC 940 has $\lambda_{e} \sim 0.6$, therefore NGC 7619 (with a fairly low $\lambda_{e} \sim 0.14$) is the only BGE that can potentially be similar to the Class D galaxies in \citet{Naab2014}. 

These anti-correlations between $h_{3}$ and $V/\sigma$ ($\xi_{3} < -3$) are typically found in remnants of gas-rich mergers \citep{Bendo2000, Gonzalez2006, Jesseit2007, Naab2007, Hoffman2009, Rottgers2014} or in simulations without strong AGN feedback (e.g. \citealt{Dubois2016, Frigo2019}). However, it should be kept in mind that the predictions of simulations regarding AGN feedback strongly depend on the particular models adopted by the simulations \citep{Naab2017}, as well as other limiting factors e.g.\ the resolution of the simulations. 

\subsubsection{Galaxy shapes}
\label{simulations}

We find that all 10 of our fast rotators are consistent with an oblate shape. Three slow-rotating galaxies show ``prolate-like" rotation, although the uncertainties on the kinematic position angle are large for the slow rotators. Four of the other slow rotators can be classified as triaxial, and one as oblate.

Simulations suggest galaxy shape depends on merger history, specifically the configuration of the most recent merger \citep{Jesseit2009, Taranu2013, Moody2014, Bassett2019, Lagos2020}. In particular, slow rotators are star-bursting disks at high redshift \citep{Dekel2009, Keres2009}, but the late evolution is dominated by dry mergers \citep{Delucia2007, Kaviraj2015}. This implies the creation of triaxial or prolate-like systems \citep{Krajnovic2018}. \citet{Ebrova2017} identified a few examples of other possible formation channels, but basically, all massive prolate (slow) rotators were created in major mergers during the last 6 Gyr in the Illustris simulation. Other cosmological simulations also agree that many massive galaxies are prolate-like, slow-rotating and formed by major mergers with radial orbits \citep{Li2018sim, Schulze2018}. This is broadly consistent with observations that kinematic misalignment between photometry and kinematics only happens in slow rotators, and that slow rotators are prevalent above a characteristic stellar mass of 2 $\times 10^{11}$ M$_{\sun}$ \citep{Krajnovic2011, Emsellem2011, Cappellari2013, Li2018obs}. In particular, a late ($z < 1$) dry major merger seems to be essential to lower the angular momentum and create the prolate-like rotation. Irrespective of earlier mergers, the last major merger is the main trigger for the prolate shape \citep{Li2018sim, Lagos2018, Schulze2018, Pillepich2019, Frigo2019, Pulsoni2020}.

\subsubsection{Misalignment between gas and stars}

We indicate whether the stars and ionised gas kinematics are aligned (3/15), misaligned (8/15) or counter-rotating (4/15) in Table \ref{Auxprop}. In general, galaxies can increase their gas supply through either internal or external processes. For BGEs, internal processes can include stellar mass loss in which case one can expect the resulting gas to have the same dynamics as the stars. External accretion of gas can come from accretion of cold gas (e.g.\ \citealt{Serra2014}), hot gas from the outer X-ray halo that cools down \citep{Lagos2015}, or accretion from gas-rich mergers or interactions. In the case of external accretion, the cold gas and the ionised gas may have angular momentum that is misaligned with the stellar kinematics. 

A complicating factor is the time-scale for the gas and stars to (re)align. \citet{Raouf2021} investigate the stellar and gas kinematics of SAMI central group galaxies to evaluate the role of group dynamical states. Their results suggest that the gas accreted following a merger would settle (i.e.\ the gas relaxation) in a way that its rotation axis is aligned with that of the stellar component (co-rotating or counter-rotating) on a time-scale equal or shorter than the time since the last major merger, otherwise the gas rotation axis would deviate from that of the stars. However, this lifetime of misalignment is also dependent on physical properties of the galaxies (e.g.\ morphology and gas fraction), as shown in the analysis of Horizon-AGN simulated galaxies by \citet{Khim2021}.  

\subsection{Merger histories from multi-wavelength data}
\label{Aux}

% Questions
It is widely accepted that fast and slow rotators follow the gas-rich, gas-poor merger evolutionary path. The question is whether this is supported by, or contradictory to, the evolutionary paths inferred from the well-studied radio, X-ray and cold gas properties of these BGEs. We show the main, very diverse, properties from previous studies of the 18 BGEs in Table \ref{Auxprop}, and briefly discuss how they relate to merger histories below. 

\subsubsection*{Radio properties}
In summary, 10/18 galaxies show point-like emission, 3/18 diffuse (extended and amorphous) emission, 2/18 with large-scale and 2/18 with small-scale jet radio morphologies from their 610 and 235 MHz observations \citep{Kolokythas2019}. The four radio jet sources in the sample are all slow rotators. NGC 193 and NGC 4261 are two of the most radio-luminous AGNs in the CLoGS BGE sample \citep{Kolokythas2018, Kolokythas2019}. 

\subsubsection*{X-ray properties}
In our sample, 10/18 galaxies show group, 2/18 galaxy-like, and 3/18 point-like morphology for the hot gas component from 0.3--2.0 keV \textit{XMM-Newton} and \textit{Chandra} X-ray observations \citep{O'Sullivan2017}. The X-ray properties of the slow rotators are fairly consistent. Ten of our 18 galaxies have group-scale X-ray halos, of which seven are slow rotators. This is not unexpected if both properties are more likely to be found in higher mass systems. The only slow rotator that does not have a large X-ray halo is ESO 507-G025\footnote{If we use the approximate correction on $\lambda_{e}$ for the effects of inclination ($\lambda_{e}/\sqrt\epsilon$, see Section \ref{Stelmass}), then ESO 507-G025 is classified as a fast rotator.}. 

Studying the thermodynamical properties of the X-ray atmospheres, \citet{Olivares2022} find that the filamentary structures and compact disks (the morphology of the ionised gas) are found in systems with small central entropy values, short $t_{\rm cool}/t_{\rm ff}$ and $t_{\rm cool}/t_{\rm eddy}$ ratios (NGC 193, NGC 410, NGC 677, NGC 777, NGC 978, NGC 1060, NGC 1587, NGC 4008, NGC 4261, NGC 5846, NGC 7619). They suggest that these ionised gas features, and the associated cold gas, are possibly formed from hot halo gas condensations via thermal instabilities.

In Table \ref{Auxprop}, we also indicate the core-type classification of the groups as cool-core/non-cool-core (CC/NCC) based on their temperature profiles \citep{O'Sullivan2017}. The classification is not available for ESO 507-G025, but all the slow rotators except NGC 777 are hosted by CC groups. In particular, all the slow rotators with cold gas are in CC groups (except ESO 507-G025 which has no hot gas component).

\subsubsection*{Cold gas detections}
The majority (7) of the 10 fast rotators contain cold gas, whereas half (4/8) of the slow rotators also contain cold gas \citep{O'Sullivan2015, O'Sullivan2018}. The presence of cold gas in both kinematic classes suggest different origins for the cold gas i.e.\ deposited during gas-rich mergers or tidal interactions (fast rotators), or cooling from the intra-group medium (IGrM, slow rotators). It should be noted that some BGEs that lack cold gas detections do have ionised gas (H$\alpha$) detections \citep{Olivares2022, Lagos2022}, albeit less bright than the H$\alpha$ in the BGEs with detected cold gas, and deeper observations might reveal some cold gas in these BGEs. 

\subsubsection*{Morphology}

Some studies report evolutionary differences between S0s and Es in groups \citep{Wilman2009, Bekki2011, Deeley2021}. We use the morphological t-type parameter from the HyperLEDA catalog (see \citealt{Makarov2014}), and we indicate the apparent E or S0 morphology for our BGEs in Table \ref{Auxprop}. 

If we consider this morphological classification of the fast and slow-rotating BGEs, most of the BGEs classified as S0 are fast rotators, but two are slow rotators (NGC 1060 and ESO 507-G025, although the latter can also be classified as a fast rotator when inclination effects are taken into account). In general, most S0s are fast rotators (e.g.\ ATLAS3D, \citealt{Emsellem2011}), but slow rotating S0s can be formed through a disruptive merger event \citep{Querejeta2015, Deeley2021}. However, the separation between E and S0 morphologies can be problematic and ill-defined due to, e.g.\ non-homogeneity in the classification process from different data collected in databases, and effects of inclination. Indeed three of the seven S0 galaxies are classified as E/S0. Therefore when connecting our BGEs to different formation scenarios, we use the kinematic classification (fast vs slow rotators), rather than the morphological classification (E vs S0).  

\subsubsection*{Star forming rings}
Previous studies suggest that H$\alpha$ rings are formed from an infall of fresh gas \citep{Mapelli2015}. The ROMULUS simulation analysis of specifically brightest group galaxies by \citet{Jung2022} highlights a case study where a star formation ring can be seen 1.5 -- 2 Gyr after a gas-rich merger. We observe star formation rings in 4/18 BGEs (see \citealt{Lagos2022}, and \citealt{Olivares2022}) as also indicated in Table \ref{Auxprop} (three are fast rotators, and ESO 507-G025 is a slow rotator that can also be classified as a fast rotator when inclination is taken into account). Point-like morphology in X-ray for the hot gas component is found in three of our BGEs with H$\alpha$ rings (NGC 924, NGC 940, and NGC 4169), whereas ESO 507-G025 has no hot gas component, potentially indicating that the cold gas is more likely to be the result of gas-rich mergers or tidal interactions instead of cooling from a hot IGrM. 

\subsubsection*{FUV star formation rate}
Six (of 47 with FUV available) of all the CLoGS BGEs are found to be bluer than [FUV - K$_{\rm s}$] = 8.8, and were therefore classified as star forming by \citet{Kolokythas2022}. Of the overlap with this study, two of these star-forming BGEs are fast rotators (NGC 924 and NGC 940) and one is a slow rotator (ESO 507-G025) that can also be classified as a fast rotator when inclination is taken into account. The rest of our fast and slow rotators are not star forming according to their classification (although NGC 4169 and 6658 are unknown as their FUV photometry is unavailable).

The fact that the three star forming BGEs (NGC 924, NGC 940, and ESO 507-G025) occupy X-ray faint systems suggest that the cold gas is unlikely to have cooled from the hot IGrM, but was acquired through gas-rich mergers or tidal interactions instead.

\bigskip
Considering all of the above, we discuss the classes of fast and slow rotating BGEs separately below, indicating any inconsistencies between different observations. 

\begin{table*}
\centering
\caption{Main BGE or host group properties from previous multi-wavelength studies \citep{O'Sullivan2015, O'Sullivan2017, O'Sullivan2018, Kolokythas2018, Kolokythas2019}. Non-detections are indicated with a dash. The Richness (R) parameter refers to the richness of the group (see \citealt{O'Sullivan2017}), and M$_{*}$ is described in Section \ref{Stelmass}. We indicate the core-type classification of the groups as cool-core/non-cool-core (CC/NCC) based on their temperature profiles \citep{O'Sullivan2017}, and entries in brackets indicate the groups where only a projected temperature profile with <3 bins is available. The SFR$_{\rm FUV}$ is from \citet{Kolokythas2022}. The four BGEs with star forming (H$\alpha$) rings is from \citet{Olivares2022} and \citet{Lagos2022}, and their names are indicated in bold text. The column entitled ``F or D'' indicates whether the ionised gas has a filamentary (F) or disc (D) morphology \citep{Olivares2022}. We also indicate gas and stellar kinematic alignment (Aligned, Misaligned, or Counter-rotating) as described in Section \ref{gasvsstars}.}      
\label{Auxprop}                              
\begin{tabular}{l c c c c c c c c c c} 
\hline
Name   & E/S0 & R & M$_{*}$ &  X-ray & Core &	Radio	&	Cold gas & SFR$_{\rm FUV}$ & F & gas vs stars  \\
 & & & $\times 10^{11}$ M$_{\sun}$ & Hot gas & & & & (M$_{\sun}$ yr$^{-1}$) & or D & kinematics \\
\hline 
\multicolumn{11}{c}{\textbf{Fast rotators}} \\
NGC 584 &  E &  4 & 1.31 &  none &  --	& point &  	HI & 0.018 & F & Misaligned (30.5$\degr$) \\
\textbf{NGC 924} &  S0 &  3 & 1.52 & -- & --	& point  	&	CO+HI  & 0.152 & D & Aligned \\
\textbf{NGC 940} &  S0 &  3 & 2.43 & point & --	& point  &	CO+HI & 0.195 & D & Aligned \\
NGC 978 &  E/S0 & 7 & 1.79 & galaxy & (NCC)  &	 point 	&	--  & 0.045 & D & Aligned \\
NGC 1453  &  E & 4 & 4.64 & GROUP & NCC  & 	point	 &	HI & 0.058 & D & Misaligned \\
NGC 1587 &   E &  4 & 2.88 & GROUP & NCC  &	diffuse	 &	CO+HI & 0.038 & F & Misaligned \\
NGC 4008  &  E & 4 & 1.85 &  galaxy & (NCC)  & 	point	 &	--  & 0.020 & F & Counter-rotating \\
\textbf{NGC 4169} & S0 &  4 & 1.59 &   point & --  &	point	 &	CO+HI & -- & D & Counter-rotating \\
NGC 6658 &  S0 & 4 & 1.35 &  point & --  &	--	&	--  & -- & -- & -- \\
NGC 7619 & E &  8 & 3.83 &  GROUP &	CC  & point &		HI & 0.058 & F & Misaligned (37.5$\degr$) \\
\hline
\multicolumn{11}{c}{\textbf{Slow rotators}} \\
\textbf{ESO 507-G025} & E/S0 & 4 & 2.15 & 	-- & --  &	diffuse	&	CO+HI  & 0.221 & D & Misaligned \\
NGC 193 &  E &   7 & 2.02 &  GROUP & CC  &	jet/large &	--  & 0.047 & F & Counter-rotating \\
NGC 410 & E &   6 & 6.23  &    GROUP & CC  &	point &		-- & 0.117 & F & Counter-rotating \\
NGC 677 &   E &  7 & 2.53 &  GROUP & CC &	diffuse &		HI  & 0.068 & F & -- \\
NGC 777 &   E &  5 & 5.35  &  GROUP & NCC	& point &		-- & 0.134 & F & -- \\
NGC 1060 &   E/S0 & 8 & 7.71 &  GROUP & CC &	jet/small & 	-- & 0.097 & F & Misaligned \\
NGC 4261 &   E &  7 & 4.64 &  GROUP & CC  &	jet/large &	CO  & 0.092 & D & Misaligned \\
NGC 5846  &  E & 5 & 3.22 &  GROUP & CC  &	jet/small &	CO  & 0.057 & F & Misaligned \\
\hline                                   
\end{tabular}
\end{table*}

\subsubsection{Fast rotators}

\textit{NGC 584:} This target belongs to an X-ray faint group, indicating that the cold gas is unlikely to be the product of cooling from a hot IGrM, but acquired through gas-rich mergers or tidal interactions instead. This is consistent with our stellar kinematics, in that the BGE is classified as a fast rotator, and the misalignment between the stellar and gas kinematics suggests external accretion of gas.  

\textit{NGC 924:} The BGE is detected in HI and CO, with a high cold gas mass, and the HI line profile is suggestive of a disk. Point-like morphology in both radio and X-ray (hot gas) potentially indicates that the cold gas is more likely to be the result of gas-rich mergers or tidal interactions instead of cooling from a hot IGrM. This is consistent with our fast rotator classification. The star forming ring for this galaxy also indicates a merger event. This is also consistent with the star formation detected in FUV. The stellar and ionised gas kinematics alignment suggest a merger in an earlier epoch.

\textit{NGC 940:} The BGE is one of the most cold gas (H2 as well as CO) rich BGEs in CLoGS. It is also HI-rich and highly FIR-luminous for an early-type galaxy. The HI and H2 line profiles in NGC 940 are suggestive of a disk, and it has a large fraction of dense gas in the disk \citep{O'Sullivan2018}. Again, the point-like morphology in both radio and X-ray (hot gas) potentially indicate that the cold gas is more likely to be the result of gas-rich mergers or tidal interactions instead of cooling from a hot IGrM. This is also in agreement with our fast rotator classification. The star forming ring for this galaxy also indicates a merger event. This is also consistent with the star formation detected in FUV. The BGE is classified as an intermediate disk according to its WISE colours \citep{Jarrett2019, Kolokythas2022}. Similar to NGC 924, the stellar and gas kinematics alignment suggest a merger in an earlier epoch.

\textit{NGC 978:} No cold gas was detected for this BGE, and it is hosted by a NCC group. It is a fast-rotating BGE, and the aligned stellar and gas kinematics suggest a merger in an earlier epoch. 

\textit{NGC 1453:} Even though this target is hosted by an X-ray bright group, the NCC status of the group makes is unlikely that the cold gas originated from IGrM cooling. This is consistent with the stellar kinematics (fast rotation, and misalignment) which point towards a recent gas-rich merger origin.  

\textit{NGC 1587:} Similar to NGC 1453,  this is hosted by an X-ray bright but NCC group, making it unlikely that the cold gas originated from IGrM cooling. Again this is consistent with the stellar kinematics (fast rotation, and misalignment) which point towards a recent gas-rich merger origin. 

\textit{NGC 4008:} No cold gas was detected in this BGE. Stellar kinematics suggest a gas-rich merger (fast rotation), with the merger possibly occurring in earlier epoch (stellar and gas kinematically aligned).  

\textit{NGC 4169:} The point-like morphology in both radio and X-ray (hot gas) indicates that the cold gas is more likely to be the result of gas-rich mergers or tidal interactions instead of cooling from a hot IGrM. This is also consistent with our fast rotator classification. The H$\alpha$ ring for this galaxy also indicates a merger event. \citet{O'Sullivan2015, O'Sullivan2018} finds NGC 4169 to be one of only two (out of 53) BGEs consistent with the far-infrared version of the star formation main sequence. The stellar and gas kinematics alignment suggest a merger in an earlier epoch.

\textit{NGC 6658:} No cold gas was detected for this BGE. It is a fast rotating BGE, suggesting a gas-rich merger. 

\textit{NGC 7619:} This target is also hosted by an X-ray bright group, and the only CC group among the fast rotators with core-type classification available, making it plausible that the cold gas could originate from IGrM cooling. However, the stellar kinematics (fast rotation, and misalignment) point towards a recent gas-rich merger. Cold gas from the IGrM, and from interaction with companions do not necessarily have to be mutually exclusive. 

\medskip

Given the complexity of the processes that take place in the centres of groups, we find very consistent merger histories inferred from different observational evidence for the fast-rotating BGEs. The possible exception is the three fast-rotating BGEs with star formation rings, where the rings may indicate a recent/ongoing transformation, yet the stellar and ionised gas kinematics are aligned. However, \citet{Bryant2019} argue that not all externally accreted gas must necessarily accrete with an angular momentum axis misaligned to the stars, and equally, not all misaligned gas is necessary recently accreted because gas can form stable misaligned orbits.

For NGC 7619, hosted by a group where the thermodynamic properties are conducive to cooling from the IGrM, cold gas contribution from external sources (such as galaxy satellite interactions or gas-rich mergers) can not be discarded (see also discussion in \citealt{Olivares2022}). Overall, the observation evidence seem consistent that the fast-rotating BGEs experienced gas-rich mergers, and these are very likely the origin of the cold gas as well.

\subsubsection{Slow rotators}

\textit{ESO 507-G025:} ESO 507-G025 has a relatively rich gas disk, and is the only slow rotator with a star forming ring \citep{Lagos2022, Olivares2022}. ESO 507-G025 is detected in CO and is the HI-richest BGE in the CLoGS sample, with a double-peaked line profile indicative of a rotating disk. ESO 507-G025 is also a relatively FIR-luminous galaxy, and actively star forming from FUV photometry (see discussion in \citealt{Kolokythas2022}). It is hosted by an X-ray faint group, indicating that the cold gas is unlikely to be the product of cooling from a hot IGrM. It is likely that this BGE is incorrectly classified as a slow rotator, e.g.\ a disky/oblate galaxy that is close to face-on that appears to be a slow rotator simply because most of the rotational motions are in the plane of the sky, and that it gained some star formation after a gas-rich merger (as also suggested by the star forming ring). If we use the approximate correction on $\lambda_{e}$ for the effects of inclination ($\lambda_{e}/\sqrt\epsilon$, see Section \ref{Stelmass}), then this BGE is indeed classified as a fast rotator, consistent with all the other observed properties.

\textit{NGC 193:} No cold gas was detected. It is hosted by an X-ray bright group, and has strong radio jets (that are misaligned with the rotation axis of the stars). Its slow rotator classification indicates a gas-poor merger, while the aligned stellar and gas kinematics possibly indicates an earlier merger.  

\textit{NGC 410:} No cold gas was detected, even though it is hosted by an X-ray bright (and CC) group, and IGrM cooling could have been possible. Its slow rotator classification indicates a gas-poor merger, while the aligned stellar and gas kinematics indicates an earlier merger.   

\textit{NGC 677:} The X-ray brightness and thermodynamic properties indicate that the cold gas likely cooled from the IGrM. This is consistent with the slow rotator classification indicating a gas-poor merger.   

\textit{NGC 777:} No cold gas was detected, even though it is hosted by an X-ray bright group, it is a NCC group. The slow rotator classification suggests a gas-poor merger, consistent with the absence of gas. 

\textit{NGC 1060:} It is hosted by an X-ray bright group, and has radio jets (that are misaligned with the rotation axis of the stars), and no cold gas was detected. Misalignment of the stellar and ionised gas kinematics suggests a recent merger. 

\textit{NGC 4261:} It is hosted by an X-ray bright (and CC) group, with very strong radio jets (that are misaligned with the rotation axis of the stars). IGrM cooling is likely responsible for the cold gas (also see \citealt{O'Sullivan2011}). This BGE has a recent ALMA detection of a sub-kiloparsec scale CO disk \citep{Boizelle2021}. At the edge of the Virgo cluster, the BGE also has dust lanes along the major axis \citep{Moelenhoff1987, Mahabal1996}, a disk of cool gas surrounding the nucleus \citep{Jaffe1993}, and disturbed outer isophotes \citep{Bilek2020, Ebrova2021}. The slow rotator classification, as well as the misalignment between the stars and the gas, indicates a possible late, gas-poor merger.

\textit{NGC 5846:} It is hosted an X-ray bright and CC group, with radio jets (that are misaligned with the rotation axis of the stars). This BGE shows convincing evidence that the molecular gas has cooled from the IGrM \citep{Schellenberger2020}. The system hosts H$\alpha$ filaments resembling those in strong cool-core galaxy clusters, with the molecular gas located within these nebulae, with no sign of rotation or streaming motions (see discussion in \citealt{Kolokythas2022}). This is consistent with its slow rotator classification. There is some evidence that NGC 5846 experienced a recent interaction with its companion, NGC 5846A, or other close companions in the galaxy group \citep{Mahdavi2005}, as well as an increasing fraction of ex-situ stars beyond 1$R_{e}$ \citep{Davison2021}. The stellar kinematics (slow rotation and misalignment of stellar and ionised gas kinematics) suggest that this was a dissipation-less and recent interaction.  

\medskip

In summary, ESO 507-G025 is most likely intrinsically a fast rotator, classified as a slow rotator due to the effects of inclination. For the other slow rotators with cold gas, all observational evidence point to cold gas cooling from the IGrM. 

%-----------------------------------------------------------------------------------------------------------------------------------------------------

\section{Summary and Conclusions}
\label{Con}

We study 18 MUSE cubes of CLoGS brightest group ellipticals (BGEs) by analysing the voronoi-binned kinematic maps ($V, \sigma, h_{3}, h_{4}, \lambda$). We measure $\lambda_{e}$, a proxy for galaxy specific stellar angular momentum within one effective radius, and use it to classify the BGEs as fast or slow rotators. We quantify the anti-correlation between higher-order kinematic moment $h_{3}$ with V/$\sigma$ (using the $\xi_{3}$ parameter), and the kinematic misalignment angle between the photometric and kinematic position angles (using the $\Psi$ parameter). We discuss predictions from simulations to discern the intrinsic shapes of the galaxies and their possible merger histories. Finally, we place the results into context by combining it with known multi-wavelength properties of the BGEs and host groups. We summarise our findings below:

\begin{enumerate}

\item Through visual inspection, we classify the majority (10/18) of our BGEs as regular rotators (RR), and 8/18 BGEs as non-regular rotators (NRR), in Table \ref{table2}. Using the spin parameter ($\lambda_{e}$), we also classify our BGEs as fast or slow rotators, and we find a one-to-one correspondence between the quantitative classification of slow/fast rotator and the qualitative classification of the kinematic morphology (non-regular/regular rotator). The standard interpretation of this bimodality is that the slow rotator galaxies experience a number of gas-poor mergers that effectively decrease their angular momentum, while for fast rotators gas accretion and gas-rich mergers tend to preserve the angular momentum, e.g.\ \citet{Naab2014}.

\item All our slow rotators are above the characteristic stellar mass of 2 $\times 10^{11}$ M$_{\sun}$, and our results agree with the known result from large samples that the fraction of fast-rotating galaxies decreases towards higher stellar mass. Indeed predictions from cosmological simulations show that slow rotators are produced in merger(s), a process that occurs at all stellar masses, but it is most efficient in transforming galaxies into slow rotators if the merger is dry, which is more likely at higher masses \citep{Naab2014, Rodriguez2016, Penoyre2017, Schulze2018, Lagos2018, vandeSande2019, WaloMartin2020}.

\item We quantify the Gauss-Hermite third moment ($h_{3}$) versus V/$\sigma$ anti-correlation with the $\xi_{3}$ parameter. From Figure \ref{fig:histo}, we can see a clear difference between the $\xi_{3}$ distributions of the fast and slow-rotating BGEs, but large variations between individual fast-rotating galaxies. These anti-correlations between $h_{3}$ and $V/\sigma$ (where $\xi_{3} < -3$ for fast rotators) are typically found in remnants of gas-rich mergers \citep{Naab2014}.

\item We find that all 10 of our fast rotators are aligned between the morphological and kinematical axis ($\Psi < 15\degr$), consistent with an oblate shape. Three galaxies, NGC 193, NGC 677, and NGC 1060 show $\Psi > 75 \degr$ (``minor axis" or ``prolate-like" rotation) although the uncertainties on the kinematic position angle are large for the slow rotators. Four of the other slow rotators can be classified as triaxial, and one as oblate.

\item Using ionised gas kinematic position angles from \citet{Olivares2022}, we also derive the misalignment between the gas and stellar kinematics. We find that 3/15 BGEs are aligned ($\Psi_{\rm gas} < 30\degr$), 8/15 are misaligned ($30\degr < \Psi_{\rm gas} < 150\degr$), and 4/15 can be considered to be aligned but counter-rotating ($150\degr < \Psi_{\rm gas} < 180\degr$). This misalignment does not correspond to classifications of slow and fast rotators, but is rather an indication of the time-scale since gas accretion.  

\item Seven (out of eight) of our slow rotators are among the 10/18 galaxies that have group-scale X-ray halos. Both properties are more likely to be found in higher mass systems. We find that the four radio jet sources in the sample are all slow rotators. 

\item The majority (7) of the 10 fast rotators contain cold gas, whereas half (4/8) of the slow rotators also contain cold gas \citep{O'Sullivan2018}. The presence of cold gas in both classes suggest different origins for the cold gas i.e.\ deposited during gas-rich mergers or tidal interactions (fast rotators), or cooling from the IGrM (slow rotators). 

\end{enumerate}

The different observational evidence seem consistent that the fast-rotating BGEs experienced gas-rich mergers or interactions, and these are very likely the origin of the cold gas rather than cooling from the IGrM (with the exception of NGC 7619, where it can be both). For the slow rotators with cold gas, all observational evidence point to cold gas cooling from the IGrM. 

Even though we find fairly consistent merger histories inferred from different observational evidence for the BGEs, we caution that some properties (e.g.\ morphology, misalignment between gas and stars, or even fast/slow rotation) can not be considered in isolation to infer an evolutionary path, particularly for complex cases such as brightest group galaxies.

\section*{Acknowledgements}

We thank the reviewer, Eric Emsellem, for his thorough and constructive comments, and suggestions that improved the paper. 

This work is based on the research supported in part by the National Research Foundation of South Africa (NRF Grant Number: 120850). PL is supported in the form of work contract (DL 57/2016/CP1364/CT0010) funded by national funds through Funda\c{c}\~ao para a Ci\^encia e Tecnologia (FCT). AB acknowledges support from NSERC (Canada) through the Discovery Grant program. E.O'S. acknowledges support from NASA through XMM-Newton award 80NSSC19K1056. 

This research made use of Astropy,\footnote{http://www.astropy.org} a community-developed core Python package for Astronomy \citep{Astropy2013, Astropy2018}, GIST\footnote{https://abittner.gitlab.io/thegistpipeline} \citep{Bittner2019}, the Voronoi-binning method \citep{Cappellari2003}, the Penalized Pixel-Fitting (pPXF) method \citep{Cappellari2004, Cappellari2017} and fit\textunderscore kinematic\textunderscore pa\footnote{All available at https://www-astro.physics.ox.ac.uk/$\sim$mxc/software/} \citep{Krajnovic2006}. Based, in part, on observations obtained using MUSE (program ID 097.A-0366(A)).

Additionally, SLJ acknowledges the Ngunnawal and Ngambri people as the traditional owners and ongoing custodians of the land on which the Research School of Astronomy $\&$ Astrophysics is sited at Mt Stromlo. Similarly, AB acknowledges the l{\fontencoding{T4}\selectfont\M{e}}\'{k}$^{\rm w}${\fontencoding{T4}\selectfont\M{e}\m{n}\M{e}}n peoples on whose traditional territory the University of Victoria stands, and the Songhees, Equimalt and \b{W}S\'{A}NE\'{C} peoples whose historical relationships with the land continue to this day. 

Any opinion, finding and conclusion or recommendation expressed in this material is that of the author(s) and the NRF does not accept any liability in this regard.

\section*{Data availability}

The data underlying this article will be shared on reasonable request to the corresponding author. The MUSE data underlying this article are available in the ESO science archive facility\footnote{http://archive.eso.org/cms.html} (program ID 097.A-0366(A)).

%%%%%%%%%%%%%%%%%%%%%%%%%%%%%%%%%%%%%%%%%%%%%%%%%%


\bibliographystyle{mnras}
\bibliography{References} 

%%%%%%%%%%%%%%%%%%%%%%%%%%%%%%%%%%%%%%%%%%%%%%%%%%

%%%%%%%%%%%%%%%%% APPENDICES %%%%%%%%%%%%%%%%%%%%%

\appendix

\section{Comparison to MASSIVE}
\label{Section:comparison}

We have NGC 410, NGC 777 and NGC 1060 in common with the MASSIVE sample of \citet{Ma2014}, and we compare our measurements in Table \ref{table:comparison}. We note that the spin parameter $\lambda_{e}$ measurements were made in circular apertures in MASSIVE \citep{Veale2017b}, whereas our measurements are in elliptical apertures. The misalignment angle (Mitchell data) $\Psi_{\rm Mitchell}$ is main-body misalignment, whereas the misalignment angle (GMOS data) $\Psi_{\rm GMOS}$ is just central misalignment (see \citealt{Ene2020}). For two galaxies (NGC 410 and NGC 777), the central kinematic axis is well aligned with the photometric axis, but the main-body kinematic axis is not. In general, we find very good agreement with the previous measurements. 

\begin{table*}
\caption{Comparison with galaxies in common with the MASSIVE sample \citep{Ma2014}. $R_{e}$ (in arcsec, from 2MASS) are from \citet{Ma2014}, and M$_{K}$ is the extinction-corrected total absolute $K$-band magnitude, also from \citet{Ma2014}. PA$_{\rm phot}$ is the photometric position angle in degrees East of North from \citet{Veale2017a}. Slow/fast rotator classifications is from \citet{Veale2017b}. The average $\langle h_{4}\rangle$ within $R_{e}$ is from \citet{Veale2018}. The spin parameter $\lambda_{e}$ (unfolded) within $R_{e}$, PA$_{\rm kin, Mitchell}$ and $\Psi_{\rm Mitchell}$ are from \citet{Ene2018}. PA$_{\rm kin, GMOS}$ and $\Psi_{\rm GMOS}$ are from \citet{Ene2020}.}      
\label{table:comparison}      
\centering                         
\begin{tabular}{l l | c c | c c | c c } 
\hline
  &   & \multicolumn{2}{c|}{NGC 410} & \multicolumn{2}{c|}{NGC 777} & \multicolumn{2}{c}{NGC 1060} \\  
   &  & (MASSIVE) & (This study) & (MASSIVE) & (This study) & (MASSIVE) & (This study) \\
\hline 
$R_{e}$ & (arcsec) & 16.8 &  16.77 &   14.6  & 14.57 &     16.8   &  16.76   \\
M$_{K}$ & (mag) & --25.90 & --25.76 $\pm$ 0.02  & --25.94 & --25.61 $\pm$ 0.02 &  --26.00 &  --25.97 $\pm$ 0.02  \\
PA$_{\rm phot}$ & (deg) & 34.9 & 40.0 $\pm$ 3.0  &  148.4 & 145.0 $\pm$ 3.0 & 74.0 &   70.0 $\pm$ 3.0   \\
$\langle h_{4}\rangle$ & & 0.041  & 0.065  &   0.051 & 0.078 &  0.055    & 0.065   \\
$\lambda_{e}$ (unfolded) & & 0.048 & 0.046 &  0.060 & 0.064 &    0.048  &   0.039  \\
Rotation &  & Slow rotator &  Slow rotator  &  Slow rotator & Slow rotator & Slow rotator &  Slow rotator     \\
PA$_{\rm kin, Mitchell}$ & (deg) & $161\pm19$ & $\downarrow$ &   $8.0\pm10.0$ & $\downarrow$ &   $342.0\pm13.5$   &  $\downarrow$  \\
PA$_{\rm kin, GMOS}$ & (deg) & $211\pm9$  & 189.1 $\pm$ 20.0  &   $311\pm22$ & 329.1 $\pm$ 20.9 &  $351\pm10$   & 343.6 $\pm$ 23.6  \\
$\Psi_{\rm Mitchell}$ & (deg) & 53.9  & $\downarrow$ &   39.6 & $\downarrow$ &      88.0   &     $\downarrow$        \\
$\Psi_{\rm GMOS}$ & (deg) & $4.8\pm9.3$ & 30.9 $\pm$ 20.2  &  $18.1\pm21.5$ & 4.1 $\pm$ 21.1 &   $83.8\pm10.0$   &   86.4 $\pm$ 23.8  \\
\hline
\end{tabular}
\end{table*}

\section{Kinematics maps}
\label{OtherKINmaps}

\begin{figure*}
\centering
\includegraphics[width=6.2cm,height=5cm]{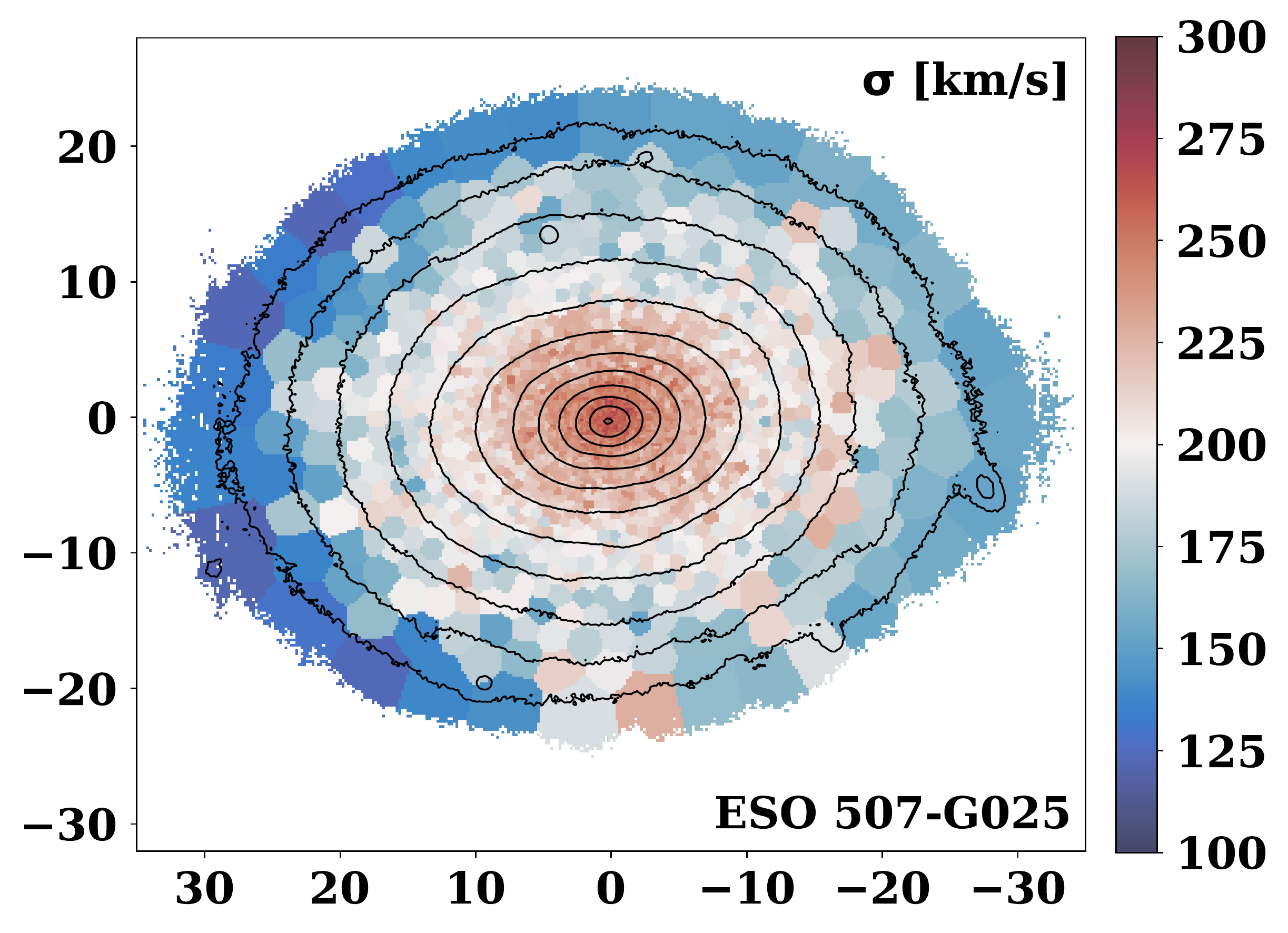}
\includegraphics[width=6.4cm,height=5cm]{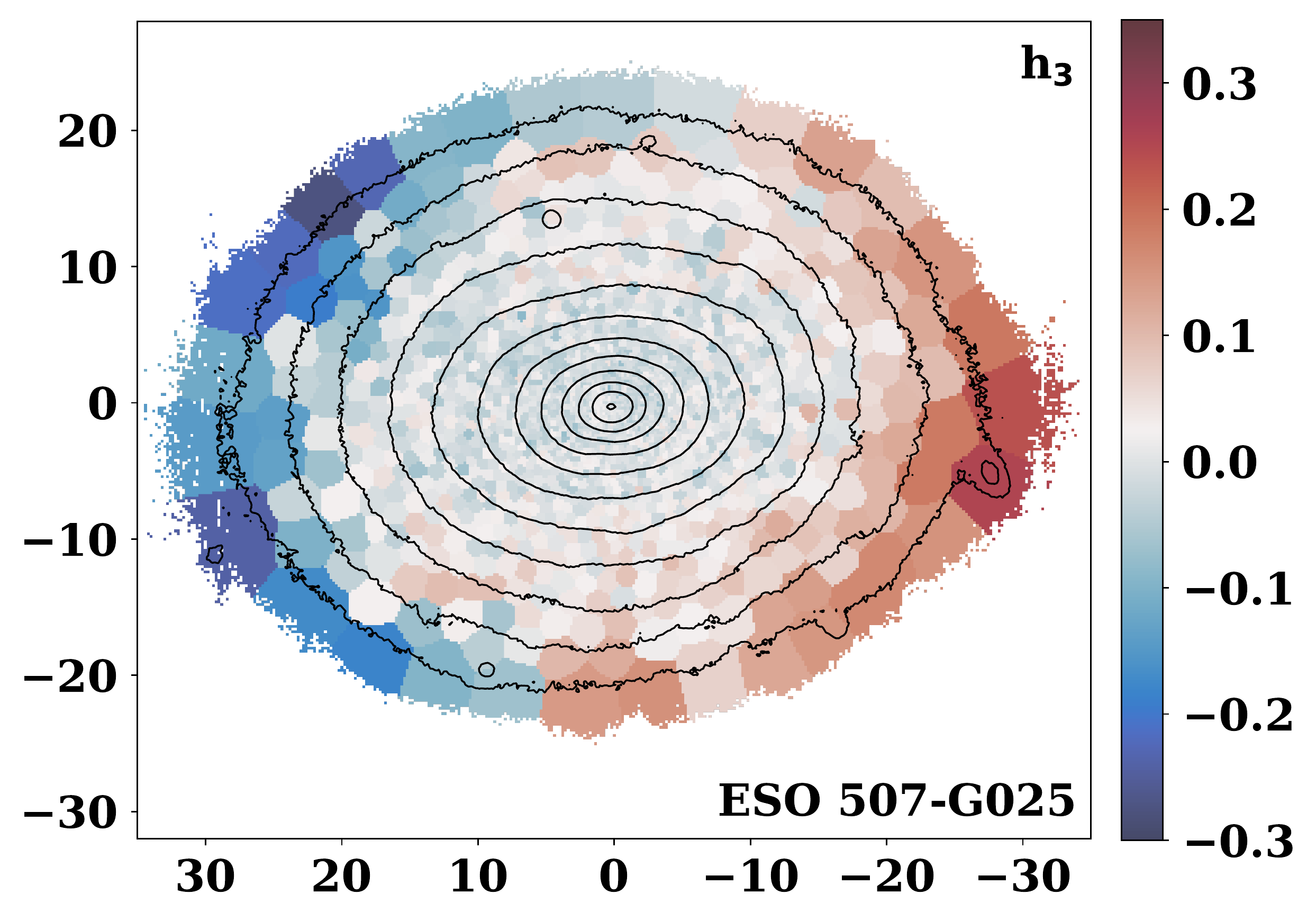}\\
\includegraphics[width=6.2cm,height=5cm]{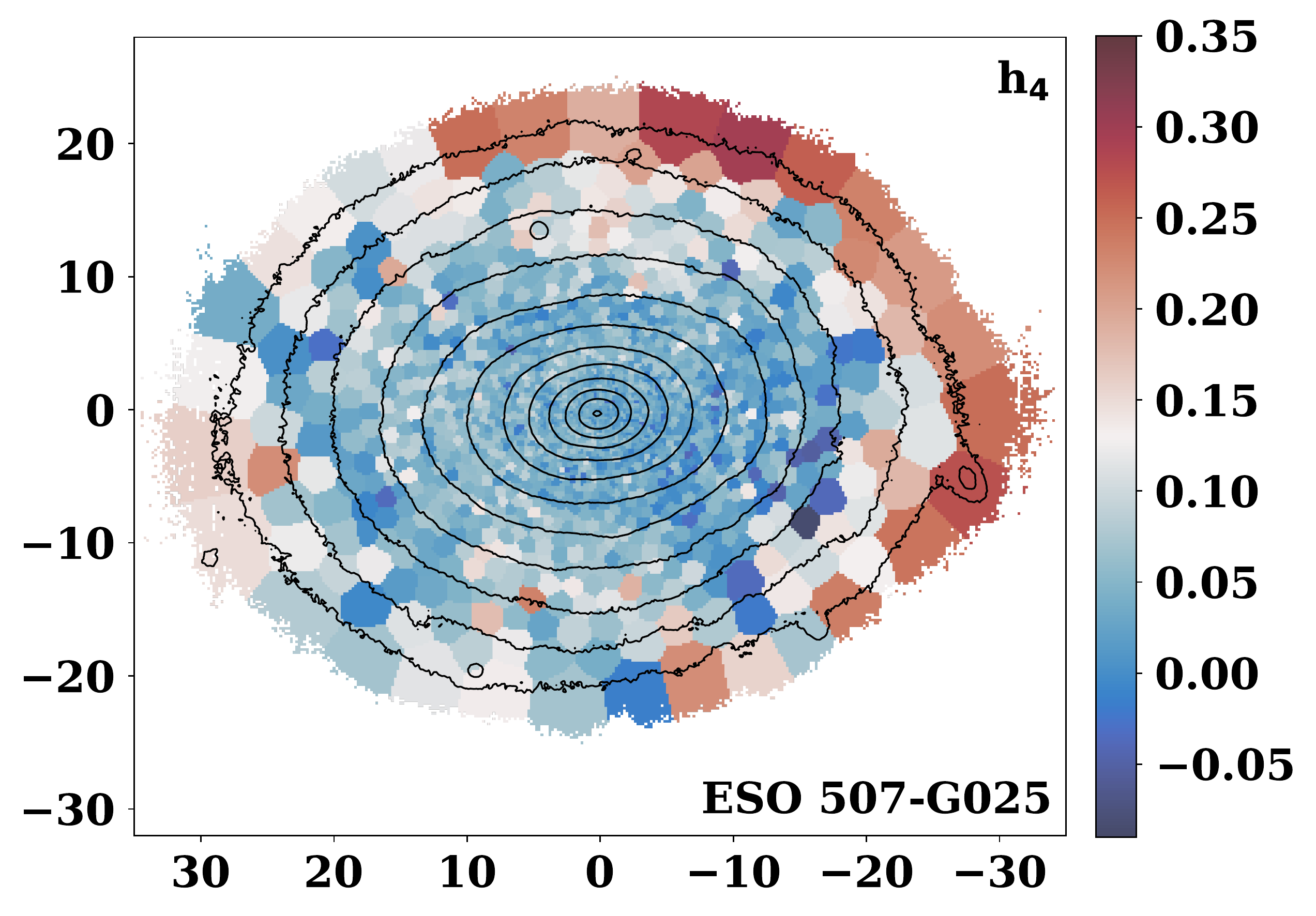}
\includegraphics[width=6.2cm,height=5cm]{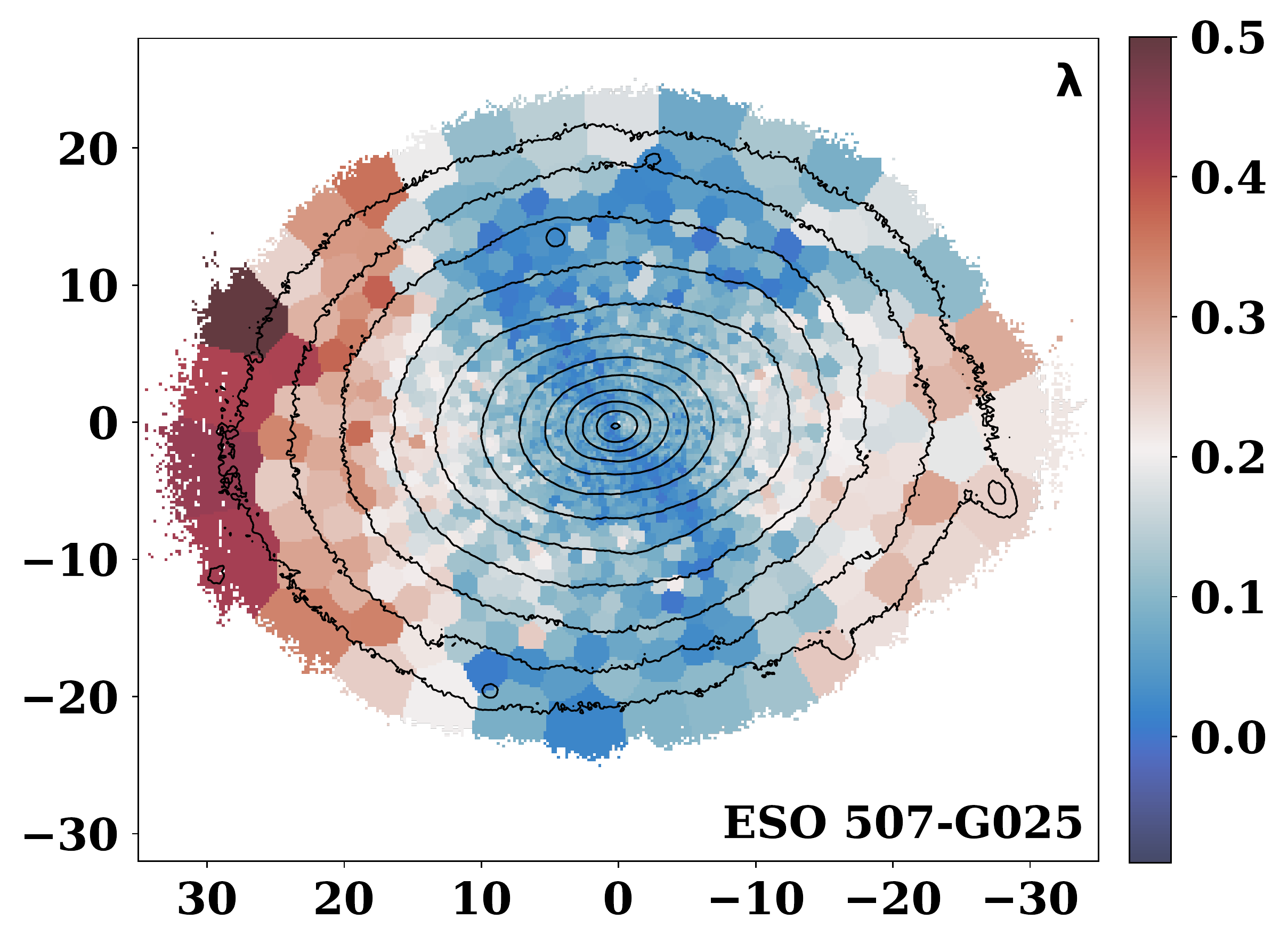}\\
\vspace{-10pt}
   \caption{ESO 507-G025 maps of kinematic measurements (velocity dispersion $\sigma$, Gauss-Hermite third moment $h_{3}$, Gauss-Hermite third moment $h_{4}$, and angular momentum $\lambda$). The spatial scale is in arcseconds.}
\label{fig:OtherKin1}
\end{figure*}

We show the kinematic maps ($\sigma, h_{3}, h_{4}, \lambda$) for ESO 507-G025 as an example in Figure \ref{fig:OtherKin1}. The kinematics maps for the other 17 BGEs are  included as supplementary information.

%%%%%%%%%%%%%%%%%%%%%%%%%%%%%%%%%%%%%%%%%%%%%%%%%%

% Don't change these lines
\bsp % typesetting comment
\label{lastpage}

\end{document}